\documentclass[12pt]{article}
\pdfoutput=1
\usepackage{graphics}
\usepackage[dvips]{graphicx}
\usepackage{mathrsfs}
\usepackage{amssymb}
\usepackage{verbatim}
\usepackage{float}
\newcommand{\be}{\begin{equation}}
\newcommand{\ee}{\end{equation}}
\newcommand{\bea}{\begin{eqnarray}}
\newcommand{\eea}{\end{eqnarray}}

\def\4vol{{\int d^4x \sqrt{-g}}}

\def\beq{\begin{equation}}
\def\eeq{\end{equation}}
\def\bitem{\begin{itemize}}
\def\eitem{\end{itemize}}

\newcommand{\nc}{\newcommand}

\nc{\nt}{\tilde{N}}
\nc{\ra}{\rightarrow}
\nc{\lsim}{\begin{array}{c}\,\sim\vspace{-21pt}\\< \end{array}}
\nc{\gsim}{\begin{array}{c}\sim\vspace{-21pt}\\> \end{array}}
\nc{\tnt}{\tilde{N}}
\nc{\tst}{\tilde{t}}

\nc{\LL}{L}
\nc{\vv}{\tilde{v}}

\title{
\vspace*{5mm} \Large\textbf{Warped Radion Dark Matter}
\vspace*{1.0cm}
\author{\textbf{An\'ibal D.~Medina~$^a$ and Eduardo Pont\'on~$^b$}\\\\
\normalsize\emph{$^a$Department of Physics, University of California,
One Shields Ave.  Davis, CA 95616~$^a$, USA}\\
\normalsize\emph{$^c$Department of Physics, Columbia University,
538 W. 120th St, New York, NY 10027, USA}}}
\date{\today}
\begin{document}
\setcounter{page}{0}
\maketitle
\begin{abstract}

Warped scenarios offer an appealing solution to the hierarchy problem.
We consider a non-trivial deformation of the basic Randall-Sundrum
framework that has a KK-parity symmetry.  This leads to a stable
particle beyond the Standard Model, that is generically expected to be
the first KK-parity odd excitation of the radion field.  We consider
the viability of the KK-radion as a DM candidate in the context of
thermal and non-thermal production in the early universe.  In the
thermal case, the KK-radion can account for the observed DM density
when the radion decay constant is in the natural multi-TeV range.  We
also explore the effects of coannihilations with the first KK
excitation of the RH top, as well as the effects of radion-Higgs
mixing, which imply mixing between the KK-radion and a KK-Higgs (both
being KK-parity odd).  The non-thermal scenario, with a high radion
decay constant, can also lead to a viable scenario provided the reheat
temperature and the radion decay constant take appropriate values,
although the reheat temperature should not be much higher than the TeV
scale.  Direct detection is found to be feasible if the DM has a small
(KK-parity odd) Higgs admixture.  Indirect detection via a photon
signal from the galactic center is an interesting possibility, while
the positron and neutrino fluxes from KK-radion annihilations are
expected to be rather small.  Colliders can probe characteristic
aspects of the DM sector of warped scenarios with KK-parity, such as
the degeneracy between the radion and the KK-radion (DM) modes.

\end{abstract}

\thispagestyle{empty}
\newpage
\setcounter{page}{1}

\section{Introduction}
\label{Intro}

Models of electroweak symmetry breaking (EWSB) that contain a stable
particle closely tied to the EWSB sector open a window of
complementarity between astrophysical observations, DM searches, and
collider experiments.  Not surprisingly, such a prospect has attracted
considerable attention in the context of supersymmetry, little Higgs
models with T-parity~\cite{Cheng:2003ju} or models with Universal
Extra Dimensions (UED)~\cite{Appelquist:2000nn}.

The latter case is interesting in that the reason the DM candidate
is stable is closely tied to the geometry/compactification of the
extra dimensional space~\cite{Cheng:2002iz,Dobrescu:2004zi}.  In
the 5D case, this is simply due to a $Z_{2}$ reflection symmetry
(usually called KK-parity) about the center of the compactified
fifth dimension.  The viability of the lightest KK-parity odd
particle (the LKP) as a dark matter (DM) candidate was first
studied in~\cite{Servant:2002aq,Servant:2002hb}.  For a review,
see~\cite{Bertone:2004pz}.  An important assumption in the
previous construction is that the extra dimensional spacetime is
flat.  On the other hand, since the seminal work of
\cite{Randall:1999ee}, the presence of a non-negligible spacetime
curvature (although with vanishing 4D effective cosmological
constant), has been understood to have rather important and
interesting consequences.  Unfortunately, in typical
implementations, no stable new particles appear in such warped
scenarios, although it is simple to build models with a stable DM
candidate by simply imposing a $Z_{2}$ symmetry (see
e.g.~\cite{Ponton:2008zv}).  There have also been proposals where
the DM and EWSB sectors share non-trivial
properties~\cite{Agashe:2004ci,Panico:2008bx,Carena:2009yt},
within the framework of warped compactifications, as well as
studies in``warped throats" that appear in type IIB string theory
on Calabi-Yau manifolds where a suitable DM candidate can be
obtained~\cite{Frey:2009qb}. Also, Ref.~\cite{Agashe:2009ja} has
considered the possible Sommerfeld enhancements arising from
radion exchange for GUT-motivated DM candidates in warped
scenarios, while stable candidates arising from a dark sector in
warped compactifications have been proposed in
Ref.~\cite{Gherghetta:2010cq}.

It is nevertheless possible, and not difficult, to build warped
scenarios with a KK-parity symmetry, similar to UED models.  The first
attempt was done in~\cite{Agashe:2007jb}, where it was proposed to
glue together two copies of the RS model along either the UV or IR
branes.  This reference considered the possibility that the first
KK-parity odd excitation of the $Z$-gauge boson could be the DM
candidate.  Nevertheless, it is well known that the lightest state
beyond the SM in such constructions is expected to be the radion
field~\cite{Goldberger:1999uk}.  If this field has a KK-parity odd
excitation, it is more likely that this ``KK-radion'' will be the LKP.
A concrete model that stabilizes the interbrane distance, and hence
gives mass to the radion field, while at the same time leading to the
existence of a KK-parity symmetry, was proposed in
Ref.~\cite{Medina:2010mu} (see \cite{McDonald:2009md} for an
alternative proposal).  Indeed, it is found that generically the LKP
is the first KK-radion mode.

In this work, we explore the viability of the KK-radion as a DM
candidate.  Given that the radion is such an intrinsic feature of
extra-dimensional scenarios, having it constitute most of the matter
density of the universe would be a rather pleasing possibility.  The
KK-radion mass is found to be parametrically lighter than the scale of
the typical Kaluza-Klein (KK) resonances, which is set by
$\tilde{k}_{\rm eff} \equiv k_{\rm eff}\, e^{-A(L)}$, where $k_{\rm
eff}$ is the curvature at the (IR) boundaries of the metric background
$ds^{2} = e^{-2A(y)} \eta_{\mu\nu} dx^{\mu} dx^{\nu} - dy^{2}$.  In
models where the KK scale is of order TeV, so that they can be tested
at the LHC, the LKP mass is expected to be around the EW scale.  Like
the radion field itself, it couples to the SM particles (and to other
KK excitations) via higher-dimension operators, suppressed by the
``radion decay constant''~\cite{Medina:2010mu}
\bea
\Lambda_r = \sqrt{\frac{3 M_5^3}{k^{3}_{\rm eff}}} \times \tilde{k}_{\rm eff}~,
\label{rdecayconst}
\eea
where $M_{5}$ is the 5D (reduced) Planck mass.  We note that, from a
theoretical point of view, $\Lambda_{r}$ can take on a large range of
values by varying the ratio $M_{5}/k_{\rm eff}$, while keeping
$\tilde{k}_{\rm eff} \sim {\rm TeV}$.  In fact, it was pointed out in
Ref.~\cite{Medina:2010mu} that varying the ratio $M_{5}/k_{\rm eff}$
allows to interpolate between RS-like ``strong warping'' scenarios
(when $k_{\rm eff} \sim M_{5}$) and ``UED-like'' scenarios (when
$k_{\rm eff} \ll M_{5}$).  Although the former case may be considered
more natural, we will also consider the second possibility, since it
gives an interesting deformation of UED models, that shares properties
of pure UED and the RS model.  Furthermore, this is also an
interesting limit from the point of view of the KK-radion relic
density.

As is well-known, obtaining the correct relic abundance of an EW scale
thermal relic requires that its interactions be of typical EW size.
Thus, unless $\Lambda_{r}$ is sufficiently close to the EW scale, the
KK-radion interactions would be too weak, the KK-radion would
freeze-out rather early in the history of the universe, and the relic
density following from the thermal freeze-out paradigm would be too
large and lead to the ``overclosure'' of the universe.  As we will show,
however, the scenario is nevertheless viable over a wide range of
$\Lambda_r$.  First, it turns out that the self-annihilation cross
section for KK-radions can be sufficiently large even when $\Lambda_r$
is in the multi-TeV range.  Second, the presence of other relatively
light colored resonances can lead to a sufficiently strong depletion
of KK-radions through coannihilation effects, and make the scenario
viable for $\Lambda_r$ as large as ${\cal O}(10^{4}~{\rm TeV})$.
Mixing with a KK excitation of the Higgs can have a similar effect.
For much larger values of the radion decay constant, it is possible to
imagine that the KK-radions never reached thermal equilibrium in the
early universe.  This is the case in the ``UED-like'' (or
``small-warping'') scenario mentioned above.  It is then necessary
that the non-thermal production of KK-radions be sufficiently
suppressed.  This restricts the range of allowed $\Lambda_{r}$ for a
given reheat temperature, $T_{R}$.  For instance, if $T_{R} \sim1~{\rm
TeV}$, one must have $\Lambda_{r} \sim 10^{15}~{\rm GeV}$.

The outline of this paper is as follows.  In Section~\ref{Basic}, we
summarize the relevant properties of the scenario proposed in
Ref.~\cite{Medina:2010mu}.  In Section~\ref{sec:HighT}, we comment on
the expected finite temperature properties of our setup.  In
Section~\ref{sec:DM}, we consider the KK-radion relic density in the
thermal freeze-out scenario, while we consider the effects of
coannihilations in Section~\ref{sec:coannihilation}.  The case of
non-thermal production and the related constraints are treated in
Section~\ref{sec:sWIMP}.  We also comment on the prospects for direct
and indirect DM searches for KK-radion DM, as well as on the collider
signals that can help in distinguishing this scenario from other
possibilities.  We conclude in Section~\ref{sec:conclusions}.  We
provide two appendices where we discuss the KK-radion couplings and
the effects of EWSB (Appendix~\ref{App:RadionFermionGeneral}), and
where we summarize the relevant cross sections and decay rates
(Appendix~\ref{App:Processes}).

\section{Models with Warped KK-Parity}
\label{Basic}

We start by summarizing the salient properties of the scenario
presented in Ref.~\cite{Medina:2010mu}.  This is a 5D model with
the fifth dimension compactified to an interval, and parameterized
by $y \in [-L,L]$.  The size of the extra dimension is stabilized
at tree-level by a bulk (GW) scalar field, $\Phi$, similar to the
Goldberger-Wise proposal~\cite{Goldberger:1999uk}.  However, it is
assumed that the gravitational backreaction of the stabilizing
scalar is non-negligible, and that it is the main source for a
non-trivial warping of the 5D spacetime.  When the scalar
potential satisfies $V(-\Phi) = V(\Phi)$, and allows for a
kink-like background solution $\langle \Phi(y) \rangle = \phi(y)$,
with $\phi(-y) = - \phi(y)$, the associated warp factor,
$e^{-A(y)}$, is found to be even about $y=0$, i.e.~one finds
$A(-y) = A(y)$.  The non-trivial kink configuration is assumed to
be chosen over the trivial vacuum expectation value (VEV),
$\phi(y)=0$, by appropriate potentials on the infrared (IR)
boundaries, at $y = \pm L$.  The background scalar energy density
is concentrated around $y = 0$, thus leading to a dynamically
generated ``UV brane'' at the origin.  In general, this is a
``fat'' brane with thickness governed by the mass parameter
appearing in the bulk scalar potential, although there is a well
defined limit where the UV brane is thin and the background has a
constant, negative curvature.  In this limit, one recovers the
IR-UV-IR configuration considered in Ref.~\cite{Agashe:2007jb}.

For the purpose of the present work, the most important consequence of
the above setup is the existence of a discrete $Z_{2}$ symmetry
(KK-parity), under which the KK modes whose wavefunctions are even
about $y=0$ are assigned KK-parity $+1$, while those whose
wavefunctions are odd about $y=0$ are assigned KK-parity $-1$.  The
resulting KK-parity symmetry is a good quantum number, and the
lightest KK-parity odd excitation (the LKP) is exactly stable.

It is also important to note that a generic feature of the spectrum is
the relative degeneracy between the KK-parity even and KK-parity odd
states of a given 5D field.  The reason is that what differentiates
even from odd states are the properties of the corresponding
wavefunctions, $f_{n}(y)$, at the origin: KK-parity even states
satisfy $f'_{n}(0) = 0$, while KK-parity odd states satisfy $f_{n}(0)
= 0$.  But when these wavefunctions are localized near the IR
boundaries, so that their values are (sometimes exponentially)
suppressed at the origin, only a small deformation is required to
convert an even wavefunction into an odd one.  Therefore, the
associated KK masses are very close.  This observation is particularly
interesting when applied to IR localized (would-be) zero-modes.  We
highlight three important examples.

First, the lightest physical scalar excitation of the 5D metric/bulk
scalar system is a KK-parity even field that is typically referred to
as the ``radion''.  Its wavefunction is given by $F_0(y) \approx
e^{2[A(y)-A(L)]}$.  This state has a mass that is parametrically
lighter than the typical KK scale.  From the argument above, the first
KK-parity odd scalar excitation has a mass that is exponentially close
to the radion mass (at tree-level).  This ``KK-radion'' is expected to
be the LKP, and shares properties closely connected to those of the
radion field.  We will refer to it either as $r'$ of $r_{-}$.

Second, since a solution to the hierarchy problem requires the Higgs
field to be highly localized near the IR boundaries, one finds light
KK-parity even and KK-parity odd $SU(2)$ Higgs doublets.  The
KK-parity even doublet is fully responsible for EWSB (giving rise to a
SM-like Higgs and the would-be Goldstone modes eaten by the $W^{\pm}$
and $Z$).  The KK-parity odd Higgs is ``inert'' in the sense that it
does not participate in EWSB, but it can be parametrically lighter
than the KK scale since, were it not for EWSB, it would be
exponentially degenerate with the KK-parity even Higgs doublet, at
tree-level.  However, we will assume that it is somewhat heavier than
the KK-radion, since it is a generic scalar whose mass is not
protected by any symmetry.  Annihilation of KK-radions into the
KK-parity even Higgs degrees of freedom will be found to play a
relevant role in setting the DM relic density.

Third, fermion zero-modes can be localized either near the ``UV
brane'' (at the center of the fifth dimension) or near the IR
boundaries.  An appealing possibility is to use this localization
properties to understand the observed fermion mass
hierarchies~\cite{Grossman:1999ra,Gherghetta:2000qt}.  Thus, assuming
that the 5D Yukawa couplings are all of the same order, the lighter SM
fermions (which include the RH tau and bottom) are localized in the UV
region, while the top quark is localized closer to the IR branes.  In
particular, one might expect that the RH top is localized close to the
IR boundaries, and therefore its first KK-parity odd excitation is
expected to be the lightest of the fermionic resonances.  How light
this state is depends on how closely localized to the IR branes the RH
top is.  We will refer to this KK-parity odd particle as $t'$.  The
next-to-lightest KK-parity odd particle (NLKP) is either $t'$ or the
inert Higgs doublet above.

It should be pointed out that in the anarchic scenarios, where the 5D
Yukawa couplings are sufficiently large to generate the top mass, with
the $(t,b)$ $SU(2)$-doublet as close to the first two generations as
possible (to minimize potentially large corrections to the
$Zb_{L}\bar{b}_{L}$ coupling), the effects from EWSB are
non-negligible.  We take these into account by diagonalizing the KK
mass matrix in the presence of the EWSB contributions.  A consequence
of this is that the lightest KK-parity odd fermion resonances are
pushed downward in mass (``level repulsion'' in the absence of a
zero-mode for the KK-parity odd subtower).  As a result, we find that
the $t'$ cannot be heavier than about a factor of 4 above the radion
mass, and is typically much closer to $m_{r} \approx m_{r'}$.
However, one might also wish to consider non-anarchic scenarios where
the Yukawa couplings themselves are
hierarchical~\cite{Delaunay:2010dw,Delaunay:2011vv}, and EWSB effects
are much less important.

To summarize, the states lighter than the KK scale, $\tilde{k}_{\rm
eff}$, are the radion, its first KK-parity odd mode (which is stable),
a light KK-parity odd $t'$ not much heavier than the previous two, and
an additional Higgs doublet, on top of the SM field content.  Our main
objective in this work will be to study the viability of this scenario
from the point of view of DM constraints.  If the KK radion is indeed
the LKP, these are strongly dependent on the radion decay constant
given in Eq.~(\ref{rdecayconst}), and we will explore several
possibilities.

\section{High-Temperature Phase Transitions}
\label{sec:HighT}

Before presenting the analysis of the KK-radion relic density, we
comment on the expected high-temperature properties of the present
scenario.  In the well-studied case of the RS background with the
radion stabilized via a Goldberger-Wise scalar, the transition from a
high-temperature ``deconfined phase'' to a low-temperature
``confined'' one has been discussed in
Refs.~\cite{Creminelli:2001th,Randall:2006py,Kaplan:2006yi,Nardini:2007me,Hassanain:2007js,Konstandin:2010cd,Konstandin:2011dr}.
The terminology arises from the AdS/CFT
correspondence~\cite{Maldacena:1997re,Witten:1998zw}, where the phase
transition maps into a deconfinement/confinement phase transition in
the dual, large $N$ gauge theory (this phase transition is strongly
first-order).  The analysis has been performed on the gravity side,
and in the limit that the backreaction of the stabilizing GW scalar is
small, by comparing two finite-temperature solutions obeying the same
boundary conditions on the UV brane (however,
Ref.~\cite{Konstandin:2010cd} considered the case where the
backreaction is large).  These are: the Euclidean RS background, with
an IR brane and Euclidean time (as measured on the UV brane)
compactified on $[0,\beta = 1/T]$; and the AdS-Schwarzschild (AdS-S)
solution, where the IR brane is replaced by a black-hole horizon.  The
free energy associated with the RS solution vanishes as a result of
the requirement that the effective 4D cosmological constant be zero at
the minimum of the radion potential.  The free energy of the AdS-S
solution is then given by $F_{\rm dec} \sim E_{0} - c N^{2} T^{4}$,
where $E_{0} \sim m_{r}^{2} \Lambda_{r}^{2}$ is a vacuum energy
contribution, and the second term corresponds to a plasma of
deconfined ``gluons'' with $N^{2}$ degrees of freedom at temperature
$T$ ($c$ is an order one number).  Recalling that under the
correspondence, $\Lambda_{r} \sim N \tilde{k}$, it follows that the
transition temperature is of order $T_{c} \sim (m_{r}^{2} \,
\tilde{k}^{2})^{1/4}$, where $\tilde{k}$ is the warped-down curvature
scale.

Our setup differs from the above RS scenario in two ways: there is no
``UV boundary'', and the backreaction of the stabilizing dynamics is
not a perturbation (but rather the major source of the warping).
Also, the analog of the AdS-S solution is not known in this setup,
although it is plausible that such a solution --with black-hole
horizons symmetrically located with respect to $y = 0$-- exists, and
has similar properties to the one mentioned above (recall that there
is a limit where our scenario reduces to two copies of RS glued
together at the UV brane).  By analogy, one might expect that the
scaling relations of the previous paragraph also hold in the present
case, with $\tilde{k} \to \tilde{k}_{\rm eff}$.  In addition, below
$T_{c}$, the rate of true-vacuum bubble nucleation is expected to be
suppressed by $e^{-{\cal O}(N^{2})}$.  For
moderate values of $N$, i.e. $\Lambda_{r}$ near the KK scale, one can
then expect that, as the temperature drops below $T_{c}$, there will
be a period of supercooling, and an associated inflationary era while
the universe remains trapped in the false vacuum, as discussed
in~\cite{Creminelli:2001th,Nardini:2007me} for the RS case.
Eventually, bubbles nucleate and the prompt decay of the radion
reheats the universe to a temperature of order TeV. From then on the
history of the universe is standard, and the KK-radion relic density
is set by the dynamics of states that are parametrically lighter than
the KK scale, and are in thermal equilibrium.  The result is
unambiguous and independent of the precise reheat temperature,
$T_{R}$.

However, we will also be interested in scenarios with small warping
($k_{\rm eff} \ll M_{5}$), hence $\Lambda_{r} \gg \tilde{k}_{\rm
eff}$.  Since this corresponds to the very large $N$ limit, we expect
that the rate of bubble nucleation will be extremely suppressed.  To
avoid an empty universe, we therefore need to assume that the highest
reheat temperature never exceeded $T_{c}$, and that the universe was
in the ``confined'' phase throughout its history.  Note that in our
scenario $m_{r} \sim \tilde{k}_{\rm eff}/\sqrt{k_{\rm eff}L}$ is at
the EW scale for any $\Lambda_{r}$.\footnote{In terms of the dual 4D
picture language, the \textit{explicit} breaking of the conformal
symmetry is in no sense small at the scale where this symmetry is
spontaneously broken.} Therefore, $T_{c}$ is expected to be at the TeV
scale, and is \textit{independent} of $N$.  Nevertheless, to account
for possible uncertainties in the above analysis that may allow
$T_{c}$ to be slightly larger, we will also study, at least formally,
the case in which the reheat temperature is as large as an order of
magnitude above $\tilde{k}_{\rm eff}$ (roughly the 4D cutoff of the KK
theory).  With this in mind, we proceed to describe several scenarios
that can lead to KK-radion DM.

\section{The KK-Radion as a WIMP}
\label{sec:DM}

We first consider the $r'$ relic density in the standard thermal
freeze-out scenario.  As mentioned in the introduction, all the
interactions of the KK-radion, like those of the radion itself, are
non-renormalizable and controlled by the radion decay constant,
$\Lambda_{r}$.  An important exception arises from a possible mixing
with the Higgs, and we will consider this case separately at the end
of this section.  It is therefore a question whether these
non-renormalizable interactions are sufficiently strong to deplete the
thermal $r'$ density down to acceptable levels, which will set an
upper bound on $\Lambda_{r}$.  A possible loophole arises when
$\Lambda_{r}$ is so large that it is possible to imagine that the
KK-radions never reached thermal equilibrium in the first place.  We
postpone the discussion of such a scenario to Section~\ref{sec:sWIMP}.
Assuming then that the $r'$ are in thermal equilibrium at a high
temperature, the self-annihilation into SM particles are responsible
for maintaining a thermal number distribution when the temperature
falls below the KK-radion mass, $m_{r'}$.

\begin{figure}
\centerline{
\resizebox{10.0cm}{!}{\includegraphics{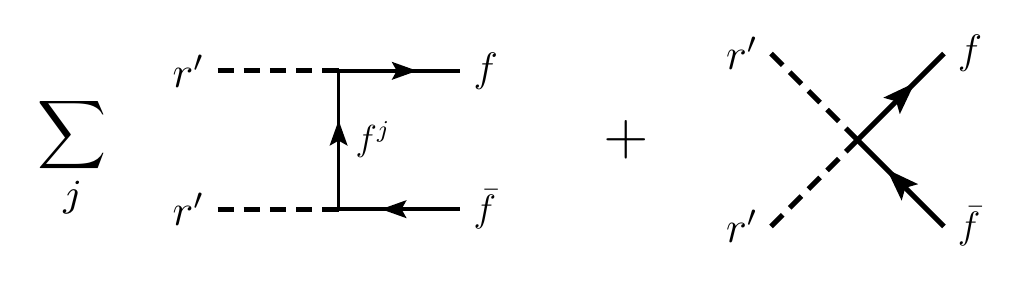}}
}
\caption{\em Self-annihilation of $r'$ into fermion pairs.
The $u$-channel diagram is understood.}
\label{fig:rprpAnnihilationff}
\end{figure}
\begin{figure}
\centerline{
\resizebox{9.0cm}{!}{\includegraphics{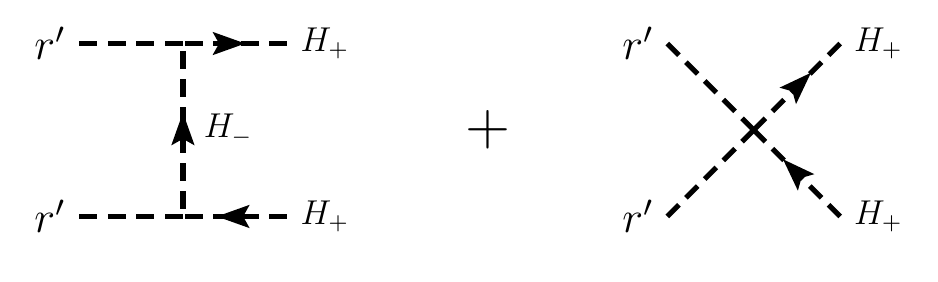}}
}
\caption{\em Self-annihilation of $r'$ into Higgs pairs.  The
$u$-channel diagram is understood.}
\label{fig:rprpAnnihilationHH}
\end{figure}

The KK-radions can annihilate into fermion pairs, as shown in
Fig.~\ref{fig:rprpAnnihilationff}.  The annihilation into massless
gauge bosons are suppressed at tree-level by $(m_{r'}/m_{V^j})^4$,
where $m_{V^j}$ is the mass of a gauge KK mode, as well as by a small
(volume suppressed) coupling (see Eq.~(\ref{sigmarprpVV}) of
Appendix~\ref{Annihiations} in Ref.~\cite{Medina:2010mu}).  We shall
therefore ignore these channels in the following.  The annihilation
into the $W$ and $Z$ gauge bosons proceed through their longitudinal
polarizations, and can be roughly taken into account through the
annihilation into Higgses shown in Fig.~\ref{fig:rprpAnnihilationHH}
(this approximation will be sufficient for our purpose).

We give the full annihilation cross sections in the non-relativistic
limit in Appendix~\ref{App:Processes}
[Eqs.~(\ref{sigmarprptt})--(\ref{sigmarprpHH})].  Note that, as
indicated in Fig.~\ref{fig:rprpAnnihilationff}, the annihilation into
fermion pairs involves a sum over the fermion KK tower in the $t$ and
$u$-channels.  Including, for illustration, a single state $f^{j}$,
the total annihilation cross section into fermion pairs can be well
approximated by
\bea
\left.  {\rm v} \sigma_{r' r' \rightarrow f \bar{f}} \right|_{j} &\sim&
\frac{N_{c} m_{r'}^2}{4 \pi  \Lambda_{r}^4} \left(1 - \epsilon_{f,r'}^2 \right)^{3/2}
\left[ G_{1100} \epsilon_{f,r'} - 2 G_{1j0}^2 \frac{\epsilon_{j,r'}^2}{1 + \epsilon_{j,r'}^2 - \epsilon_{f,r'}^2} \, \left.(\epsilon_{j,r'} + \epsilon_{f,r'} \right) \right]^2~,
\label{Simplerprpff}
\eea
where $N_{c}=3$ ($N_{c}=1$) for quarks (leptons), $\epsilon_{f,r'} =
m_{f}/m_{r'}$, $\epsilon_{j,r'} = m_{j}/m_{r'}$, and $G_{1j0}$
characterizes the $r'$-$f_{j}\bar{f}$ vertex, while $G_{1100}$
characterizes the $r' r'$-$f\bar{f}$ vertex.~\footnote{To obtain this
simplified expression we took $g^{LL}_{1j0} \ll 0$ (assuming $t_{L}$
is nearly flat), and we also neglected $g^{RR}_{1j0}$ (for the top,
this one is not small, but ends up giving a subdominant contribution).
We further took $G^{LR}_{1j0} \approx G^{RL}_{1j0}$ ($\equiv G_{1j0}$)
and defined $G_{1100} \equiv G^{LR}_{1100} = G^{RL}_{1100}$ (the
notation is defined in Appendix~\ref{App:Processes}).} It would seem,
from the second term in the square brackets, that the cross section
increases like $m^{2}_{j}$ for heavy KK states.  However, in the same
limit, the coupling constants $G_{1j0}$ decrease even faster, and one
can check that the sum over $j$ is typically dominated by the first
couple of terms.  The effects from EWSB, which are sizable for the top
tower in anarchic scenarios, are important in enhancing the above
effective couplings.  In particular, we find that for the $t\bar{t}$
channel, $G_{1100} \approx -6$ due to numerical factors characteristic
of the two-KK-radion couplings [e.g.~the factor of $4$ in
$\tilde{X}^{RL}_{i_{1} i_{2} jk}$, given after Eq.~(\ref{EWSBXX})], as
well as due to EWSB mixing with the KK-tower.  For the first top KK
resonance, although $G_{110} \approx 0.3$, the enhancement due to the
larger KK mass makes the second term in the square brackets of
Eq.~(\ref{Simplerprpff}) non-negligible (but still subdominant).  It
is also important to note that $G_{1j0}$ is proportional to EWSB. In
fact, the full expression (\ref{Simplerprpff}) is proportional to
$m_{f}^2$, although this fact is somewhat hidden inside $G_{1j0}$ for
the second term in square brackets.  Therefore, the annihilation rate
of KK-radions into the light families is negligible compared to the
annihilation into top quarks.

The annihilation of KK-radions into Higgses proceeds through the
diagrams in Fig.~\ref{fig:rprpAnnihilationHH}.  In the limit of strong
IR localization of the Higgs field, there is a KK-parity even Higgs
doublet, $H_{+}$, that is responsible for EWSB, and a KK-parity odd
doublet, $H_{-}$, that does not participate in EWSB, but that can be
parametrically lighter than the KK scale (see comments in
Section~\ref{Basic}).  We assume in this work that this KK-parity odd
Higgs doublet is heavier than the KK-radion (at the end of this
section we comment on the possibility that the CP-even component of
the ``inert'' Higgs doublet is the LKP).  In principle, there can be
heavy Higgs resonances contributing in the $t$ and $u$-channels, as
for the annihilation into fermions, but these resonances decouple in
the limit that the Higgs is strongly IR localized.  For simplicity, we
will restrict ourselves to this limit, which will be sufficient to
illustrate our point.  If the SM-like Higgs is lighter than the
KK-radion, while the inert Higgs doublet is heavier, we have [see
Eq.~(\ref{sigmarprpHH}) in Appendix~\ref{App:Processes}]
\bea
{\rm v} \sigma_{r' r' \rightarrow HH} &\sim&
\frac{N_{H} m_{r'}^2}{\pi \Lambda_{r}^4}~.
\label{sigmarprpHHSimple}
\eea
where, for large $m_{r'}$, $N_{H} = 4$ takes into account the SM-like
Higgs as well as the would-be Goldstone modes, and therefore also
annihilation into $W^{\pm}$, $Z$.

In order to estimate the total KK-radion annihilation cross
section, we assume that the RH top tower is localized near the IR
boundaries, that the LH top is approximately uniform along the
extra dimension, and that the lighter families are localized near
$y = 0$ (the UV region).  For concreteness, and using the
parameterization discussed in Ref.~\cite{Medina:2010mu} to
characterize fermion localization, which is analogous to the one
used in the pure Randall-Sundrum AdS$_{5}$ background, we take
$c_{t_{R}} = -0.2$ and $c_{t_{L}} = 0.52$ (these reproduce the top
quark mass for $\tilde{k}_{\rm eff} \approx 1.2~{\rm TeV}$, after
including EWSB effects).  As explained before, we can neglect
annihilations into the light families, but need to take into
account the annihilation into the Higgs and massive gauge bosons.
For reference, in the ``strong warping'' benchmark scenario
defined in Ref.\cite{Medina:2010mu} and used in this section, the
branching fractions are: ${\rm BR}(r'r' \to t\bar{t}) \approx
0.70$ and ${\rm BR}(r'r' \to HH) \approx 0.30$, where $H$ denotes
the four real d.o.f.~in the Higgs doublet.  It is expected that
these are distributed as ${\rm BR}(r'r' \to hh) \approx {\rm
BR}(r'r' \to ZZ) \approx \frac{1}{2} {\rm BR}(r'r' \to W^+ W^-)
\sim 7.5\%$.

Given the total thermally averaged annihilation cross section,
$\langle {\rm v} \sigma_{r'r'} \rangle$, the $r'$ relic density is
given by~\cite{Kolb:vq}
\bea
\Omega_{r'} h^2 \approx \frac{1.04 \times 10^9~{\rm GeV}^{-1}}{M_{Pl}}
\frac{x_F}{\sqrt{g_\ast} \langle {\rm v}  \sigma_{r'r'} \rangle}~,
\label{relicdensity}
\eea
where $g_{\ast}$ is the effective number of relativistic degrees of
freedom at the time of $r'$ decoupling, $M_{Pl} = 1.22 \times
10^{19}~{\rm GeV}$ is the Planck mass, and $x_{F} = m_{r'}/T_{F}$ gives
the freeze-out temperature by solving for
\bea
x_{F} &=& \log \! \left. \left\{ \zeta (\zeta + 2) \sqrt{45\pi} \left( \frac{g}{2\pi^3} \right)
\frac{m_{r'} M_{P} \langle {\rm v}  \sigma_{r'r'} \rangle}{\sqrt{g_{\ast} x}} \right\}
\right|_{x=x_{F}}~,
\label{xF}
\eea
where one can take $\zeta \sim 0.5$, and in this case $g = 1$.

\begin{figure}[t]
\centerline{
\resizebox{7.5cm}{!}{\includegraphics{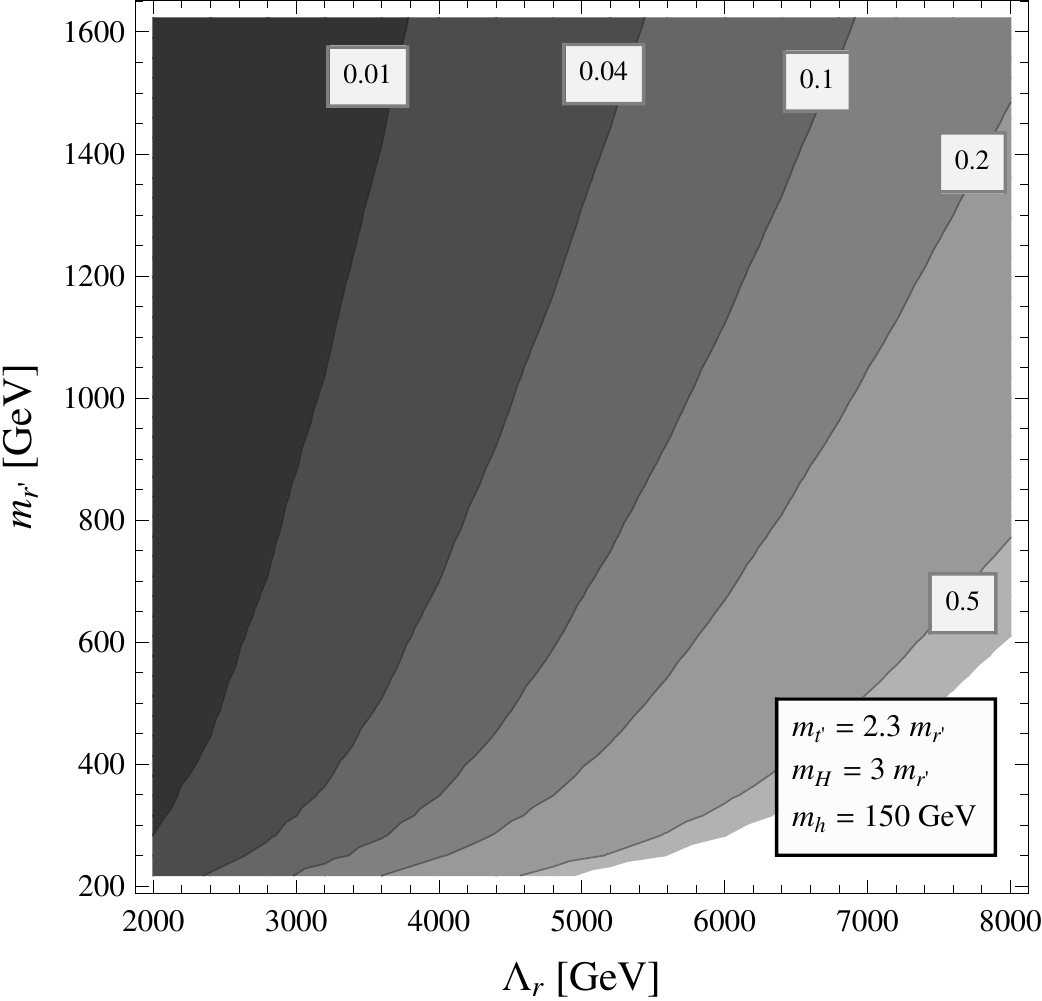}}
}
\caption{\em Contours of constant $\Omega_{r'}h^{2}$ in the
$m_{r'}$--$\Lambda_{r}$ plane, in models of flavor anarchy, where EWSB
effects are significant.
We use the ``strong warping'' benchmark relation $m_{t'} \approx 2.3
m_{r'}$, so that coannihilations give a small contribution, and take a
SM-like Higgs mass $m_{h} \approx 150~{\rm GeV}$, while the KK-parity
odd doublet has mass $m_{H} = 3m_{r'}$.  }
\label{fig:RelicDensiyGeneral}
\end{figure}
In the left panel of Fig.~\ref{fig:RelicDensiyGeneral}, we show
contours of constant $\Omega_{r'} h^2$ in the $m_{r'}$-$\Lambda_{r}$
plane, including the previously described annihilation channels.
Since we are assuming a flavor anarchy scenario, we take into account
EWSB effects, which are non-negligible in that case.  We take a
SM-like Higgs mass of $m_{h} = 150~{\rm GeV}$, while the mass of the
KK-parity odd Higgs doublet is taken as $m_{H} = 3m_{r'}$.  As
mentioned before, we take into account the annihilation into gauge
bosons by using a multiplicative factor $N_{H}=4$ for the KK-parity
even Higgs channel.  We vary $m_{r'}$ by scaling $\tilde{k}_{\rm
eff}$, i.e. keeping all mass ratios in the spectrum fixed.  However,
we force the SM fermions masses to remain fixed at their observed
values as $\tilde{k}_{\rm eff}$ is changed.  The dimensionless
couplings arising from overlap integrals are fairly insensitive to
$\tilde{k}_{\rm eff}$, and are kept fixed to the ones corresponding to
the ``strong warping scenario'' defined in Ref.~\cite{Medina:2010mu}.
The radion decay constant can be controlled via the ratio
$M_{5}/k_{\rm eff}$, which takes values between about $1.0$ and $2.4$
in the plot.  We note that in this example $m_{t'}/m_{r'} \approx 2.3$
with $m_{r'} \approx 0.22 \, \tilde{k}_{\rm eff}$, so that the NLKP is
sufficiently heavier than $r'$ that coannihilation effects are
negligible.  As can be seen from the figure, the correct relic density
of $\Omega_{r'} h^2 \approx 0.1$ can be obtained for natural values of
the parameters.  In the ``strong warping'' benchmark scenario with
$\tilde{k}_{\rm eff} = 1.2~{\rm TeV}$, we have $m_{r'} = 270~{\rm
GeV}$ and $\Lambda_{r} = 4.4~{\rm TeV}$, which leads to $\Omega_{r'}
h^2 \approx 0.24$.  Reducing $\Lambda_{r}$ to $3.6~{\rm TeV}$ gives
agreement with the WMAP constraint.  One finds $x_{F} \approx 22$, so
that the freeze-out temperature is about $10~{\rm GeV}$.

Thus, we see that in spite of the non-renormalizable interactions of
the radion, the ``WIMP miracle'' can be operative.  This is due to
sizable radion couplings to fermion pairs (associated with EWSB) and
can be qualitatively understood from our
expression~(\ref{Simplerprpff}).  Taking $\tilde{k}_{\rm eff} =
1.2~{\rm TeV}$, $m_{r'} = 0.25 \, \tilde{k}_{\rm eff}$, $m_{j} =
m_{t'} \approx 0.5 \, \tilde{k}_{\rm eff}$, $G_{1100} \approx -6$ and
$G_{110} \approx 0.3$, one estimates $\langle \sigma_{r'r'} {\rm v}/c
\rangle \sim 0.84~{\rm pb} \times \left( 3~{\rm TeV}/\Lambda_{r}
\right)^{4}$, where we recall that WMAP requires a cross section of
about $0.8~{\rm pb}$.

\begin{figure}[t]
\centerline{
\resizebox{15.0cm}{!}{\includegraphics{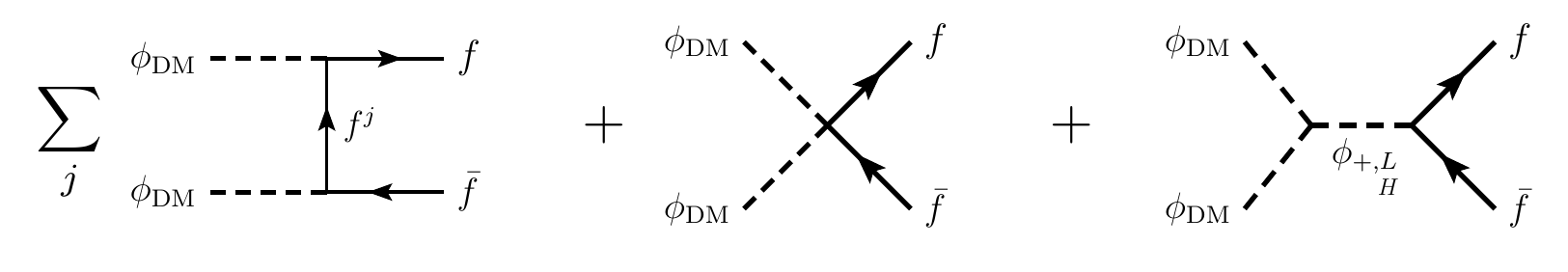}}
}
\vspace*{5mm}
\centerline{
\hspace{5mm}
\resizebox{15.0cm}{!}{\includegraphics{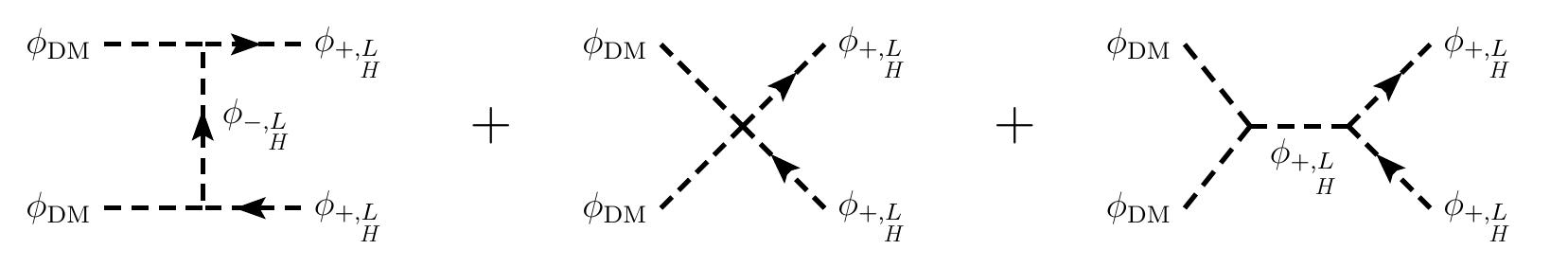}}
}
\caption{\em Self-annihilation of $\phi_{\rm DM} = \phi_{-,L}$ in the
general case where the radion/Higgs and KK-radion/inert Higgs mix.
The subscripts $L$ ($H$) refer to the lighter (heavier) state, while
the subscripts $+$ and $-$ refer to KK-parity.  These generalize the
diagrams of Figs.~\ref{fig:rprpAnnihilationff} and
\ref{fig:rprpAnnihilationHH}.}
\label{fig:rprpAnnihilationMixing}
\end{figure}

One should also keep in mind that the radion interactions can be
enhanced in the presence of the operator $R H^{\dagger} H$, where $R$
is the Ricci scalar.  Such a term leads to kinetic mixing between the
radion and CP-even, KK-parity even
Higgs~\cite{Csaki:1999mp,Giudice:2000av}, as well as between the KK
radion and the CP-even, KK-parity odd Higgs.  As a result,
renormalizable interactions to the SM particles are induced (although
the mixing angles are formally suppressed by $v/\Lambda_{r}$).  The
presence of such a mixing offers the best chance at direct detection
of KK-radion DM, and we defer the associated technical discussion to
Section~\ref{sec:Detection}.  However, here we comment on the effects
for the relic density computation in the presence of such mixing.  The
two lightest KK-parity odd mass eigenstates are in general admixtures
of $r' \equiv r_{-}$ and $h \equiv h_{-}$, which we call $\phi_{-,L}$
and $\phi_{-,H}$.  The subscripts $L$ ($H$) refer to the lighter
(heavier) of the two mass eigenstates.  The DM candidate is $\phi_{\rm
DM} = \phi_{-,L}$.  Only when the mixing is small, or the unperturbed
$m_{r'}$ is much smaller than $m_{h_{-}}$, can this state be thought
as mostly KK-radion (our assumption in the previous analysis).  There
is a similar mixing in the $(r,h) \equiv (r_{+},h_{+})$ sector, thus
defining mass eigenstates $\phi_{+,L}$ and $\phi_{+,H}$.  Both of
these KK-parity even states contribute in the final state when they
are lighter than the DM particle, although it may happen that only one
is sufficiently light to contribute.  The Feynman diagrams of
Figs.~\ref{fig:rprpAnnihilationff} and \ref{fig:rprpAnnihilationHH}
are generalized to those in Fig.~\ref{fig:rprpAnnihilationMixing}.
Note the presence of an s-channel diagram, which can lead to a
resonant enhancement if any one of the KK-parity even masses is close
to about twice $m_{\rm DM} \equiv m_{-,L}$.  The final states include
also the Goldstone modes, although those processes are not changed by
the mixing with the radion tower.

The mixing with the Higgs field can significantly enhance the DM
annihilation cross section, and therefore allow for larger values of
$\Lambda_{r}$.  As a point of reference, in the limit that $m_{h_{-}}
< m_{r'}$ and $\Lambda_{r} \to \infty$, the DM becomes pure $h_{-}$,
and its annihilation cross section, which is dominated by the
$t\bar{t}$ channel, reduces to
\bea
{\rm v} \sigma_{\phi_{DM} \phi_{DM} \rightarrow \bar{t}t} &\sim& \frac{N_{c} X_{t_{R}t^1_{L}}^4}{2\pi} \frac{m_{t'}^2 }{(m_{\rm DM}^2 + m_{t'}^2)^2}~,
\eea
where $X_{t_{R}t^1_{L}}$ is the coupling of $h_{-}$ to the RH top and
the first KK resonance of the LH top.~\footnote{The third top
resonance has a larger coupling that compensates for its larger mass,
so that it ends up giving a comparable contribution to the $t'$
exchange.  The heavier KK modes give a much smaller effect.  The Higgs
annihilation channels involve the Higgs quartic coupling, $\lambda$.
For a heavy Higgs (i.e.~$h_{+}$), these channels, if kinematically
open, can give a contribution comparable to the $t\bar{t}$ one.} This
coupling is relatively large, and therefore the result is dominated by
the $t$- (and $u-$) channel diagrams of the annihilation into
fermions.  With $X_{t_{R}t^1_{L}} \approx 2$ (which includes EWSB
effects), and taking $m_{t'} \sim m_{DM} \approx m_{h_{-}}$ (but still
assuming that $m_{t'} > m_{h_{-}}$), we find that the observed relic
density is obtained for $m_{\rm DM} \sim 30~{\rm TeV}$.  Such large
masses are not well motivated in the present scenario, and it follows
that for $m_{h_{-}} \sim 1~{\rm TeV}$, a too large $h_{-}$ component
will deplete very effectively the final DM density.  We conclude that
the DM should be mostly KK-radion, although a small $h_{-}$ component
that is interesting from the point of view of direct DM searches is
allowed.

\section{Coannihilations with $t'$}
\label{sec:coannihilation}

We have seen in the previous subsection that the $r'$
self-annihilation cross section can be large enough to avoid
the overproduction of KK-radions, provided the radion decay constant is
sufficiently low.  Does this mean that models with much larger radion
decay constants are ruled out?  Here we emphasize the point made in
Section~\ref{Basic} that the first KK top excitation is expected to be
relatively close to $m_{r'}$.  The reason is two-fold: first, the
$SU(2)$ doublet and singlet top fields cannot be too far from the IR
brane, or else $m_{t}$ cannot be reproduced.  This means that the
first KK-parity odd excitation of \textit{either} $t_{L}$ or $t_{R}$
\textit{must} be parametrically lighter than the KK scale.  Second,
when EWSB effects are included this state is pushed down even further
due to the level repulsion of the second KK-parity odd
state.~\footnote{Assuming that the 5D top Yukawa coupling saturates
the NDA~\cite{Chacko:1999hg} estimate of $Y_{5D}/L \lesssim 3/(4\pi)$,
and taking both top zero-modes as far away as possible from the IR
boundaries ($c_{t_{R}} \approx 0.45$ and $c_{t_{L}} \approx 0.48$), we
find that before EWSB the lightest KK-parity odd eigenvalue is about
$\sim 2.31 \, \tilde{k}_{\rm eff}$, which goes down to $\sim 0.89 \,
\tilde{k}_{\rm eff}$ when EWSB is taken into account.  In the ``strong
warping'' scenario considered in Ref.~\cite{Medina:2010mu}, with
$c_{t_{R}} \approx -0.2$ and $c_{t_{L}} \approx 0.52$, we have $m_{t'}
\approx 0.5 \, \tilde{k}_{\rm eff}$.  The KK-radion mass is $m_{r'}
\approx 0.22 \, \tilde{k}_{\rm eff}$.  }

\begin{figure}[t]
\centerline{
\resizebox{16cm}{!}{\includegraphics{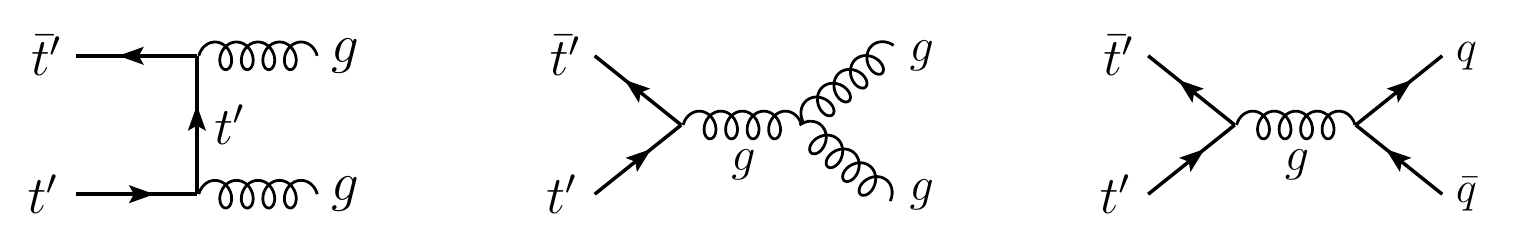}}
}
\caption{\em Self-annihilation of $t'$ into gluon and fermion pairs
via the QCD interactions.}
\label{fig:tptpAnnihilation}
\end{figure}
\begin{figure}
\centerline{
\resizebox{9.5cm}{!}{\includegraphics{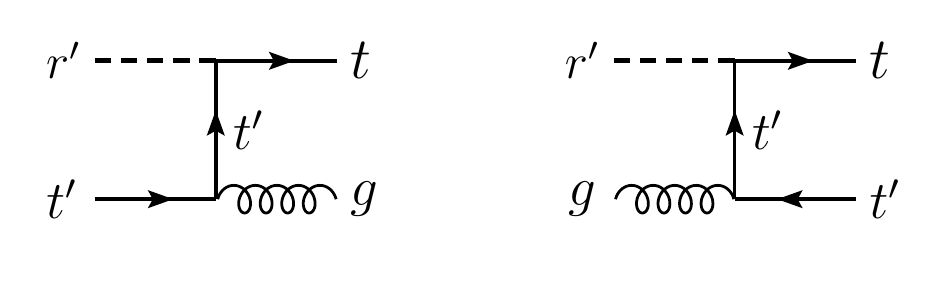}}
}
\caption{\em The left diagram (as well as a diagram where the gluon is
replaced by a Higgs) can play a role in the co-annihilation scenario.
The right diagram is an example of the processes responsible for
maintaining the $r'$ in thermal equilibrium until $t'$ freeze-out.}
\label{fig:rptpCoAnnihilation}
\end{figure}

If it turns out that $m_{t'}$ is relatively degenerate with the
KK-radion (but heavier), coannihilations with the strongly interacting
$t'$ can become dominant.  In fact, when $\Lambda_{r}$ is large, the
final $r'$ relic density can be completely controlled by the $t'$
self-annihilation cross section, and hence by QCD (see
Fig.~\ref{fig:tptpAnnihilation}).  In this case, the coannihilation
channel shown in the left diagram of Fig.~\ref{fig:rptpCoAnnihilation}
gives a negligible contribution to the effective annihilation cross
section.  However, it is important that the crossed diagram, $r'g
\rightarrow \bar{t}'t$, although suppressed by
$(m_{t}/\Lambda_{r})^2$, can be effective in depleting the KK-radion
number density, and thus maintain the $r'$ in equilibrium until the
time the $t'$ decouple.  Indeed, for $m_{t'} \sim m_{r'}$ and at the
temperature associated with the decoupling of $t'$, i.e. $m_{t'}/T_{F}
\sim 25$, one has $n_{g} \langle {\rm v} \sigma(r'g \rightarrow
\bar{t}'t) \rangle/n_{t'}\langle {\rm v} \sigma(t'\bar{t}' \rightarrow
gg,q\bar{q}) \rangle \sim (T_{F}/m_{t})^{3/2} \, e^{m_{t'}/T_{F}}
\times (m_{t}/\Lambda_{r})^2$.  Thus, even with $\Lambda_{r}$ as large
as $\sim 10^{4}~{\rm TeV}$, the $r'$ number density maintains an
equilibrium distribution by processes such as $r'g \rightarrow
\bar{t}'t$ (or the crossed reactions $t'g \rightarrow r't$) until $t'$
freeze-out.

The final relic density, after the KK tops have decayed into the
KK-radions, is determined by an effective cross
section~\cite{Griest:1990kh}:
\bea
\sigma_{\rm eff} &=& \frac{1}{g_{\rm eff}^{2}}
\left[ \rule{0mm}{7mm}
\sigma_{r' r' \rightarrow f \bar{f}, HH}
+ g_{t'} \, e^{-x\Delta} (1+\Delta)^{3/2} \, (\sigma_{r' t' \rightarrow g t} + \sigma_{r' \bar{t}' \rightarrow g \bar{t}} + \sigma_{r' t' \rightarrow h t} + \sigma_{r' \bar{t}' \rightarrow h \bar{t}} )
\right. \nonumber \\
& & \left. \hspace{7mm} \mbox{}
+ g_{t'}^{2}  e^{-2x\Delta} (1+\Delta)^{3} \left( \sum_{q} \sigma_{t' \bar{t}'
\rightarrow q \bar{q}} + \sigma_{t' \bar{t}' \rightarrow g g} \right)
\right]~,
\label{crosssectionSingletBidoublet}
\\
g_{\rm eff} &=& 1 + g_{t'} \, e^{-x \Delta} (1+\Delta)^{3/2}~,
\nonumber
\eea
where $g_{t'} = 2 \times N_{c}$ is the number of degrees of freedom
associated with $t'$ (and $\bar{t}'$).  Here $x \equiv m_{r'}/T$,
while
\bea
\Delta\equiv \frac{m_{t'}-m_{r'}}{m_{r'}}
\label{Delta}
\eea
parameterizes the degree of degeneracy between $r'$ and $t'$.
Although it is straightforward to include them, we neglect EW
processes involving annihilations through a $Z$ or a photon, since
they give a small contribution compared to those involving gluons.

We give the expressions for $\sigma_{r' t' \rightarrow g t} $ and
$\sigma_{r' t' \rightarrow h t}$ in Eqs.~(\ref{vsigmaritjTogtk}) and
(\ref{vsigmaritjTohtk}) of Appendix~\ref{App:Processes}, respectively.
Here we focus on the case where $\Lambda_{r}$ is sufficiently large
that all processes suppressed by powers of $\Lambda_{r}$ can be
ignored (except, implicitly, for the purpose of maintaining thermal
equilibrium, as discussed above).  As a result, and provided that $t'$
and $r'$ are sufficiently close in mass, the final relic density is
largely determined by the QCD-controlled cross sections
\bea
{\rm v} \sigma_{t' \bar{t}' \rightarrow q \bar{q}} &\approx& \frac{2\pi \alpha_{s}^2}{9m_{t'}^{2}}~,
\hspace{1cm}
{\rm v} \sigma_{t' \bar{t}' \rightarrow g g} ~\approx~  \frac{7\pi \alpha_{s}^2}{27m_{t'}^{2}}~,
\label{vsigmatpQCD}
\eea
where, for simplicity, we neglected the SM quark masses.  It is
straightforward to keep the full dependence on $m_{t}$, and we will do
so in the numerical analysis, but the effects are relatively minor
(much smaller than the errors involved in the approximation of
Eq.~(\ref{relicdensityCoannihilations}) below to the full integration
of the Boltzmann equations).  Similarly, the velocity-dependent terms
in all the cross sections make a negligible difference.  In summary,
we have the thermally averaged effective cross section
\bea
\langle {\rm v} \sigma_{\rm eff} \rangle &\approx&  \frac{172\pi \alpha_{s}^2}{3m_{t'}^{2}} \, \frac{(1+\Delta)^3 }{[ e^{x \Delta} + 6 (1+\Delta)^{3/2} ]^{2}}~.
\label{tpQCD}
\eea
The relic density is then approximately given by~\cite{Griest:1990kh}
\bea
\Omega_{r'} h^2 \approx \frac{1.04 \times 10^9~{\rm GeV}^{-1}}{M_{Pl}}
\frac{x_F}{\sqrt{g_\ast} I_{a}}~,
\label{relicdensityCoannihilations}
\eea
where $g_{\ast} = 86.25$ is the effective number of relativistic
degrees of freedom at the time of $t'$ decoupling, the annihilations
after freeze-out are taken into account by
\bea
I_{a} &\approx&  x_{F} \int_{x_{F}}^\infty \! dx \, \frac{\langle {\rm v} \sigma_{\rm eff} \rangle}{x^2}~,
\eea
and $x_{F} = m_{t'}/T_{F}$ gives the freeze-out temperature by solving
Eq.~(\ref{xF}) with the replacements $\langle {\rm v} \sigma_{r'r'}
\rangle \to \langle {\rm v} \sigma_{\rm eff} \rangle$ and $g \to
g_{\rm eff}$.

\begin{figure}
\centerline{
\resizebox{10.cm}{!}{\includegraphics{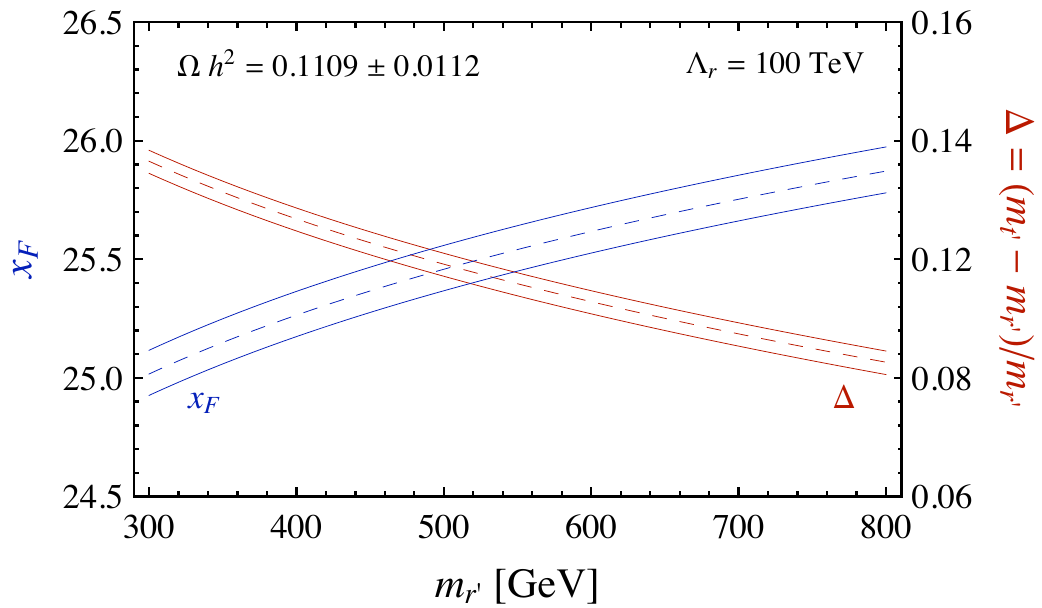}}
}
\caption{\em WMAP constraint on the degree of degeneracy, $\Delta$, as
defined in Eq.~(\ref{Delta}), as a function of $m_{r'}$ (red band,
right axis).  We also show the associated $x_{F}$ (blue band, left
axis).  We assume that $\Lambda_{r} =100~{\rm TeV}$, which is large
enough for the relic density to be controlled by the QCD $t'$
annihilation cross section, as well as $\Delta$.}
\label{fig:RelicDensiyDegenerate}
\end{figure}
Imposing the 7-year WMAP constraint, $\Omega h^2 = 0.1109 \pm
0.0112$~\cite{Komatsu:2010fb}, determines $\Delta$ as a function
of $m_{t'}$ (or $m_{r'}$).  In
Fig.~\ref{fig:RelicDensiyDegenerate}, we show this constraint, as
a function of $m_{r'}$.  We see that the relic density can be
accommodated for a wide range of $r'$ masses, provided the
required degeneracy with $t'$ is present.  We also show in the
same figure the freeze-out temperature, as parameterized by
$x_{F}$.

\section{The KK-Radion as a Non-Thermal Relic}
\label{sec:sWIMP}

We now wish to consider a qualitatively different scenario: one where
the KK-radion never reaches thermal equilibrium.  This can happen when
the radion decay constant is very large.  As was explained in
Section~\ref{Basic}, this can be realized in scenarios that share
properties of the Randall-Sundrum models and UED's.  Hence they can be
motivated outside the context of the DM relic density.  We assume that
the mixing with the Higgs is negligible.

As discussed in Section~\ref{sec:HighT}, when $\Lambda_{r} \gg
\tilde{k}_{\rm eff}$ we must assume that the reheat temperature is
lower than the critical temperature for the deconfinement/confinement
phase transition.  Otherwise, the universe would effectively be
trapped eternally in the false vacuum.  Even if a bubble of true
vacuum nucleated, the ensuing universe would be dominated by radion
oscillations until their decay, which would happen rather late.  The
resulting reheat temperature may not be high enough to allow for a
successful BBN. If instead, after a period of inflation, the inflaton
decay reheats the universe only to $T_{R} < T_{c}$ we can assume that
the universe was always in the ``confined'' phase.  Due to the
superweak couplings of the radion field, we can assume that it remains
at its zero-temperature minimum, so that no energy is stored in the
radion tower.  Note also that, since the radion mass is of order the
EW scale, it is easy to suppress the deSitter fluctuations of the
radion field during the inflationary epoch.  Thus, radion oscillations
are expected to play no role, unlike in the case of UEDs with a very
light radion~\cite{Kolb:2003mm}.  Provided the reheat temperature is
not much lower than the EW scale, KK-radions can be produced in the
scatterings or decays of KK-parity odd particles in the plasma, but at
a slow rate due to the suppressed couplings.  Nevertheless, it is well
known that such production can produce a significant number of
superweakly interacting particles~\cite{Moroi:1993mb}, which can
translate into a bound on the reheat temperature $T_{R}$ if
overclosure of the universe is to be avoided.

As we will see, this requires the reheat temperature to be relatively
low, around the cutoff of the extra-dimensional theory or lower, even
if we overlook the upper bound set by $T_{c}$.  In addition, it bounds
$\Lambda_{r}$ from below to ensure that the KK-radions are
sufficiently weakly coupled.  An upper bound on $\Lambda_{r}$ can also
be obtained by requiring that the late decays of the NLKPs or the
non-thermally produced radions do not conflict with Big-Bang
nucleosynthesis (BBN) constraints.

\subsection{Non-Thermal Production in Scatterings and Decays}
\label{sec:ThermalsWIMP}

Our point of departure is the Boltzmann equation
\bea
\frac{dn_{r'}}{dt} + 3 H n_{r'} = C~,
\label{Boltzmann}
\eea
where $H$ is the Hubble constant and $n_{r'}$ is the number density of
$r'$.  The collision operator on the r.h.s. contains several $r'$
production channels that are our main concern in this section.
However, since we are assuming that the $r'$ are produced very
inefficiently, so that they never reach thermal equilibrium, we
neglect the inverse processes that would tend to deplete the $r'$
number density.  This assumption will be checked self-consistently at
the end.

The production channels can be divided into scattering and decay
processes.  The former include scatterings of thermal KK fermions off
the plasma, such as $g t^j \to r' t^{k}$, where $t^{j}$ and $t^{k}$
are KK excitations of the top quark (including EWSB effects, these are
admixtures of the doublet and singlet $SU(2)_{L}$ top towers).  The
towers of the other SM fermions (the ``light families'') can also
contribute, and will be included in our analysis.  In addition, at
sufficiently high temperatures, the KK excitations of the gluons,
$g^{n}$, can also be active and lead to $r'$ production, e.g.~via
$g^{n} g^{j} \to r' g^{k}$ or $g^{n} q^j \to r' q^{k}$.  Nevertheless,
as we will see below, the scattering processes represent a subdominant
source of thermal $r'$ production, and therefore we will be content
with obtaining an estimate for the $r'$ production via scattering
based on the processes with a zero-mode gluon and KK fermions given
above.  For the same reason we also do not consider scattering
processes involving the weak gauge bosons.\footnote{These are
suppressed compared to the processes involving gluons by the
multiplicity factor of 8 gluons versus 4 weak gauge bosons, as well as
by a weak gauge coupling squared.} We should point out that the
production of higher KK resonances of the radion can also eventually
produce $r'$.  However, as was emphasized in~\cite{Medina:2010mu}, the
interactions of such states are controlled by a larger decay constant
[$\Lambda^{n}_{r} \sim x_{n} \Lambda_{r}$, where $m_{n} = x_{n} \,
k_{\rm eff} \, e^{-A(L)}$ is the $n$-th KK-radion mass], and therefore
this contribution is negligible.

KK-radions can also be produced in (rare) decays of KK tops, $t^j \to
r' t^{k}$, as well as of the KK excitations of the light families, and
from gauge boson decays, $V^j_{\mu} \to r' V^{k}_{\mu}$, with $V_{\mu}
= G^a_{\mu}, W^{\pm}_{\mu}, Z_{\mu}, A_{\mu}$.  We also include the
decays of the inert Higgs doublet, $h_{-} \to h_{+} r'$, $a \to G^0
r'$ and $H^{\pm} \to G^{\pm} r'$, where $h_{+}$ is the SM-like Higgs
and $G^{0}$, $G^{\pm}$ are the would-be Nambu-Goldstone bosons eaten
by the $Z$ and $W^\pm$.  Again, it is safe to neglect the production
of heavier KK-radion resonances.

During a radiation dominated epoch the Boltzmann equation can be
written in terms of $x\equiv m_{t'}/T$ as
\begin{equation}
\frac{d \tilde{n}_{r'}}{dx}=\frac{x^4}{H(m_{t'})} \, C[x]~,
\label{boltzmann}
\label{Boltzmann2}
\end{equation}
where $\tilde{n}_{r'}(x) \equiv x^3 n_{r'}(x)$ and
$H(x)=H(m_{t'})x^{-2}=1.66 \, g_{*}^{1/2}m_{t'}^2/M_{Pl} \times
x^{-2}$, with $g_{*}$ the effective number of relativistic degrees of
freedom at temperature $T$.  Note that we choose to measure the ``time
variable'' $x$ in units of the $t'$ mass, $m_{t'}$ (we are assuming
that $t'$ is the NLKP).  Eq.~(\ref{Boltzmann2}) can be integrated
immediately from $x_{R} = m_{t'}/T_{R}$, where $T_{R}$ is the
reheating temperature, up to any given $x$, thus obtaining
$\tilde{n}_{r'}(x)$ and thereby $n_{r'}(x)$.

Note that most of the heavy particles involved remain in thermal
equilibrium until they disappear (by decaying) from the bath: the KK
fermion resonances can decay via $t^{j} \to t^{k} V^{l}_{\mu}$, while
the KK gauge bosons can decay into fermion pairs, $V^{l}_{\mu} \to
t^{j} \bar{t}^{k}$ (always preserving KK parity).  These processes are
controlled by renormalizable interactions and the rates are very large
compared to the Hubble scale at the times of interest.  The exceptions
to the above are $t'$ which must decay into $r'$ with a rate
suppressed by $\Lambda_{r}$, and the radion KK states whose
interactions are also controlled by $\Lambda_{r}$. Of these $r'$ is
stable and by assumption remains out of thermal equilibrium.  Since,
as we will see, this assumption requires that $\Lambda_{r}$ be very
large, the $t'$ are quasi-stable, and freeze-out when their
self-annihilation cross section becomes ineffective at $x_{F} =
m_{t'}/T_{F}$.  By this time, the number density of all the heavier KK
particles is highly Boltzmann suppressed, and any $r'$ from decays of
the few remaining heavy particles give a negligible contribution.
Therefore, we can integrate the Boltzmann equation including
\textit{all} processes until the time of $t'$ freeze-out, assuming
thermal equilibrium distributions.  This will give a $r'$ number
density that we will call $n^{(1)}_{r'}$.  At a somewhat later time,
the remaining $t'$ will decay out of equilibrium, giving an additional
contribution to the $r'$ number density.  Assuming that there is no
significant entropy production between the time of $t'$ freeze-out and
when the $t'$ decay into $r'$, we can effectively include this
contribution at $x=x_{F}$ by an additive term $n^{(2)}_{r'} =
n_{t'}(x_{F})$.

The total $r'$ relic density today is then given by $\Omega_{r'} =
m_{r'} n_{r'}(x_{0})/\rho_{c}$ where $\rho_c = 1.05 \times 10^{-5} \,
h^2~{\rm GeV~cm}^{-3} = 8.06 \times 10^{-47}~h^{2}~{\rm GeV}^4$ is the
critical density of the universe today, $x_{0}$ corresponds to the
temperature of the universe today, and $n_{r'}(x_{0})$ can be
connected to $n_{r'}(x_{F}) = n^{(1)}_{r'} + n^{(2)}_{r'}$ from
conservation of entropy.  Thus,
\bea
\Omega_{r'}^{(1)} &=& \frac{m_{r'}}{\rho_c}\frac{s_0}{s(x_{F})}\frac{1}{x_{F}^3}\frac{1}{H(m_{t'})} \int^{x_{F}}_{x_{R}} \! dx \, x^{4} \, C[x]~,
\label{omegar1}
\eea
where the entropy is given by
\bea
s(x) &=& \frac{2\pi^2}{45} g_{*}(x) \frac{m_{t'}^3}{x^3}~,
\eea
and $s_0 = 2889.2~{\rm cm}^{-3} \approx 2.22 \times 10^{-38}$ ${\rm
GeV}^3$ is the entropy density today.  Since $T_{F} \sim m_{t'}/30$
will be at most ${\cal O}(100~{\rm GeV})$, the relativistic degrees of
freedom are those of the SM, e.g. $g_{*}(T = 100~{\rm GeV}) = 92.25$.

The contribution coming from the decays of $t'$ after freeze-out, as
in the SWIMP scenario~\cite{Feng:2003xh,Feng:2003uy}, takes the form
[see Eq.~(\ref{relicdensity})]
\bea
\Omega_{r'}^{(2)} h^2 &=& \frac{m_{r'}}{m_{t'}} \times \frac{1.04 \times 10^9~{\rm GeV}^{-1}}{M_{Pl}} \frac{x_{F}}{\sqrt{g_*}}\frac{1}{\langle {\rm v} \sigma_{t' \bar{t}' \rightarrow q \bar{q},gg} \rangle}~,
\label{Omega2}
\eea
where ${\rm v} \sigma_{t' \bar{t}' \rightarrow q \bar{q},gg}$ is
determined by the QCD interactions of $t'$, as given in
Eq.~(\ref{vsigmatpQCD}) (see also Fig.~\ref{fig:tptpAnnihilation}).
Neglecting all the quark masses (even the top quark mass gives a small
correction for typical parameters), we have
\bea
\langle {\rm v} \sigma_{t' \bar{t}' \rightarrow q \bar{q},gg} \rangle &\approx& \frac{43}{27}\frac{\pi\alpha_s}{m_{t'}^2}~,
\label{tpQCDSimple}
\eea
where $43/27 = 7/27 + (2/9) N_{f}$, with $N_{f} = 6$.  The
subleading corrections in the relative velocity, ${\rm v}$, give
also a negligible effect.  The freeze-out temperature,
parameterized by $x_{F}$, is given by Eq.~(\ref{xF}) with
$\langle{\rm v} \sigma_{r'r'} \rangle \to \langle{\rm v}
\sigma_{t' \bar{t}' \rightarrow q \bar{q},gg} \rangle$, $g = 2
\times N_{c} = 6$, and $m_{r'} \to m_{t'}$.

\subsubsection{Decay Processes Before $t'$ Freeze-out}
\label{sec:decays}

%
\begin{figure}[t]
\centerline{
\resizebox{10.5cm}{!}{\includegraphics{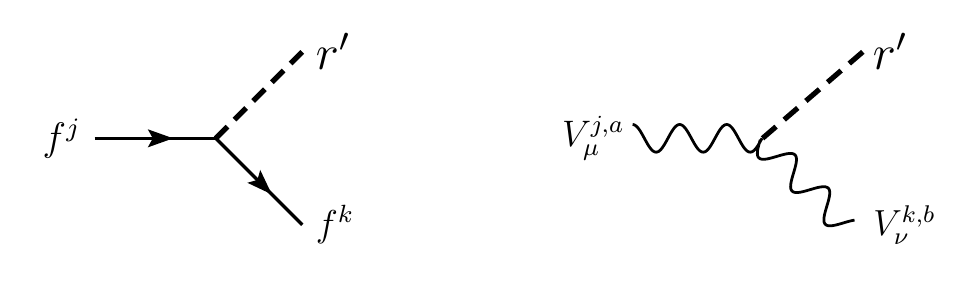}}
}
\caption{\em Production of $r'$ in the decays of (KK-parity odd) KK
fermions and KK gauge bosons.  Reversing the fermion arrows gives the
production from anti-fermion decays.}
\label{fig:DecayDiagrams}
\end{figure}
We now discuss the different contributions to the collision operator
in Eq.~(\ref{omegar1}), starting with the production of $r'$ in the
decays of KK-parity odd fermions and gauge bosons (see
Fig.~\ref{fig:DecayDiagrams}).  Using decays of KK quarks (which
includes the top) for illustration, the contribution to the collision
operator, per spin and color degree of freedom of the $q^{j}$, is
\bea
C^{\rm Decay}_{j,k} &\equiv&\int e^{-E_{q^{j}}/T} \, \Gamma_{q^{j} \to r' q^{k}} \,
\frac{d^3p_{q^{j}}}{(2\pi)^3}~,
\label{ColOpDecay}
\eea
where the decay rate is
\bea
\Gamma_{q^{j} \to r' q^{k}} = \frac{1}{ 2 E_{q^{j}}} \int (2\pi)^4 \delta(p_{q^{j}} - p_{r'} - p_{q^{k}}) \overline{\left| {\cal M}_{q^{j} \to r' q^{k}} \right|^2} \, \frac{d^3p_{r'}}{(2\pi)^3 2 E_{r'}} \frac{d^3p_{q^{k}}}{(2\pi)^3 2 E_{q^{k}}}~,
\eea
which is related to the decay rate in the CM via the time dilation
factor $\gamma = E_{q^{j}}/m_{j}$ by $\Gamma_{q^{j} \to r' q^{k}} =
\Gamma^{\rm CM}_{q^{j} \to r' q^{k}}/\gamma$.  Since the CM partial
decay width is simply a constant, we can do the integral in
Eq.~(\ref{ColOpDecay}) to obtain
\bea
C^{\rm decay}_{j,k}[x_{j}] &=& \frac{4\pi}{(2\pi)^3} \,  m_{j}^3 \, \Gamma^{\rm CM}_{j,k} \, \frac{K_{1}(x_j)}{x_{j}}
\label{ColDecay} \\[0.5em]
&\approx& \frac{4\pi}{(2\pi)^3} \, m_{j}^3 \, \Gamma^{\rm CM}_{j,k} \times
\left\{\begin{array}{lcl}
\displaystyle
\sqrt{\frac{\pi}{2}} \, x_{j}^{-3/2} e^{-x_j}
& &
\displaystyle
x_j \gsim 1 \\ [1em]
\displaystyle
x_j^{-2}
& &
\displaystyle
x_j \ll 1
\end{array}\right.~,
\eea
where $K_{1}(x)$ is a Bessel function, $x_j \equiv m_{j}/T =
(m_{j}/m_{t'}) \, x$, and now $\Gamma^{\rm CM}_{j,k}$ stands for any
of $\Gamma^{\rm CM}_{f^{j} \to r' f^{k}}$, $\Gamma^{\rm
CM}_{V^{j}_{\mu} \to r' V^{k}_{\mu}}$, $\Gamma^{\rm CM}_{h_{-} \to
h_{+} r'}$, $\Gamma^{\rm CM}_{a \to G^{0} r'}$ or $\Gamma^{\rm
CM}_{H^{\pm} \to G^{\pm} r'}$, according to the case (we give their
expressions in Eqs.~(\ref{GammaCMtj}), (\ref{GammaCMgj}),
(\ref{Gammahm}) and (\ref{GammaHpma}) of
Appendix~\ref{App:Processes}).  We see that when $T_{R} < m_{j}$ (i.e.
$x_{j} > 1$ for all relevant times), the contribution from $C^{\rm
decay}_{j,k}$ to the r.h.s. of Eq.~(\ref{omegar1}) is exponentially
small.  If, on the other hand, $T_{R} \gg m_{j}$, and taking into
account that $m_{j} \geq m_{t'} \gg T_{F}$, the relevant integral in
Eq.~(\ref{omegar1}) gives
\bea
\int_{x_{R}}^{x_{F}} \! dx \, x^4 \frac{K_{1}(x_{j})}{x_{j}} &=& \epsilon^{5}_{t',j} \int_{m_{j}/T_{R}}^{m_{j}/T_{F}} \! dx_{j} \, x^4_{j} \frac{K_{1}(x_{j})}{x_{j}} ~\approx~ \epsilon^{5}_{t',j} \left[ \frac{3\pi}{2} - \frac{m_{j}^{3}}{3T_{R}^3} \right]~,
\label{ColIntDecay}
\eea
where $\epsilon_{t',j} \equiv m_{t'}/m_{j}$, and therefore the
contribution from decays is rather insensitive to the reheat
temperature in this case.  The support of the integral is $1/2
\lesssim x_{j} \lesssim 7$, and we see that the bulk of $r'$ produced
in decays of heavy particles occurs for temperatures of order the mass
of the decaying particle, $m_{j}$.  In particular, states with masses
much heavier than the reheat temperature give only an exponentially
small contribution.  Note also that, since parametrically $\Gamma^{\rm
CM}_{j,k} \sim m^{3}_{j}/\Lambda_{r}^{2}$, for states lighter than
$T_{R}$ we have $\int_{x_{R}}^{x_{F}} \!  dx \, x^4 C^{\rm
decay}_{j,k} \sim m_{t'}^{5} m_{j}/\Lambda_{r}^{2}$, and the
individual contributions scale linearly with the KK mass, $m_{j}$.  It
follows that the contribution to the $r'$ relic density from KK
particle decays grows linearly with the reheat temperature, which sets
the effective cutoff for how many KK states give a non-negligible
effect.

\subsubsection{Scattering Processes}
\label{sec:scattering}

%
\begin{figure}[t]
\centerline{
\resizebox{15.5cm}{!}{\includegraphics{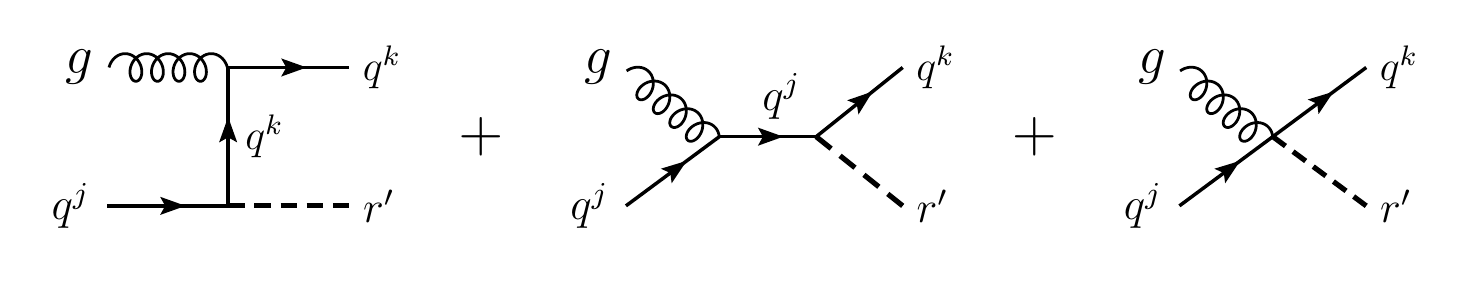}}
}
\caption{\em Tree-level scattering diagrams for $g q^j \to r' q^{k}$.
Reversing the fermion arrows one obtains the diagrams for $g \bar{q}^j
\to r' \bar{q}^{k}$.  The contact diagram is required by gauge
invariance.  }
\label{fig:ScatteringDiagrams}
\end{figure}
We consider now the scattering processes.  The contribution to the
collision operator from each individual process $g q^j \to r' q^{k}$
and for each spin/polarization/color degree of freedom in the initial
state takes the form
\bea
C^{\rm scatt.}_{j,k} &\equiv& \int e^{-(E_{q^{j}} + E_{g})/T} \left( 4F\sigma_{g q^j \to r' q^{k}}\right) \frac{d^3p_{g}}{(2\pi)^3 2 E_{g}} \frac{d^3p_{q^{j}}}{(2\pi)^3 2 E_{q^{j}}}~,
\label{ColOpScatt}
\eea
where the scattering cross section is
\bea
\sigma_{g q^j \to r' q^{k}} = \frac{1}{4F} \int (2\pi)^4 \delta(p_{g} + p_{q^{j}} - p_{r'} - p_{q^{k}}) \overline{\left| {\cal M}_{g q^{j} \to r' q^{k}} \right|^2} \frac{d^3p_{r'}}{(2\pi)^3 2 E_{r'}} \frac{d^3p_{q^{k}}}{(2\pi)^3 2 E_{q^{k}}}~,
\eea
and the flux factor is $F = p_{q^{j}}\cdot p_{g}$ ($= \sqrt{s}
|\vec{p}|$, with $|\vec{p}|$ the three-momentum of one of the initial
state particles in the CM).  The relevant diagrams are shown in
Fig.~\ref{fig:ScatteringDiagrams}, and the analytic expression for the
cross section is given in Eq.~(\ref{scatteringXS}) of
Appendix~\ref{App:Processes}.  Following Ref.~\cite{Gondolo:1990dk},
it is convenient to write $d^{3} p_{g} d^{3}p_{q^{j}} = 2\pi^{2} E_{g}
E_{q^{j}} dE_{+} dE_{-} ds$, where $E_\pm \equiv E_g \pm E_{q^j}$, and
$s = m^{2}_{j} + 2 E_{g} (E_{q^{j}} - p_{q^{j}} \cos\theta)$ is the
total CM energy squared, while $\theta$ is the angle between the
momenta of the initial state particles.  It is assumed that $s >
(m_{r'} + m_{k})^{2}$.  One finds that the physical region corresponds
to $E_{+} \geq \sqrt{s}$, while $|E_{-} - E_{+} m_{j}^{2}/s| \leq
\sqrt{E_{+}^{2} - s} \, (1 - m_{j}^{2}/s)$.  The integrand is
independent of $E_{-}$, so that the $E_{-}$ integral can be trivially
done.  The cross section (times the flux factor) depends only on $s$,
and the integral over $E_{+}$ gives $\int_{\sqrt{s}}^{\infty} dE_{+}
\sqrt{E_{+}^{2} - s} \, e^{-E_{+}/T} = \sqrt{s} \, T
K_{1}(\sqrt{s}/T)$, where $K_{1}$ is a modified Bessel function of the
second kind.  Thus, we can write
\bea
x_j^{4} \, C^{\rm scatt.}_{j,k}[x_{j}] &=& m_{j}^4 \int_{x_{j}}^{\infty}
\frac{8\pi^2}{(2\pi)^6}K_{1}(u)(u^2-x_{j}^2) (F\sigma_{g q^j \to r' q^{k}}) du~,
\label{master2}
\eea
where $u \equiv \sqrt{s}/T$ and $x_j \equiv m_{j}/T$ are dimensionless
variables.  The full collision operator associated with scattering of
KK fermions is then~\footnote{Although the couplings of the KK-radion
to fermion zero-modes are suppressed by the corresponding fermion
mass, its couplings to the KK states are similar for all the light
family KK towers.  Since the zero-mode contribution gives a negligible
effect, we can describe the effect of the light families by a
multiplicity factor, which would be appropriate in anarchic
scenarios.}
\bea
C^{\rm scatt.}[x] &=& \sum_{j,k} {\cal N} C^{\rm scatt.}_{j,k}[x_{j}] ~,
\label{FullColOpScattering}
\eea
where ${\cal N} = 2 \textrm{ (gluon pol.)} \times (N_{c}^2-1) \textrm{
(gluons)} \times 2 \textrm{ ($q^{j}$ spin)} \times N_{c} \textrm{
($q^{j}$ color)} \times 2 \textrm{ ($q^{j}$ and $\bar{q}^{j}$)} = 8
N_{c} (N_{c}^2-1)$, with $N_{c} = 3$.  For leptons and weak gauge
bosons the multiplicities are different, but as we explained at the
beginning of Section~\ref{sec:ThermalsWIMP}, we neglect such
contributions.  We consider each SM quark flavor separately in the
sum.  Also, the allowed processes are of the form $g^{+} q^{-}
\rightarrow r^{-} q^{+}$ or $g^{+} q^{+} \rightarrow r^{-} q^{-}$,
with $\pm$ referring to a KK-parity even/odd state.

\medskip
In the remainder of this subsection, we provide analytic
approximations for the elementary building blocks, $C^{\rm
scatt.}_{j,k}$.  The reader mostly interested in the physics results
may wish to jump to Section~\ref{sec:results}, where we discuss them.

In the non-relativistic regime ($T \lesssim m_{j}$, which implies that
the integral in Eq.~(\ref{master2}) is dominated by the region $u \sim
x_{j}$), we have the approximate result
\bea
F \sigma^{\rm non-rel}_{gq^{j} \to r' q^{k}} &\approx& \frac{\alpha_{s}}{8 N_{c}} \, \frac{m_{j}^{2}}{\Lambda_{r}^{2}} \, \frac{x^{4}_{j}}{(u^{2}-x^{2}_{j})^2 + x^{2}_{j} \, (\Gamma_{j}/T)^{2} }
\left[ (1 + \Delta \epsilon) \, \frac{\tan^{-1} \! \left[ \frac{\theta(\Delta \epsilon, \epsilon_{k,j})}{1 + \Delta \epsilon} \right]}{\theta(\Delta \epsilon, \epsilon_{k,j})} - 1 \right]
\nonumber \\[0.5em]
& & \mbox{} \times \theta(\Delta \epsilon, \epsilon_{k,j}) \,
\left\{ \rule{0mm}{5mm}
\left[ (G_{1jk}^{RL})^{2} + (G_{1jk}^{LR})^{2} \right] (1 + \Delta \epsilon) + 4 G_{1jk}^{RL} G_{1jk}^{LR} \, \epsilon_{k,j} \right\}~,
\label{ScatteringNR}
\eea
where $\theta(\Delta \epsilon, \epsilon_{k,j}) = \sqrt{(1 + \Delta
\epsilon)^2 - 4 \epsilon_{k,j}^{2}}$, $\Delta \epsilon \equiv
\epsilon_{k,j}^{2} - \epsilon_{r',j}^{2}$, $\epsilon_{k,j} =
m_{k}/m_{j}$, and $\epsilon_{r',j} = m_{r'}/m_{j}$, i.e. we normalize
with respect to the incident $t^{j}$ mass.  The $G^{RL}_{ijk}$ and
$G^{LR}_{ijk}$ are the relevant couplings of the KK-radion (with $i=1$
for $r'$) to the fermion KK modes [see Eqs.~(\ref{GRL}), (\ref{MRL})
and (\ref{Lrff})].  This result is dominated by the s-channel diagram
of Fig.~\ref{fig:ScatteringDiagrams}, and we have explicitly included
the $q_{j}$ width, $\Gamma_{j}$, which regulates the limit where the
incoming gluon is ultrasoft and the s-channel $q^{j}$ is nearly on
shell.

In the relativistic regime (i.e.~$T \gg m_{j}$) one can instead use
\bea
F \sigma^{\rm rel}_{gq^{j} \to r' q^{k}} &\approx& \frac{\alpha_{s}}{64 N_{c}} \, \frac{m_{j}^{2}}{\Lambda_{r}^{2}} \left[ (G_{1jk}^{RL})^{2} + (G_{1jk}^{LR})^{2} \right] \left[ 4 \log \left( \frac{u}{x_{k}} \right) - 3 \right]~,
\label{ScatteringRel}
\eea
where we took the limit $\sqrt{s} \gg m_{j}$ of the exact cross
section given in Eq.~(\ref{scatteringXS}) of
Appendix~\ref{App:Processes}, and defined $x_k \equiv m_{k}/T$.  This
result is dominated by the t-channel diagram in
Fig.~\ref{fig:ScatteringDiagrams}.  The logarithmic divergence when
$m_{k} \to 0$ is similar to the usual forward scattering singularity
associated with the t-channel exchange of a massless particle.
\begin{figure}[t]
\centerline{
\resizebox{9.cm}{!}{\includegraphics{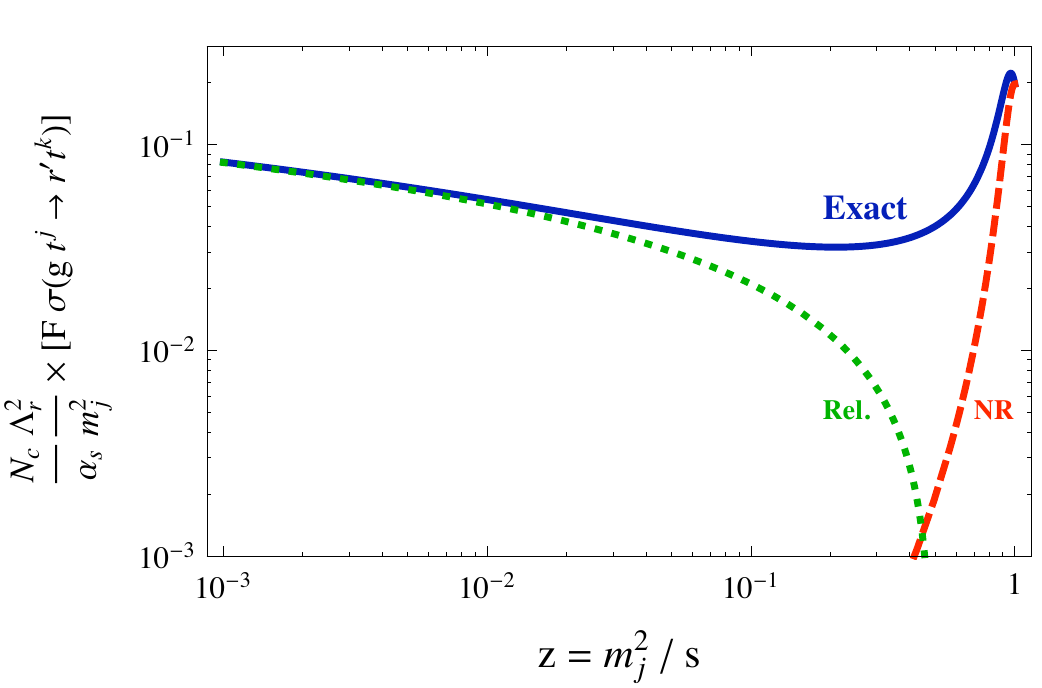}}
}
\caption{\em $F \times \sigma(gt^{j} \to r' t^{k})$ as a function of
$z = m_{j}^{2}/s$, assuming $m_{r'}/m_{j} = 0.20$, $m_{k}/m_{j} =
0.67$, $G_{1jk}^{RL} = 0.50$ and $G_{1jk}^{LR} = 0.41$, which
correspond to the decay $gt^{5} \to r' t^{4}$ of the ``small warping''
benchmark scenario.  The solid (blue) curve is the exact result given
in Eq.~(\ref{scatteringXS}) of Appendix~\ref{App:Processes}, the
dashed (red) curve is the non-relativistic approximation of
Eq.~(\ref{ScatteringNR}), and the dotted (green) curve is the
relativistic approximation of Eq.~(\ref{ScatteringRel}).  We also
assumed $\Gamma_{j}/m_{j} = 1/10$, which is important only for $z
\approx 1$.}
\label{fig:ScatteringXS}
\end{figure}
In Fig.~\ref{fig:ScatteringXS}, we show the exact scattering cross
section as given in Eq.~(\ref{scatteringXS}) of
Appendix~\ref{App:Processes}, together with its non-relativistic and
relativistic approximations, Eqs.~(\ref{ScatteringNR}) and
(\ref{ScatteringRel}), respectively.  We choose typical parameters in
order to illustrate the precision of the approximations.  We will use
the exact result in our numerical analysis, but the approximate limits
provide a more transparent analytic understanding.

For instance, in the non-relativistic limit, the $u$-dependence of the
integrand in the collision term of Eq.~(\ref{master2}) is given by
\bea
g_{\rm NR} du &=& \frac{K_{1}(u)(u^2-x_{j}^2) }{(\frac{u^2}{x_j^2}-1)^2+\gamma_j^2} \, du~,
\eea
where $\gamma_{j} \equiv \Gamma_{j}/m_{j}$.  It is instructive to
rewrite $g_{\rm NR}$ as a function of
\bea
w&=&\frac{u^2-x_j^2}{x_j^2\gamma_{j}}~,
\hspace{1cm}
du~=~\frac{x_j\gamma_{j}}{2\sqrt{1+w\gamma_{j}}} \, dw~,
\label{wdef}
\eea
so that
\bea
g_{\rm NR} du &=& \frac{1}{2} \, x_j^{3} K_{1} \! \left(x_j\sqrt{1+w\gamma_{j}}\right) \,  \frac{w}{w^2+1} \frac{dw}{\sqrt{1 + w \gamma_{j}}}~.
\eea
This function determines the support of the scattering collision
integral as follows.  For $x_j\gg 1$ and writing $u = x_{j} + \Delta
u$, we see that in this ultra non-relativistic regime ($T \ll m_{j}$),
we have $K_1(x_j\sqrt{1+w\gamma_{j}}) = K_1(u) \approx
\sqrt{\frac{\pi}{2 x_{j}}} \, e^{-(x_{j} + \Delta u)}$, where the use
of the asymptotic form of the Bessel function is well justified.  The
exponential damping $e^{-\Delta u}$ ensures that the support of the
integral satisfies $\Delta u \lesssim {\cal O}(1)$, and we can
truncate the region of integration at $w_{\rm max} \approx 2/(x_j
\gamma_{j})$.  Thus, using $g_{\rm NR} du \approx \frac{1}{2}
\sqrt{\frac{\pi}{2}} \, x_j^{5/2} e^{-x_j} \, \frac{w \, dw}{w^2+1}$
for $0 < w < w_{\rm max}$ (and 0 otherwise), we can write
\bea
x^{4}_{j} \, C_{j,k}^{\rm non-rel}[x_{j}] &\approx& m_{j}^4 \, \frac{2\pi^2}{(2\pi)^6}
\sqrt{\frac{\pi}{2}} \, x_j^{5/2} \, e^{-x_j} f_{NR}(m_{j},m_{k},m_{r'}) \log\left[1 + \frac{4}{x^2_j\gamma^2_{j}}\right]~,
\label{CNR}
\eea
where one should recall that $x_{j} = (m_{j}/m_{t'}) \, x$, and we
defined
\bea
f_{NR}(m_{j},m_{k},m_{r'}) &=& \left[ \frac{x^{4}_{j}}{(u^{2}-x^{2}_{j})^2 + x^{4}_{j} \, \gamma_{j}^{2} } \right]^{-1} \times F \sigma^{\rm non-rel}_{gt^{j} \to r' t^{k}}~,
\eea
a dimensionless function of the masses and couplings only (not
temperature) that can be read from Eq.~(\ref{ScatteringNR}).

In the relativistic limit, $x_{j} \ll 1$, the $u$-dependence of the
integrand in the collision term of Eq.~(\ref{master2}) is given by
\bea
g_{\rm R} &=& K_{1}(u)(u^2-x_{j}^2) \times \left[ 4 \log \left( \frac{u}{x_{k}} \right) - 3 \right]~.
\eea
This function has a maximum approximately at $u_{max} \approx
\frac{3}{2}+ 1/\log\left[\frac{3}{2x_{k}}\right]$.  We may then
replace $u$ by $u_{max}$ inside the logarithm in $g_{\rm R}$.  Using
$\int_{x_{j}}^{\infty} K_{1}(u)(u^2-x_{j}^2) = 2 x_{j} K_{1}(x_{j})
\approx 2$ for $x_{j} \ll 1$, we can then write
\bea
x_j^{4} \, C^{\rm rel}_{j,k}[x_j] &\approx& m_{j}^4 \,\frac{16\pi^2}{(2\pi)^6} \,
f_{R}(m_{j},m_{k},m_{r'}) \, \left[ 4 \log \left( \frac{u_{max}}{x_{k}} \right) - 3 \right]~,
\label{CR}
\eea
where $x_{j} = (m_{j}/m_{t'}) \, x$ and $x_{k} = (m_{k}/m_{t'}) \, x$, while
\bea
f_{R}(m_{j},m_{k},m_{r'}) &=& \frac{\alpha_{s}}{64 N_{c}} \, \frac{m_{j}^{2}}{\Lambda_{r}^{2}} \left[ (G_{1jk}^{RL})^{2} + (G_{1jk}^{LR})^{2} \right]
\label{fR}
\eea
is temperature-independent.  Thus, the contribution to $r'$ production
via scattering of highly relativistic particles to the l.h.s.~of
Eq.~(\ref{omegar1}) scales like $x \sim 1/T$ (up to logarithms).
Unlike the decay processes, scatterings that occur all the way up to
the reheat temperature, $T_{R}$, contribute uniformly in the
$x$-variable.  These results follow essentially from the phase space
for both cases (taking into account the time dilation factor in the
decay widths).  Eqs.~(\ref{CNR}) and (\ref{CR}) provide the basic
analytic results to understand the production of $r'$ via scattering
of KK fermions off the plasma.

\begin{figure}[t]
\centerline{
\includegraphics[height=5.2cm]{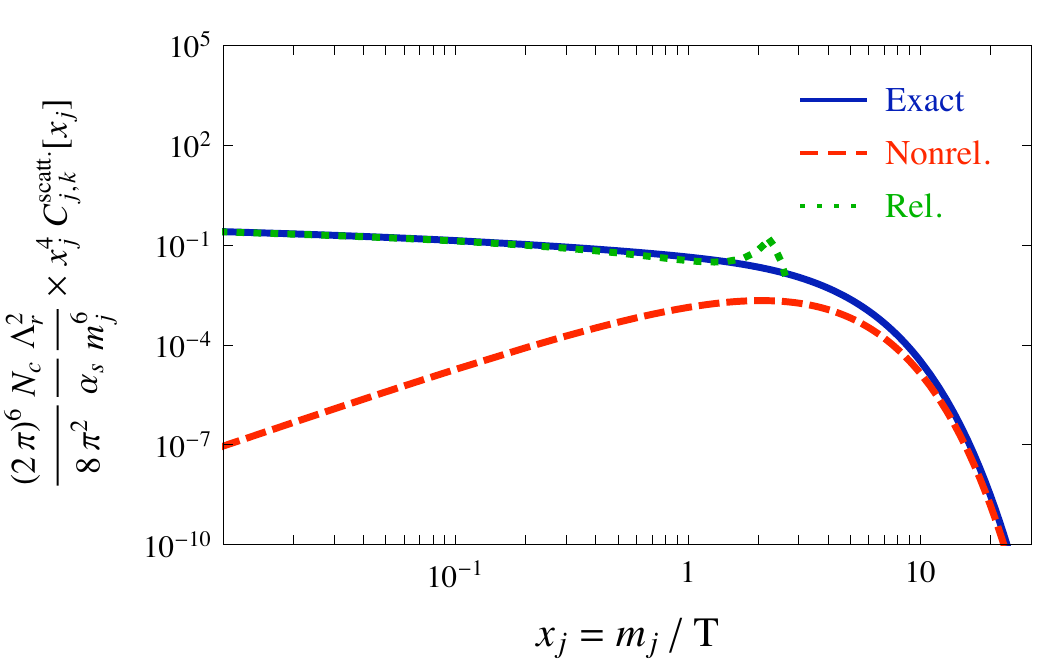}
\hspace*{3mm}
\includegraphics[height=5.2cm]{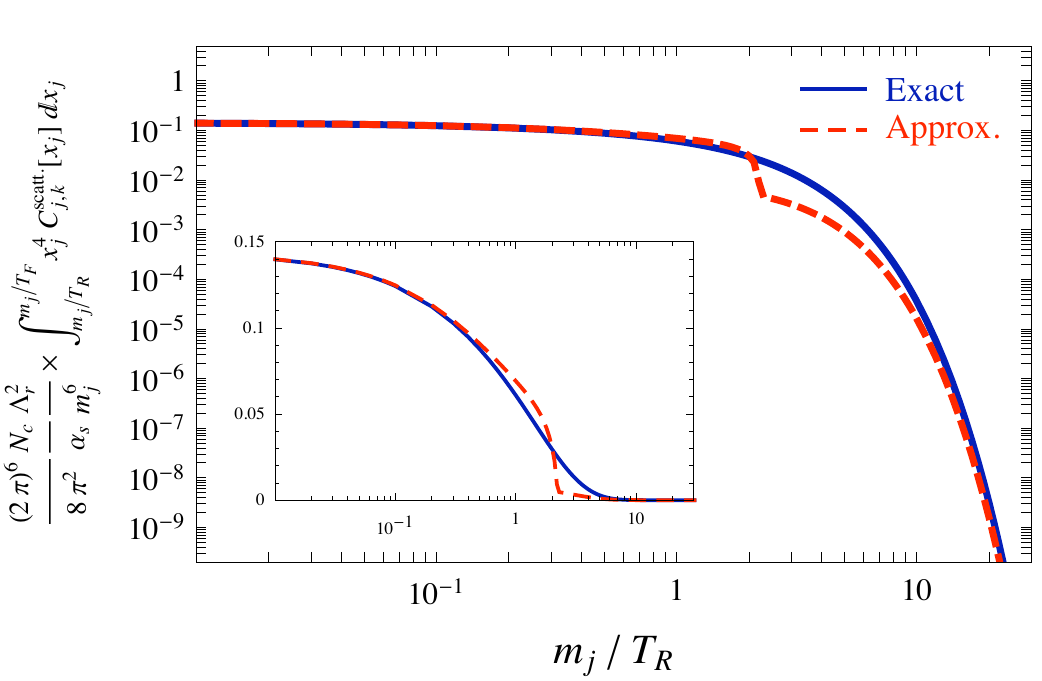}
}
\caption{{\em Left panel: Collision operator for the process $gt^{j}
\to r' t^{k}$ as a function of $x_{j} = m_{j}/T$.  The solid (blue)
curve is the exact result with the full cross section,
Eq.~(\ref{master2}), the dashed (red) curve is the non-relativistic
approximation, Eq.~(\ref{CNR}), and the dotted (green) curve is the
relativistic approximation, Eq.~(\ref{CR}).  Right panel: the
integral w.r.t. $x_{j}$ of the previous quantity, as a function of
$m_{j}/T_{R}$ (see discussion in the main text).  It is assumed that
$m_{j}/T_{F} \gg 1$, where $T_{F}$ is the $t'$ freeze-out temperature.
The solid (blue) curve corresponds to the exact result.  The dashed
(red) curve is obtained by using the relativistic approximation when
$x_{j} < x^{\rm tr}_{j,k}$ and the non-relativistic approximation when
$x_{j} \geq x^{\rm tr}_{j,k}$, where $x^{\rm tr}_{j,k} =
3m_{j}/(2m_{k})$.  The inset shows the same plot with a linear
vertical scale.  We have assumed the same parameters as in
Fig.~\ref{fig:ScatteringXS}.  }}
\label{fig:CollisionOP}
\end{figure}
In the left panel of Fig.~\ref{fig:CollisionOP} we show $x_j^{4} \,
C^{\rm scatt.}_{j,k}[x_{j}] $ as given in Eq.~(\ref{master2}) for the
process $gt^{j} \to r' t^{k}$, as a function of $x_{j} = m_{j}/T$, and
factoring out the overall parametric dependence
$\frac{8\pi^2}{(2\pi)^6} \frac{\alpha_{s}}{N_{c}}
\frac{m^{6}_{j}}{\Lambda^{2}_{r}}$ (solid, blue curve).  We also show
the non-relativistic (dashed, red) and relativistic (dotted, green)
approximations given in Eqs.~(\ref{CNR}) and (\ref{CR}), respectively.
We see that the relativistic approximation works well up to $x_{j}
\sim {\cal O}(1)$, while for larger $x_{j}$ the non-relativistic
approximation may be used.  In the right panel of
Fig.~\ref{fig:CollisionOP} we show the integral of the previous
quantity with respect to $x_{j}$ from $x_{j} = m_{j}/T_{R}$ to
$x_{j}=m_{j}/T_{F}$ (solid, blue curve).  Here $T_{F}$ is the
temperature when $t'$ freezes-out, and it is assumed that $m_{j}/T_{F}
\gg 1$.  Note that this integral differs from the integral appearing
in Eq.~(\ref{omegar1}) for the relic density by an overall factor of
$\epsilon^{5}_{t',j} \equiv (m_{t'}/m_{j})^{5}$.  The dashed, red
curve corresponds to an approximation where the simpler relativistic
and non-relativistic expressions are used for $x_{j} < x^{\rm
tr}_{j,k}$ and $x_{j} \geq x^{\rm tr}_{j,k}$, respectively, for a
conveniently chosen transition point $x^{\rm tr}_{j,k}$.  We find that
$x^{\rm tr}_{j,k} = 3m_{j}/(2m_{k})$ gives a reasonably good
approximation throughout the whole range, except in a region where
$m_{j}/T_{R} \sim {\rm few}$.  In most of this region, however, the
contribution to the final $r'$ relic density is suppressed since
$t^{j}$ becomes non-relativistic.  We also see from the rather weak
logarithmic dependence on $x_{j}$ of the relativistic expression given
in Eq.~(\ref{CR}) that, for $m_{j}/T_{R} \ll 1$, the contribution to
the integral in the l.h.s.~of Eq.~(\ref{omegar1}) becomes essentially
independent of $T_{R}$ [see also inset in the right panel of
Fig.~\ref{fig:CollisionOP}, and discussion after Eq.~(\ref{fR})].
This is similar to the decay process discussed in
Subsection~\ref{sec:decays}.  We can therefore easily compare the
contributions form scattering and decays for reheat temperatures that
are large compared to the given particle masses.  Using the result of
Eq.~(\ref{ColIntDecay}) together with Eq.~(\ref{ColDecay}) and
$\Gamma^{\rm CM}_{j,k} \sim m^{3}_{j}/(16\pi \Lambda_{r}^{2})$ [see
Eqs.~(\ref{GammaCMtj}) or (\ref{GammaCMgj})], we have
\bea
\frac{\int^{x_{F}}_{x_{R}} \! dx \, x^{4} \, C^{\rm decay}_{j,k}[x]}{\int^{x_{F}}_{x_{R}} \! dx \, x^{4} \, C^{\rm scatt.}_{j,k}[x]} &\sim& \frac{[4\pi/(2\pi)^{3}] \, [m_{j}^{6}/(16\pi \Lambda_{r}^{2})] \, (3\pi/2)}{[8\pi^2/(2\pi)^6] \, (\alpha_{s}/N_{c}) \, (m^{6}_{j}/\Lambda^{2}_{r}) \times 10^{-1}}
~\sim~ \frac{75\pi^{2} N_{c}}{20\alpha_{s}}~,
\eea
where the $10^{-1}$ is read from the right panel of
Fig.~\ref{fig:CollisionOP}.  This estimate indicates that the
contribution to $\Omega_{r'}$ from the decay processes dominates over
the one due to scatterings.

\subsection{Constraints on a Superweakly Interacting KK-Radion}
\label{sec:results}

\begin{figure}[t]
\centerline{
\includegraphics[height=6cm]{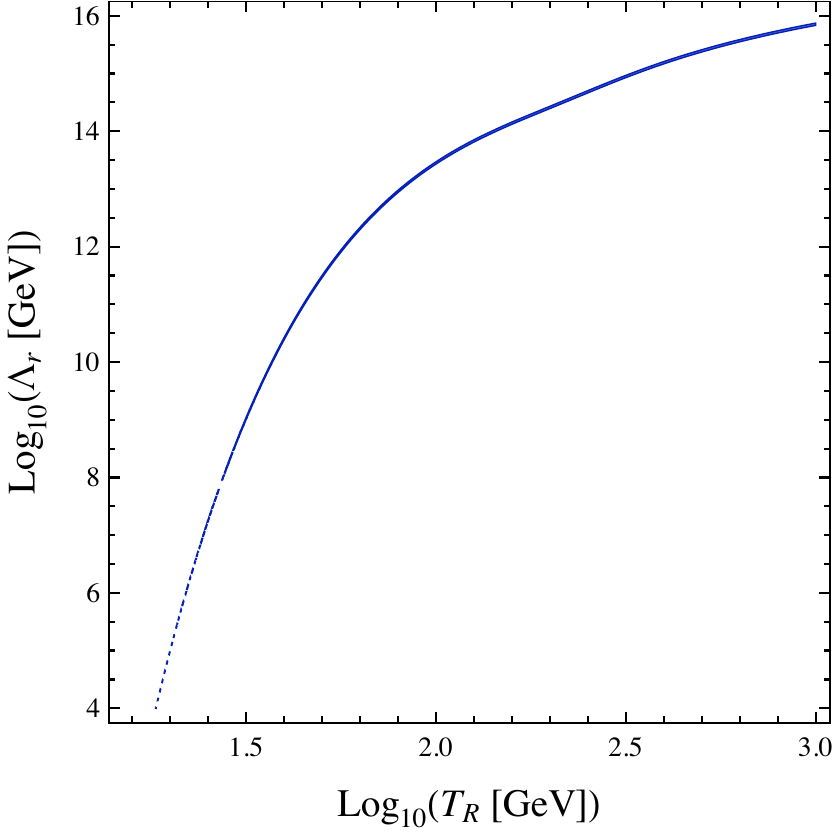}
\hspace*{8mm}
\includegraphics[height=6cm]{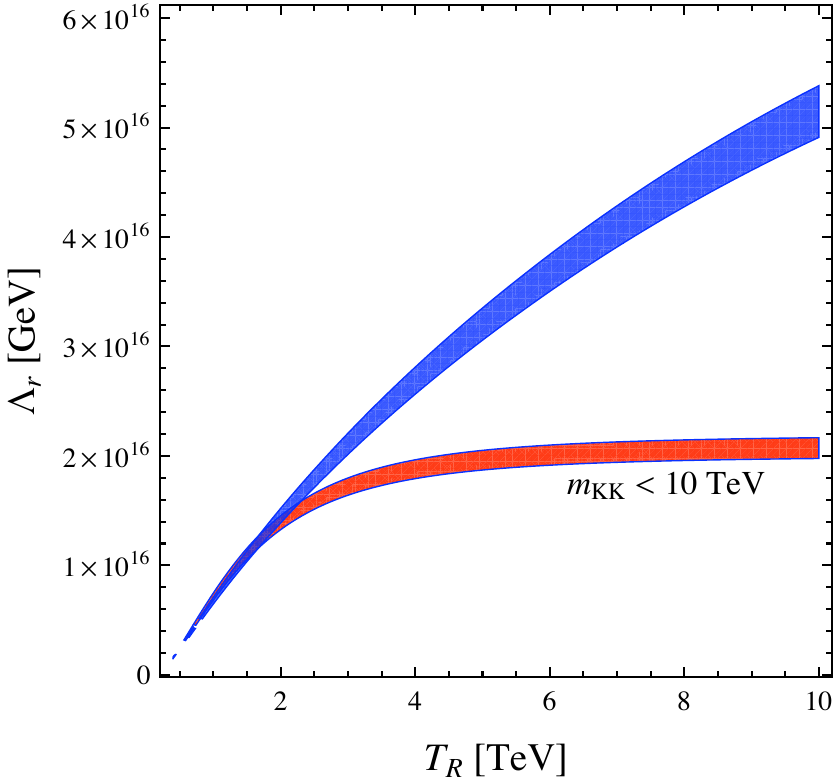}
} \caption{{\em The WMAP constraint in the $\Lambda_{R}$-$T_{R}$
plane.  The left panel shows the allowed region on a logarithmic
scale for $15~{\rm GeV} < T_{R} < \tilde{k}_{\rm eff} \sim 1~{\rm
TeV}$ and $10^4~{\rm GeV} < \Lambda_{r} < 10^{16}~{\rm GeV}$.  In
the right panel we show, on a linear scale, the WMAP allowed
region up to $T_{R} = 10~{\rm TeV}$ (the cutoff of the KK theory),
and illustrate the effect of the number of KK modes included.  The
blue (upper) region includes contributions from KK states up to an
order of magnitude above the maximum reheat temperature shown. The
red (lower) region shows the result of truncating the KK tower at
$10~{\rm TeV}$.}} \label{fig:nonthermalrpdensity}
\end{figure}

In the previous subsection we have setup the formalism to compute the
production of $r'$ in decays of heavy particles, and in scatterings
against the plasma.  We established that the decay processes dominate
over the scattering ones, essentially as a result of phase space
considerations, and the fact that at any given energy there are KK
states with masses of that order, whose decays are important.  Thus,
we can focus on the production of $r'$ via heavy KK mode decays.  For
a given reheat temperature, $T_{R}$, decays of particles sufficiently
heavier than $T_{R}$ give an exponentially small contribution, while
decays of particles lighter than $T_{R}$ give a contribution to the
integral in the l.h.s.~of Eq.~(\ref{omegar1}) of order
$(4\pi)/(2\pi)^{3} m_{j}^{3} \, \Gamma^{\rm CM}_{j,k}
(m_{t'}/m_{j})^{5} (3\pi/2) = (3G^{2}/64\pi^{2}) m_{t'}^{5} m_{j}
/\Lambda_{r}^{2}$, where $G$ is a dimensionless effective coupling
constant for the corresponding decay vertex.  The linear dependence on
$m_{j}$ translates into a linear dependence on the reheat temperature,
which sets the effective cutoff for how many states give a
non-negligible contribution to the $r'$ relic density.  Thus, the
contribution from each KK tower is dominated by the states whose
masses are around $T_{R}$.  In fact, our discussion in
Subsection~\ref{sec:decays} on the $r'$ production due to decays of
non-relativistic particles, shows that this contribution can be
significant even for masses an order of magnitude (or so) above
$T_{R}$.  One should also include a multiplicity factor for the
internal degrees of freedom for the decaying particles.  For instance,
for each of the $N_{f} = 6$ quark towers, the multiplicity is $2
\textrm{ ($q$ and $\bar{q}$)} \times 2 \textrm{ (spin)} \times N_{c}
\textrm{ (color)}$.  Thus, from the quark towers alone, setting $m_{j}
\sim T_{R}$, and using Eq.~(\ref{omegar1}), we can estimate the KK
radion density arising from decays of KK modes as
\bea
\Omega_{r'}^{(1)} h^{2} &\sim& 0.06 \, G^{2} \times \left( \frac{ 10^{16}~{\rm GeV}}{\Lambda_{r}} \right)^2 \left( \frac{T_{R}}{10~{\rm TeV}} \right) \left( \frac{m_{r}}{300~{\rm GeV}} \right)~,
\label{NonThermalEstimate}
\eea
where we took $g_{*} = 86.25$ Note that Eq.~(\ref{NonThermalEstimate})
is independent of $m_{t'}$.  A more precise determination is shown in
Fig.~\ref{fig:nonthermalrpdensity}, were we show the region selected
by WMAP in the $T_{R}$-$\Lambda_{r}$ plane, by treating the top towers
separately from the light fermion towers (as suggested by the anarchic
picture of flavor), and including the decays of KK gauge bosons, as
well as the much smaller $r'$ production by scatterings.  The internal
degrees of freedom associated with the top tower give a factor
$2\times 2\times N_{c}$, while for the light generation towers we have
a factor $5\times 2\times 2\times N_{c} + (3+3/2)\times 2\times 2$,
which includes the five lighter quarks, three charged leptons and
three LH neutrinos.  The corresponding factor for KK gauge boson
decays is $3(N_{c}^{2} - 1) + 3 \times 4$.  We also include the $r'$
produced in decays of $t'$ after its freeze-out, which we previously
denoted by $\Omega_{r'}^{(2)} h^{2}$, although this contribution is
very small (see below).

>From our discussion in Section~\ref{sec:HighT}, one might expect that
if the reheat temperature is too large, the correct metric background
would correspond to a blackhole solution, and the KK modes, including
the radion tower cease to have meaning (in the 4D dual picture, the
theory would be in a deconfined phase).  Naively, we expect the
critical temperature to be of order $\tilde{k}_{\rm eff}$, and
therefore we show in the left panel of
Fig.~\ref{fig:nonthermalrpdensity} reheat temperatures up to
$\tilde{k}_{\rm eff} \sim 1~{\rm TeV}$.  We also show reheat
temperatures as low as $T_{R} = 15~{\rm GeV}$ (which is about the $t'$
freeze-out temperature), where the KK-radion density is still small
compared to its equilibrium distribution.  The radion decay constant
that reproduces the observed DM density ranges from
$10^{4}-10^{16}~{\rm GeV}$, depending on $T_{R}$.

However, given that we do not have the corresponding blackhole
solution that would allow us to compute the critical temperature, we
also show in the right panel of Fig.~\ref{fig:nonthermalrpdensity} the
result of assuming that the KK theory is the correct description up to
temperatures of order its 4D cutoff (which we estimate to be of order
$5-10~{\rm TeV}$).  Even in this case, WMAP selects a $\Lambda_{r}$ not
much larger than $10^{16}~{\rm GeV}$.  For reheat temperatures above a
TeV, lower values of $\Lambda_{r}$ would lead to overproduction of
$r'$, while larger values would make the $r'$ species a subdominant
component of the observed DM density.  However, note that, within the
model, $\Lambda_{r}$ cannot be arbitrarily large, but is bounded from
above by the Planck mass.~\footnote{Recall that these scales are only
effective scales, characterizing the strength of radion and graviton
interactions.  In the present ``small warping'' scenario, the physical
cutoff of the theory on the UV brane is in fact significantly smaller,
of order the 5D Planck mass, $M_{5}$.} Thus, quite aside from the
likely difficulty in nucleating to the true vacuum from a
``deconfined'' one if the temperature is above the critical
temperature, this result suggests that $T_{R}$ cannot be much higher
than the cutoff of the effective theory, or else the KK-radions would
be overproduced.

As mentioned above, states somewhat heavier than $T_{R}$ in general
cannot be neglected.  This raises the question of how to treat these
states when $T_{R}$ is close to the cutoff of the effective theory,
even assuming that the theory remains in the ``confined phase'', so
that a KK radion can be properly identified.  Nevertheless, even if
the relevant degrees of freedom above the cutoff are not accurately
described by the KK theory, one might expect that their couplings to
the KK-radion are still suppressed by $\Lambda_{r}$.  One can
therefore get an idea by including the contributions from KK states up
to about one order of magnitude above the cutoff.  The result
corresponds to the upper, blue region in the right panel of
Fig.~\ref{fig:nonthermalrpdensity}, which exhibits the linear
dependence on $T_{R}$ previously discussed.  However, we also show the
result of truncating the KK tower at a cutoff of $10~{\rm TeV}$
(lower, red region), as might be appropriate if for some unknown
reason the relevant degrees of freedom above that scale are
significantly more weakly coupled to the KK-radion than expected from
the KK picture.  The comparison between the two curves gives an idea
of the effect of the heavy states at larger reheat temperatures.

\begin{figure}[t]
\centerline{
\includegraphics[height=5.2cm]{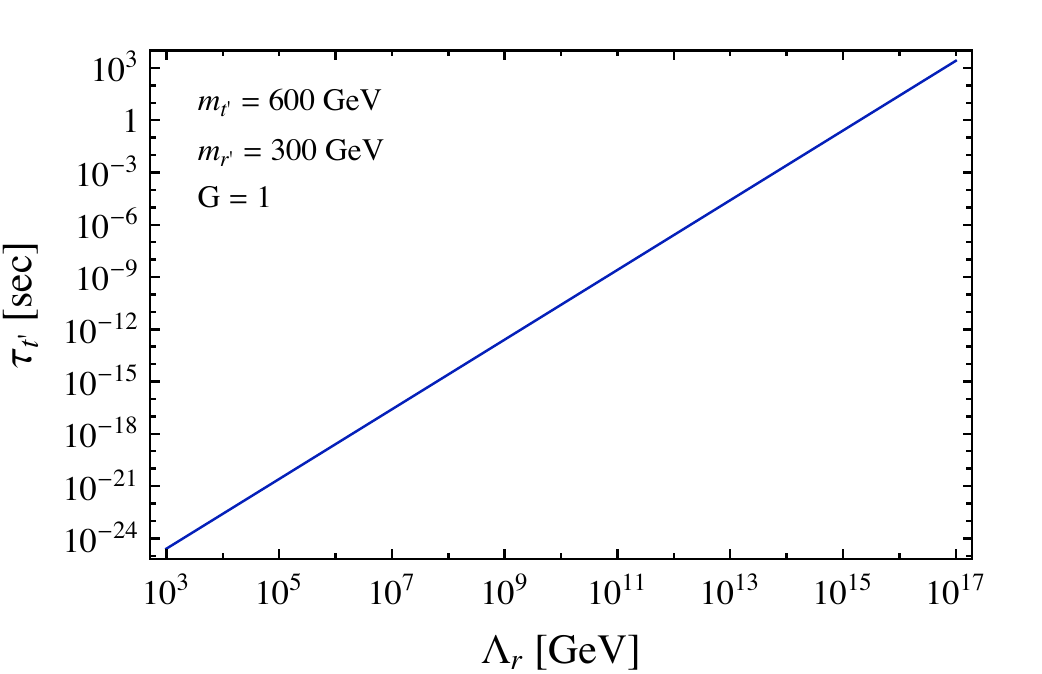}
\hspace*{0mm}
\includegraphics[height=5cm]{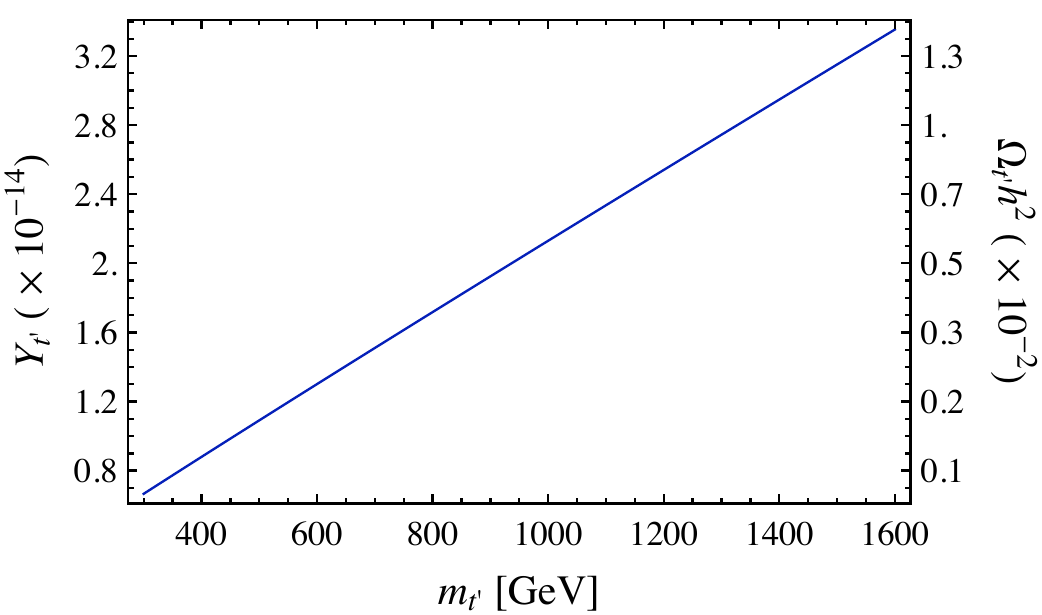}
} \caption{{\em Left panel: $t'$ lifetime as a function of $\Lambda_r$
for $m_{t'} = 600~{\rm GeV}$, $m_{r'} = 300~{\rm GeV}$ and $G = 1$
(see text).  Right panel: $t'$ yield, $Y_{t'}(x_F)$, as a function of
$m_{t'}$}.  For reference, in the right axis we show $\Omega_{t'} h^2
\equiv (m_{t'}/m_{r'}) \Omega^{(2)}_{r'} h^2$ corresponding to the
yield shown in the left axis and the mass $m_{t'}$ given by the
curve.}
\label{fig:BBN}
\end{figure}

In this scenario, the large value of $\Lambda_{r'}$ would make the
NLKP long-lived.  For instance, if the NLKP is $t'$, and assuming
$m_{t'} > m_{r'} + m_{t}$, it decays via $t' \to r' t$ with a decay
width given by
\bea
\Gamma_{t'} &=&
\frac{G^2m^3_{t'}}{16 \pi \Lambda_{r}^2} \, \sqrt{ \rule{0mm}{3.5mm}
\left[1 - \left(\frac{m_{t}}{m_{t'}} + \frac{m_{r'}}{m_{t'}}\right)^2 \right]
\left[1 - \left(\frac{m_{t}}{m_{t'}} - \frac{m_{r'}}{m_{t'}}\right)^2 \right] }~,
\label{Gammatp}
\eea
where $G$ is an effective coupling constant [see
Eq.~(\ref{GammaCMtj})].  In Eqs.~(\ref{Gammahm}) and
(\ref{GammaHpma}), we give the decay widths for the KK-parity odd
Higgses in the case that one of these is the NLKP. The constraints are
similar to the case of a $t'$ NLKP, and in the following we
concentrate on the latter case.  In the left panel of
Fig.~\ref{fig:BBN}, we show the $t'$ lifetime as a function of
$\Lambda_{r}$ for $m_{t'} = 600~{\rm GeV}$, and taking $G=1$.  The
result for other values can be roughly obtained by simple rescaling,
according to the prefactor in Eq.~(\ref{Gammatp}).  These late time
decays are potentially dangerous.  When massive particles decay into
high energy quarks or gluons, the latter rapidly fragment into
hadrons, and if the decays of the mother particle occurs at times
$10^{-2}\; {\rm s}\lesssim t\lesssim 10^4\; {\rm s} $,\footnote{At
earlier times, such hadrons predominantly scatter off the background
photons and electrons, transferring most of their kinetic energy and
reaching kinetic equilibrium.} the scattering off background protons
and neutrons via the strong interactions can lead to a dangerous
interconversion of background protons and neutrons even after the
freeze-out time of the neutron-proton ratio.  The effect tends to
increase $n/p$, thus increasing the $^{4}{\rm He}$ abundance with
respect to the standard predictions of BBN. For lifetimes smaller than
about $10^{4}~{\rm s}$, EM energy deposition is weakly
constrained~\cite{Feng:2004zu}, hence the left panel of
Fig.~\ref{fig:BBN} suggests that the constraints arise primarily from
the hadronic energy injection.

In the right panel of Fig.~\ref{fig:BBN} we show the $t'$ yield,
$Y_{t'}(x_{F})\equiv n_{t'}(x_F)/s(x_F)$, which is set by the QCD
processes of Eq.~(\ref{tpQCDSimple}), and depends only on $m_{t'}$.
In the right axis of the same figure we show the associated
$\Omega_{t'} h^{2}$, which leads to $\Omega^{(2)}_{r'} h^{2} =
(m_{r'}/m_{t'}) \Omega_{t'} h^{2}$, according to Eq.~(\ref{Omega2}).
We can see that the latter one is typically of order $10^{-2}$ or
less, and therefore much smaller than the total $\Omega_{r'} h^{2}$.
This suppression in the $t'$ yield allows the picture to be roughly
consistent with BBN constraints even when the $t'$ decay during or
somewhat after BBN. The yield can be constrained as a function of the
lifetime of the long-lived relic and its mass, which determine the
amount of injected energy and when this injection
occurs~\cite{Kohri:2001jx,Kawasaki:2004yh}.  We can therefore bound
$\Lambda_{r}$ as a function of $m_{t'}$ as follows.  For a given
$m_{t'}$ the right panel of Fig.~\ref{fig:BBN} gives the yield,
$Y_{t'}$.  From the results of Ref.~\cite{Kohri:2001jx} we determine
how late the relic can decay without being in conflict with BBN
constraints (typically around $200~{\rm s}$).  This can then place an
upper bound on $\Lambda_{r}$ via the $t'$ decay width.  Assuming for
illustration that $G = 1$ and $m_{r'} = 300~{\rm GeV}$ in
Eq.~(\ref{Gammatp}), we obtain in this way
\begin{table}[h]
\begin{center}
\begin{tabular}{|c|c|c|c|c|}
\hline
\rule{0mm}{4mm}
$m_{t'}$ [GeV] & 500 & 800 & 1000 & 1200 \\ [0.3em]
\hline
\rule{0mm}{5mm}
$\Lambda_{r}~{\rm [GeV]} \lesssim$
& $2 \times 10^{16}$ & $5 \times 10^{16}$ & $7 \times 10^{16}$ & $10^{17}$ \\ [0.3em]
\hline
\end{tabular}
\end{center}
\label{mtpvsLambda}
\end{table}

\vspace{-6mm}

\noindent which gives an upper bound on $\Lambda_{r}$ for a few
typical values of $m_{t'}$ (or, equivalently, a lower bound on
$m_{t'}$ for several values of $\Lambda_{r}$).  Of course, if
$m_{t'} < m_{r'} + m_{t}$, so that we have a three-body
decay, the BBN constraint becomes more stringent.

There is a stronger constraint from the late decays of the
\textit{radions} (produced in decays of KK states, analogously to the
production of the KK-radions described above).  In fact, since the
masses and wavefunctions of the radion and KK-radion are nearly
identical, their couplings to KK modes are very similar (with
appropriate replacements of the semi-degenerate even and odd modes, as
dictated by KK parity).  We have checked that the number density of
radions is very similar to that of KK-radions.  The radion yield can
then be expressed as
\bea
Y_{r} &\approx& \frac{\Omega_{r'} \rho_{c}}{s_{0} m_{r'}}~,
\eea
where $\rho_{c}$ and $s_{0}$ are the critical density and entropy of
the universe today, respectively, and $\Omega_{r'}$ is the DM relic
density.  For instance, for $m_{r'} = 500~{\rm GeV}$, and assuming
$\Omega_{r'} h^2 = 0.1$, we get a radion yield of $Y_{r} \approx 7
\times 10^{-13}$.  On the other hand, the decays of the radion into
$W^\pm$, $Z$, and $h_{+}$ give a radion lifetime of
\bea
\tau_{r} &\approx& \left[\frac{m_{r}^3}{8 \pi \Lambda_{r}^2} \right]^{-1}~\approx ~ 0.1 \times \left( \frac{\Lambda_{r}}{10^{15}~{\rm GeV}} \right)^2
\left( \frac{500~{\rm GeV}}{m_{r'}} \right)^3~{\rm sec}~.
\eea
Here we assumed that $m_{W,Z},m_{h} \ll m_{r}$ and neglected the
smaller width into $t\bar{t}$ and massless gauge bosons (we give the
exact formulas in Eqs.~(\ref{Gammarff})--(\ref{Gammarhh}) of
Appendix~\ref{RadionDecays}).  For the above reference parameters
($m_{r} \approx m_{r'} = 500~{\rm GeV}$ with $Y_{r} \approx 7 \times
10^{-13}$), Ref.~\cite{Kohri:2001jx} gives an upper bound on the
radion lifetime of about $0.5~{\rm sec}$.  Referring to the left panel in
Fig.~\ref{fig:nonthermalrpdensity}, we infer that the reheat
temperature should be at most in the few hundred GeV range, but that
under this assumption it is possible for KK-radions to account for
the DM relic density, while being consistent with BBN
constraints (from both radion and $t'$ decays~\footnote{Heavier KK
radions are more weakly coupled and therefore decay even later.
However, they are also produced in smaller quantities, and given that
the reheat temperature is significantly below their mass, we expect
their number density to be exponentially suppressed.}).  Recall that a
maximum reheat temperature of order a TeV may be required in order to
not exceed the critical temperature for a deconfinement/confinement
phase transition, as discussed in Section~\ref{sec:HighT}.

\section{Direct and Indirect Detection}
\label{sec:Detection}

We now turn to the feasibility of KK-radion detection in either direct
or indirect DM searches, as well as collider experiments.


\medskip
\noindent
\underline{\textit{Direct detection:}}

\noindent As has been mentioned before, most of the interactions
between the KK-radion and the SM particles are non-renormalizable, and
suppressed by $\Lambda_{r}$.  After EWSB, renormalizable interactions
with the SM can be induced by terms of the form $\frac{1}{2} \left[
\delta(y - L) +\delta(y + L) \right] \sqrt{g_{\rm ind}} \, \xi {\cal
R}_{4} H^{\dagger} H$ (subject to the KK-parity symmetry).  For
simplicity, we restrict here to an exactly IR localized Higgs field,
so that the previous operators are IR localized and involve the Ricci
scalar constructed from the induced metric.  These induce kinetic
mixing between the radion and the SM-like
Higgs~\cite{Csaki:1999mp,Giudice:2000av}.  In the present context they
also induce kinetic mixing between the KK-radion and the KK-parity odd
(CP-even) $h_{-}$ that resides in the inert Higgs doublet, as already
mentioned at the end of Section~\ref{sec:DM}.  Canonical normalization
is achieved by the field redefinitions $r_{\pm} = \bar{r}_{\pm}/Z$ and
$h_{\pm} = \bar{h}_{\pm} + (\delta/Z) \bar{r}_{\pm}$, where $\delta =
6\sqrt{2} \xi v/\Lambda_{r}$ and $Z^2 = 1 + 12 \xi (1 - 6 \xi)
v/\Lambda_{r}$, with $v = 174~{\rm GeV}$.  In the following, we will
use the notation $r_{+} = r$ and $r_{-} = r'$ to denote the radion and
KK radion states. After the previous field redefinitions, the mass
matrix becomes non-diagonal.  The mass eigenbasis is obtained by the
orthogonal transformations:
\bea
\left(
\begin{array}{ccc}
r_{\pm}  \\
h_{\pm}
\end{array}
\right)
&=&
\left(
\begin{array}{ccc}
U_{r_{\pm},L}  & U_{r_{\pm},H}   \\
U_{h_{\pm},L}  & U_{h_{\pm},H}
\end{array}
\right)
\left(
\begin{array}{ccc}
\phi_{\pm,L}  \\
\phi_{\pm,H}
\end{array}
\right)~,
\eea
where the subscript $L$ ($H$) refers to the lighter (heavier) mass
eigenstate within the KK-parity even and odd sectors, which we denote
with the $+$ and $-$ subscripts, as usual.  The matrix elements above
are explicitly given by
\bea
U_{r_{\pm},L} = \frac{a_{\pm} - c_{\pm} - \Delta_{\pm}}{\sqrt{(a_{\pm} - c_{\pm} - \Delta_{\pm})^2 + 4 b_{\pm}^2}}~,
\hspace{1cm}
U_{r_{\pm},H} = \frac{a_{\pm} - c_{\pm} + \Delta_{\pm}}{\sqrt{(a_{\pm} - c_{\pm} + \Delta_{\pm})^2 + 4 b_{\pm}^2}}~,
\nonumber \\[0.5em]
U_{h_{\pm},L} = \frac{2 b_{\pm}}{\sqrt{(a_{\pm} - c_{\pm} - \Delta_{\pm})^2 + 4 b_{\pm}^2}}~,
\hspace{1cm}
U_{h_{\pm},H} = \frac{2 b_{\pm}}{\sqrt{(a_{\pm} - c_{\pm} + \Delta_{\pm})^2 + 4 b_{\pm}^2}}~,
\eea
where $\Delta_{\pm} = \sqrt{(a_{\pm} - c_{\pm})^2 + 4 b_{\pm}^2}$,
and~\footnote{Note that, when $a_{\pm} < c_{\pm}$,
$(U_{r_{\pm},L},U_{h_{\pm},L}) \to (-1,0)$ and
$(U_{r_{\pm},H},U_{h_{\pm},H}) \to (0,1)$ as $b_{\pm} \to 0$.  When
$a_{\pm} > c_{\pm}$, one has instead $(U_{r_{\pm},L},U_{h_{\pm},L})
\to (0,1)$ and $(U_{r_{\pm},H},U_{h_{\pm},H}) \to (1,0)$ as $b_{\pm}
\to 0$.}
\bea
a_{\pm} = \frac{1}{Z^2} \left( m^2_{r_{\pm}} + \delta^2 m^2_{h_{\pm}} \right)~,
\hspace{1cm}
b_{\pm} = \frac{\delta}{Z} \, m^2_{h_{\pm}}~,
\hspace{1cm}
c_{\pm} = m^2_{h_{\pm}}~.
\eea
Here $m^2_{r_{\pm}}$ and $m^2_{h_{\pm}}$ are the mass parameters in
the $(r_{\pm}, h_{\pm})$ basis.  The physical masses in the KK-parity
even and odd sectors are given by
\bea
m^2_{\pm,L} &=& \frac{1}{2} \left( a_{\pm} + c_{\pm} - \Delta_{\pm} \right)~,
\hspace{1cm}
m^2_{\pm,H} ~=~ \frac{1}{2} \left( a_{\pm} + c_{\pm} + \Delta_{\pm} \right)~.
\eea

The DM particle is the lightest KK-parity odd state above, $\phi_{\rm
DM} \equiv \phi_{-,L}$, with mass $m_{\rm DM} \equiv m_{-,L}$.  Now we
can identify the interactions relevant for scattering of the DM
candidate against nuclei.  As shown in~\cite{Medina:2010mu}, there are
$r_{-} \, h_{+} h_{-}$ and $r_{+} \, h_{-} h_{-}$ terms of the form
\bea
- \frac{2}{\Lambda_{r}} r_{-} \left[\partial_{\mu} h_{+} \partial^{\mu} h_{-} - 4 m^{2}_{h_{+}} h_{+} h_{-} \right]
&\to&
\sum_{\alpha = L,H} \frac{1}{\Lambda_{r}} \, U_{r_{-}, L} U_{h_{+}, \alpha} U_{h_{-}, L}  \left( \Box \phi_{+,\alpha} + 8 m^{2}_{h_{+}} \phi_{+,\alpha} \right) \phi_{\rm DM}^{2} ~,
\nonumber \\[0.5em]
- \frac{2}{\Lambda_{r}} r_{+} \left[\partial_{\mu} h_{-} \partial^{\mu} h_{-} - 2 m^{2}_{h_{-}} h_{-} h_{-} \right] &\to&
\nonumber \\
&& \hspace{-1.5cm}
-\sum_{\alpha = L,H} \frac{1}{\Lambda_{r}} \, U_{r_{+}, \alpha} U_{h_{-}, L}^2  \left[ \Box \phi_{+,\alpha} + (2 m^2_{\rm DM} - 4 m^{2}_{h_{-}}) \phi_{+,\alpha} \right]\phi_{\rm DM}^{2} ~,
\nonumber
\eea
where, on the r.h.s, we isolated the terms quadratic in $\phi_{\rm
DM}$.  To obtain the above expressions we integrated by parts and
assumed that $\phi_{\rm DM}$ is on-shell: $\Box \phi_{\rm DM} =
-m_{\rm DM}^2 \phi_{\rm DM}$.  There is also a contribution from the
Higgs potential
\bea
- \frac{3}{\sqrt{2}} \frac{m^2_{h_{+}}}{2v} h_{+} h_{-}^2
&\to&
- \frac{3}{\sqrt{2}} \frac{m^2_{h_{+}}}{2v}  \sum_{\alpha = L,H} U_{h_{+}, \alpha}  U_{h_{-}, L}^2  \phi_{+,\alpha} \phi_{\rm DM}^{2} ~.
\nonumber
\eea
The exchanged KK-parity even state can be either $\phi_{+,L}$ or
$\phi_{+,H}$.  The corresponding Feynman rules for the $\phi_{+,L}
\phi_{\rm DM}^2$ and $\phi_{+,H} \phi_{\rm DM}^2$ vertices are,
respectively, $(2i/\Lambda_{r}) ({\cal Z}_{L,H} \, q^2 + {\cal
M}_{L,H}^2)$, where $q$ is the 4-momentum of the KK-parity even state,
${\cal Z}_{\alpha} = U_{r_{-}, L} U_{h_{+}, \alpha} U_{h_{-}, L} -
U_{r_{+}, \alpha} U_{h_{-}, L}^2$, and ${\cal M}_{\alpha}^2 = 8
m^{2}_{h_{+}} U_{r_{-}, L} U_{h_{+}, \alpha} U_{h_{-}, L} - (2
m^2_{\rm DM} - 4 m^{2}_{h_{-}}) U_{r_{+}, \alpha} U_{h_{-}, L}^2 -
(3/2\sqrt{2}) (\Lambda_r/v) m_{h_{+}}^2 U_{h_{+},\alpha}
U_{h_{-},L}^2$.  Identifying also the Higgs (i.e. $h_{+}$) component
in $\phi_{+,L}$ and $\phi_{+,H}$, we obtain the DM-nucleon scattering
cross-section
\bea
\sigma_{\phi_{\rm DM} N \to \phi_{\rm DM} N} &\approx&
\frac{g_{\rm hNN}^2}{\pi \Lambda_{r}^2} \,
\frac{m_{N}^2}{m^4_{+,L} m^4_{+,H} (m_{N} + m_{\rm DM})^2} \,
\left( U_{h_{+},L} {\cal M}_{L}^2 m^2_{+,H} + U_{h_{+},H} {\cal M}_{H}^2 m^2_{+,L} \right)^2~.
\label{NucleonCrossSection}
\eea
In the above, we neglected the coupling of the radion to the nucleon
compared to the Higgs-nucleon coupling, $g_{\rm hNN} \approx 340~{\rm
MeV}/246~{\rm GeV}$ \cite{Burgess:2000yq}.  Note, however, that for
$\Lambda_{r} \sim 2~{\rm TeV}$, the radion-nucleon coupling, of order
$m_{N}/\Lambda_{r}$, need not be completely negligible (it can be
easily incorporated if wanted).
\begin{figure}[t]
\centerline{
\includegraphics[height=7cm]{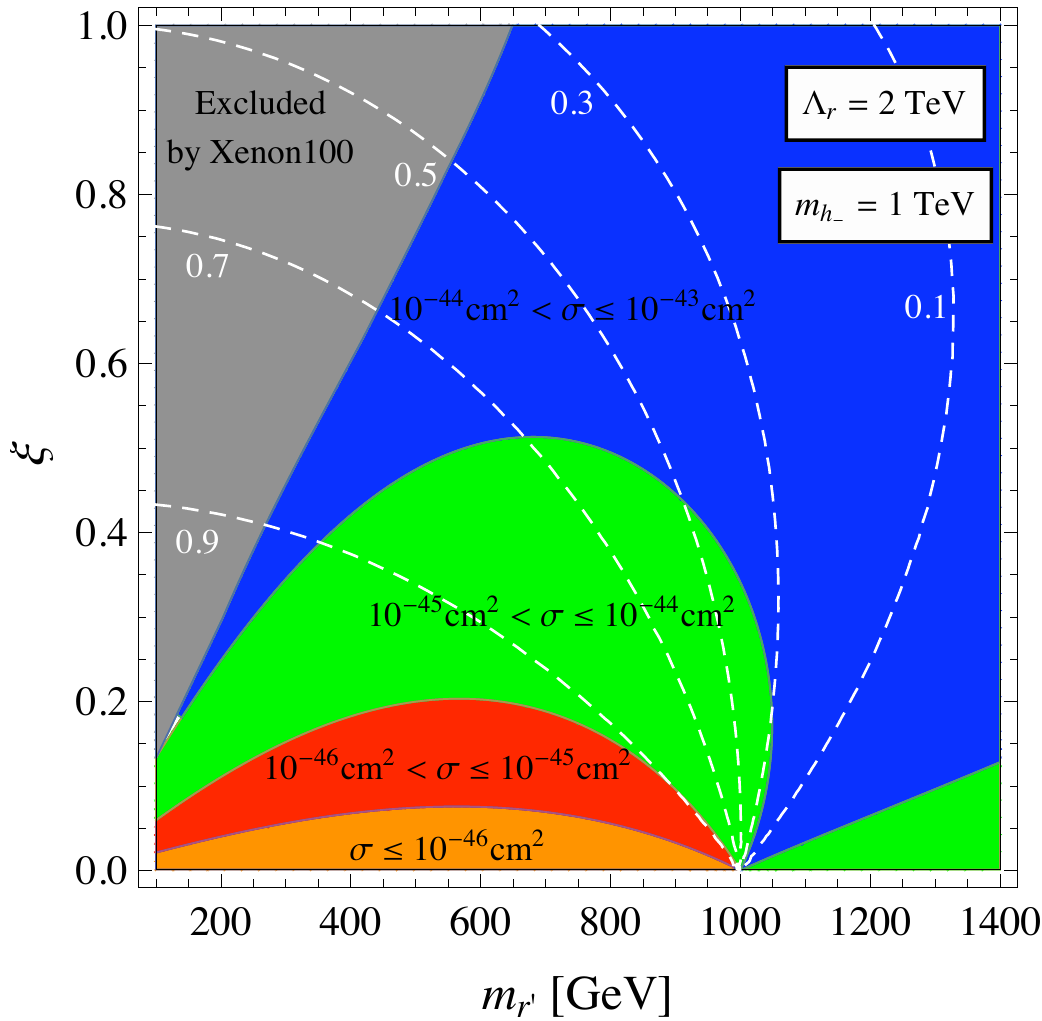}
\hspace*{5mm}
\includegraphics[height=7cm]{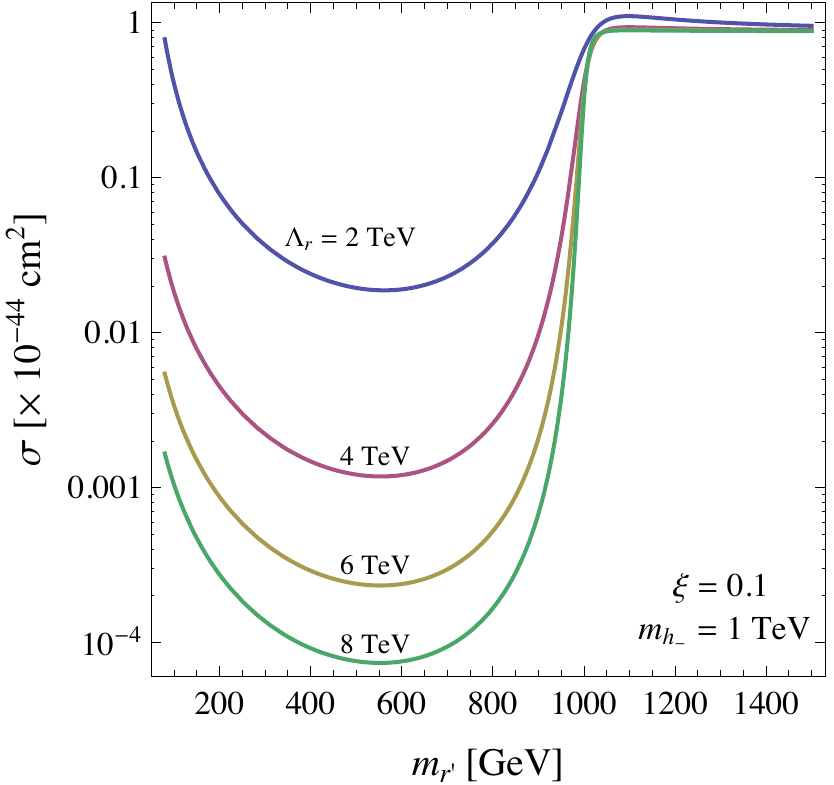}
}
\caption{{\em Left panel: Contours of constant $\sigma_{\phi_{\rm DM}
N \to \phi_{\rm DM} N}$ in the $\xi$--$m_{r_{-}}$ plane, as given in
Eq.~(\ref{NucleonCrossSection}), for $\Lambda_{r} = 2~{\rm TeV}$.
Also shown is the region currently excluded by XEONO100 (gray).  The
dashed, white lines represent the $r_{-}$ content of the DM candidate,
$|U_{r_{-},L}|^2$ (with $|U_{h_{-},L}|^2 = 1 - |U_{r_{-},L}|^2$ being
the $h_{-}$ content).  Right panel: Nucleon cross section as a
function of $m_{r_{-}}$, for different values of $\Lambda_{r}$, and
for constant $\xi = 0.1$.  Both figures correspond to $m_{h_{-}} =
1~{\rm TeV}$.}}
\label{fig:DirectDetection}
\end{figure}

In the left panel of Fig.~\ref{fig:DirectDetection}, we show contours
of constant $\sigma_{\phi_{\rm DM} N \to \phi_{\rm DM} N}$ in the
$\xi$--$m_{r_{-}}$ plane, taking $\Lambda_{r} = 2~{\rm TeV}$ and
$m_{h_{-}} = 1~{\rm TeV}$.  Note that, for practical purposes, the DM
physical mass is very close to $m_{r'} \equiv m_{r_{-}}$ for $m_{r'}
\lesssim m_{h_{-}}$, and very close to $m_{h_{-}}$ otherwise.  Thus,
the horizontal axis in the plot can be readily interpreted in terms of
$m_{\rm DM}$.  The dashed (white) lines correspond to constant
$|U_{r_{-},L}|^2$ (the $r_{-}$ content of the DM matter).  As
discussed at the end of Section~\ref{sec:DM}, the $h_{-}$ content,
given by $1 - |U_{r_{-},L}|^2$ should be small, or else the DM relic
density will be too small due to efficient annihilation via the Higgs
couplings.  However, we see that there is a region where the DM is
mostly $r_{-}$, with a sizable nucleon cross section (e.g. below the
dashed line of $|U_{r_{-},L}|^2 = 0.9$).  In fact, the current direct
DM searches may have a sensitivity to this region, as shown by the
gray region corresponding to the most recent exclusion by the XENON100
experiment~\cite{Aprile:2011hi}.

In the right panel of Fig.~\ref{fig:DirectDetection}, we show curves
of the nuclear cross section for several values of $\Lambda_{r}$ for
fixed $\xi = 0.1$ and $m_{h_{-}} = 1~{\rm TeV}$.  The plateau for
$m_{r'} > m_{h_{-}}$ corresponds to the limit in which the DM is pure
$h_{-}$, where Eq.~(\ref{NucleonCrossSection}) reduces to
\bea
\sigma_{\phi_{\rm DM} N \to \phi_{\rm DM} N} &\approx& \frac{9g_{\rm hNN}^2}{8\pi v^2}
\, \frac{m_{N}^2}{(m_{N} + m_{\rm DM})^2}~,
\eea
with $m_{\rm DM} = m_{h_{-}} = 1~{\rm TeV}$ (we also assumed
$\Lambda_{r} \gg v$, so that the exchanged $h_{+}$ has no radion
admixture).


\medskip
\noindent
\underline{\textit{Indirect detection:}}

\noindent Regarding indirect detection, the most interesting
annihilation channels are those that lead to positrons, neutrinos or
photons.  The direct annihilation cross section into $e^+ e^{-}$ pairs
takes the form ${\rm v} \sigma_{r' r' \rightarrow e^{+} e^{-}} \approx
(G_{1100}^{2}/4\pi) \,(m^{2}_{e}/\Lambda_{r}^{4})$, where $G_{1100}$
is the coupling for the contact interaction between two KK-radions and
$e^+ e^{-}$ (see second diagram in Fig.~\ref{fig:rprpAnnihilationff}).
In anarchic scenarios, due to EWSB effects and the subsequent mixing
with the electron KK resonances, the coupling $G_{1100}$ can be as
large as ${\cal O}(30)$.  However, the $m_{e}^{2}$ factor makes this
channel extremely suppressed.  Mixing with the Higgs does not help
since it also couples to electrons via the small electron Yukawa
coupling.  Similarly, the direct annihilation rate into neutrinos is
negligible.  Softer positrons/neutrinos can be produced in the decay
products of the dominant $t\bar{t}$ and Higgs (or gauge boson)
annihilation channels, but these the fluxes are
expected to be small.  For instance, given that the spin-independent
DM-nucleon cross section is bounded from above by about $10^{-43}~{\rm
cm}^2 \approx 10^{-7}~{\rm pb}$ for $m_{\rm DM} \sim 300~{\rm GeV}$,
neutrinos from DM annihilations in the Sun would produce at most a
couple of neutrino events per year~\cite{Halzen:2009vu} at neutrino
telescopes such as IceCube/Deep Core.  This is well below the
atmospheric neutrino background, and the detection  in this
channel is not likely.

\begin{figure}[t]
\centerline{
\resizebox{10.5cm}{!}{\includegraphics{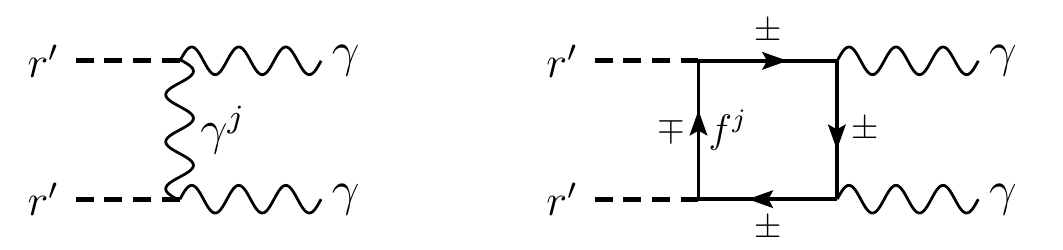}}
}
\caption{\em Tree-level (from heavy gauge boson exchange) and 1-loop
(from KK fermions) contributions to $r'r'\to \gamma\gamma$.  The $\pm$
indicate the allowed KK-parity combinations.}
\label{fig:gamma2Diagrams}
\end{figure}

We therefore focus on the photon flux from KK-radion annihilations,
starting with a possible line signal.  There are \textit{tree-level}
direct annihilations from the exchange of KK-photons, as shown in
Fig.~\ref{fig:gamma2Diagrams}.  A given KK-parity odd photon induces
the effective dimension-8 operator $-2g_{1j0}^2/(\Lambda_{r}^2
m_{j}^2) \, \partial^{\mu} [r' F_{\mu\nu}] \partial_{\alpha}[ r'
F^{\alpha\nu}]$, where $g_{1j0}$ is the coupling between the
KK-radion, the KK-photon with mass $m_{j}$, and the massless photon,
as given in Ref.~\cite{Medina:2010mu}.  Since both $\Lambda_{r}$ and
$m_{j}$ may be expected to be of order a couple TeV, this contribution
is small (we give the full cross-section for any $m_{j}$ in
Eq.~(\ref{sigmarprpVV}) of Appendix~\ref{App:Processes}).  There are
also 1-loop contributions, e.g. from fermion KK loops, as shown in the
right diagram of Fig.~\ref{fig:gamma2Diagrams}.  Imposing only
$U(1)_{\rm em}$ gauge invariance, the effect from each KK level is
logarithmically divergent by power counting, which would seem to lead
to a linear divergence after summing over KK modes.~\footnote{Although
there is a double sum in the square diagram, the ``diagonal''
couplings, i.e.~those for the pairs of KK-parity even and odd fermions
with closer masses are largest, while the corresponding couplings
rapidly decrease as the KK-fermions become more separated in
mass~\cite{Medina:2010mu}.} Nevertheless, as pointed out in
Ref.~\cite{Csaki:2007ns} for the case of the linear radion couplings
to $\gamma\gamma$ (where a triangle rather than a square diagram is
involved), these divergences simply renormalize the 5D operator
$\sqrt{g} \, F_{MN}F^{MN}$.  Unfortunately, this operator does not
lead to interactions involving two KK-radions and two gauge
zero-modes, as pointed out in Ref.~\cite{Medina:2010mu}.  The
remaining finite 1-loop pieces contribute to the dimension-8 operator
quoted above, and result in a negligible effect.

Nevertheless, one can still write 5D operators that do lead to the
$(r')^2\gamma\gamma$ interaction if the stabilizing scalar field
is involved (since the KK-radion is really an admixture of metric
and stabilizing scalar fluctuations).  However, we find that a
more important contribution can arise from an IR localized
operator involving instead the extrinsic curvature.  For instance,
consider the operator $[\delta(y+L) + \delta(y-L)] \sqrt{g_{\rm
ind}} (\eta/8\Lambda_{5}^{3}) K^{2} F_{\mu\nu}F^{\mu\nu}$, where
$\Lambda_{5}$ is the 5D cutoff, the indices are contracted with
the induced metric, and $K = 4[ A'(y) + \partial_{y}F(x,y)]$ is
the trace of the extrinsic curvature, including radion
fluctuations about the background, as parameterized in
Refs.~\cite{Csaki:2007ns,Medina:2010mu}.  Reduction to 4D with the
flat photon wavefunction, $1/\sqrt{2L}$, as well as the KK-radion
wavefunction, which satisfies $F_{0}(L) \approx 1$ and $F_{0}'(L)
\approx 2A'(L)$, leads to the dimension-6 operator~\footnote{We
point out that, in spite of being localized on the IR boundaries,
the above operator leads to a non-vanishing linear coupling
between the radion field and two photons: $\int \!  d^{5}x \,
\sqrt{g} \, F(x,y) \, T \to -\int \!  d^{4}x \, (8 \eta \, k_{\rm
eff}^{2}/L\Lambda_{5}^{3}) (r/\Lambda_{r}) F_{\mu\nu} F^{\mu\nu}$,
where $T$ is the trace of the stress-energy tensor induced by the
operator.  This result holds even if the gauge fields are strictly
IR localized, and therefore can be interpreted as an explicit
breaking of conformal invariance in a dual 4D picture.  }
\bea
- \frac{e^{2} \kappa}{8 \Lambda _{r}^{2}} \, r^{\prime2} F_{\mu\nu} F^{\mu\nu}~,
\label{HigherDimPhotons}
\eea
where $e^2 \kappa \approx 64\eta/(k_{\rm eff} L) \times (k_{\rm
eff}/\Lambda_{5})^{3}$, with $k_{\rm eff} \equiv A'(L)$.  NDA suggests
that the unknown dimensionless coefficient, $\eta$, can be as large as
$3\pi/2$~\cite{Chacko:1999hg}.  Taking also $k_{\rm eff} L \approx 60$
and $\Lambda_{5} \sim 5 k_{\rm eff}$, one estimates $e^{2} \kappa
\approx 0.04$.  Notice that the same operator contains a term without
radion insertions, which renormalizes the electric charge.  However,
this effect is very small and we neglect it in the following.

The interaction of Eq.~(\ref{HigherDimPhotons}) leads to a direct
$\gamma\gamma$ annihilation cross-section given by
\bea
\langle \sigma_{2\gamma} v/c \rangle &\approx&
\left( \frac{m_{\rm DM}}{\Lambda_{r}} \right)^{4}
\frac{3\pi\alpha^{2}\kappa^{2}}{m^{2}_{\rm DM}}~,
\label{XS-photons}
\eea
where $m_{\rm DM} = m_{r'}$, and $\alpha = e^{2}/4\pi$ is the fine
structure constant. In the following, we do not
include the additional $\gamma Z$ channel, nor any mixing with the
Higgs.

The differential photon flux from the galactic center can be written as
\bea
\frac{d\Phi_{\gamma}}{dE} &=& 5.66\times 10^{-12}~{\rm cm^{-2} s^{-1}} \, \frac{d N_{\gamma}}{d E}
\left( \frac{\langle \sigma_{2\gamma} v/c \rangle}{1~{\rm pb}} \right)
\left( \frac{1~{\rm TeV}}{m_{\rm DM}} \right)^{2} \bar{J}(\Delta\Omega) \Delta\Omega~,
\label{PhotonFlux}
\eea
where $d N_{\gamma}/d E = 2 \times \delta(E-m_{\rm DM})$ is the
differential photon yield in the $2\gamma$ annihilation channel,
$\bar{J}(\Delta\Omega) \equiv (1/\Delta\Omega) \int_{\Delta\Omega}
J(\psi) d\Omega$ integrates over the angular acceptance of the
detector $\Delta\Omega$, and $J(\psi)$ is conventionally defined as
\bea
J(\psi) = \frac{1}{8.5~{\rm kpc}} \left( \frac{1}{0.3~{\rm GeV/cm^{3}}} \right)^{2}
\int^{\infty}_{0}dl \rho^{2}(r)~,
\eea
where $r^{2} = l^{2} + r^{2}_{0} - 2 l r_{0} \cos\psi$, with $r_{0}
\approx 8.5~\rm{kpc}$ the distance from the Earth to the galactic
center.  The integration is along the line of sight, $dl$, and encodes
the information about the DM distribution, assuming a spherical DM
halo of energy density $\rho(r)$.  The quantity
$\bar{J}(\Delta\Omega)$ can vary over several orders of magnitude
depending on the halo model.  We will use here the Navarro-Frenk-White
(NFW) profile~\cite{Navarro:1995iw}, but the results for other
profiles can be obtained by simple rescaling.\footnote{It has been
claimed that cored profiles are the rule~\cite{Donato:2009ab}, in
which case the photon flux would be significantly smaller.} We have
also used the standard DM density at the Sun's location of $0.3~{\rm
GeV/cm}^3$, although it has been recently argued that it should be
somewhat larger~\cite{Salucci:2010qr}.  We take $\Delta \Omega =
10^{-5}~{\rm sr}$, as would be relevant for the HESS and Fermi-LAT
$\gamma$-ray experiments.  We then have $\bar{J}(\Delta\Omega)
\Delta\Omega \approx 0.15$.

We take into account the detector energy resolution by convoluting
\bea
\left. \frac{d\Phi_{\gamma}}{dE_{0}} \right|_{\rm measured} &=&
\int \! dE \, G(E,E_{0}) \frac{d\Phi_{\gamma}}{dE}~,
\eea
where, following~\cite{Bertone:2010fn}, we take a Gaussian kernel,
$G(E, E_{0}) = 1/(\sqrt{2 \pi} E_{0} \sigma) \, e^{-(E - E_{0})^2/(2
\sigma^2 E_{0}^2)}$ with $\sigma = \xi/2.3$, and $\xi$ is the
detector's relative energy resolution.
\begin{figure}[t]
\centerline{
\includegraphics[height=5.5cm]{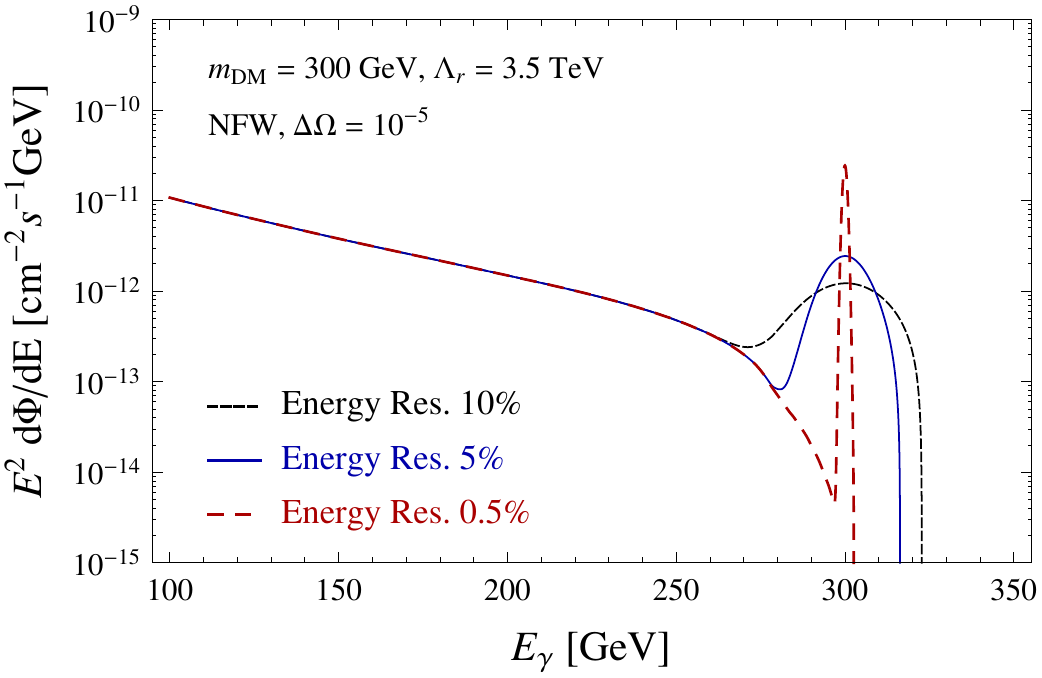}
}
\caption{{\em Continuum photon flux and $2\gamma$-line signal from
KK-radion annihilation at the center of the galaxy.  For the line
signal, we show three detector energy resolutions: $10\%$ (short
dashed, black line), $5\%$ (solid, blue line) and $0.5\%$ (dashed, red
line).}}
\label{fig:PhotonFlux}
\end{figure}
We show the line signal in Fig.~\ref{fig:PhotonFlux} for $m_{\rm DM} =
300~{\rm GeV}$ and $\Lambda_{r} = 3.5~{\rm TeV}$ (the rest of the
parameters are those for the ``strong warping'' benchmark scenario).
We also show in the figure the continuum $\gamma$-ray flux spectrum as
computed by \verb|micrOMEGAs| 2.4~\cite{Belanger:2010gh}, which also
reproduces our result for the DM relic density (and, for the previous
parameters, gives $\Omega_{r'} h^{2} \approx 0.11$).  We see that the
line feature is distinguishable from the continuum spectrum signal,
perhaps even for the current typical detector energy resolutions of
about $10\%$.

It is interesting to compare to UED scenarios~\cite{Cheng:2002ej}.
The continuum photon signal was computed in
Ref.~\cite{Bergstrom:2004cy}, where it was pointed out that the
resulting spectrum is rather flat up to the DM mass.  This is due to
the large branching fraction for direct annihilation of the KK
$B^{(1)}$ into lepton pairs, which results in a hard photon spectrum.
In our case, with annihilations mostly into top pairs, gauge bosons
and the Higgs, the resulting spectrum is softer than in the UED
scenario.  The photon line signal, which arises from a calculable
1-loop effect, was also studied in
UEDs~\cite{Bergstrom:2004nr,Bertone:2010fn}.  It was found that the
line peak could be easily seen provided the detector energy resolution
was small enough.  In the KK-radion scenario, we point out that an
incalculable 5D interaction can also lead to a prominent line signal.
One should also remark that the KK-radion can be lighter than
$B^{(1)}$, which can be very interesting from the point of view of
FERMI-LAT, which has sensitivity for photon energies between about 20
and 300 GeV. We find that the integrated photon flux from the
KK-radion signal, in the above range, is $\Phi_{\gamma} \approx
3\times 10^{-12}~{\rm cm}^{-2}~{\rm s}^{-2}$, which is somewhat more
than an order of magnitude smaller than FERMI-LAT's sensitivity of
about $10^{-10}~{\rm cm}^{-2}~{\rm s}^{-2}$.  However, one should
recall that there are large uncertainties associated with the DM halo
model profile that can change the signal by orders of magnitude.


\medskip
\noindent
\underline{\textit{Collider signals:}}

\noindent Finally, we comment briefly on the collider signals of our
setup.  Regarding the more generic KK phenomenology, which can be
expected to share features of the widely studied RS scenarios
(see~\cite{Davoudiasl:2009cd} for a review), we only note that the
generic 1st KK-level states (which are KK-parity odd) and 2nd KK level
states (which are KK-parity even) can lead to signals both with and
without missing energy.  In this respect, the situation is rather
different from supersymmetric scenarios with R-parity.  In addition,
we have emphasized that KK-parity even and odd states come in
relatively degenerate pairs, which allows to access both types of
states almost simultaneously.  This can be contrasted to the UED case
with KK-Parity: although the first KK-level UED particles always lead
to missing energy signals~\cite{Cheng:2002ab}, and the second level KK
states can lead to pure SM particle final
states~\cite{Cheng:2002ab,Burdman:2006gy}, the latter are about a
factor of 2 heavier than the former in 5D
constructions~\cite{Appelquist:2000nn} (in 6D constructions such as
the ``chiral square'' of Ref.~\cite{Dobrescu:2004zi} this factor is
just $\sqrt{2}$, which can be further diluted due to radiative
effects~\cite{Ponton:2005kx}).  Note also that there are significantly
lighter KK-parity odd fields (discussed next) that mitigate the
requirement of pair-production, so that the reach in KK-parity even
and odd states should be similar.

Our main focus here is on the DM-related collider signals.  Of
particular importance are the KK-parity odd states such as the first
excited top resonance (probably the $SU(2)$ singlet), or perhaps the
``inert'' Higgs doublet.  These states may be expected to be
parametrically lighter than the remaining 1st KK-level states, and
thus provide the most promising portal for $r'$ production.  For
instance, in the ``strong benchmark scenario'' discussed in previous
sections, we have $m_{r'} \sim 300~{\rm GeV}$ and $m_{t'} \sim
600~{\rm GeV}$.  The vector-like quarks have a strong production cross
section at the LHC of a couple
picobarn~\cite{Azuelos:2004dm,Aguilar-Saavedra:2005pv}.  Their main
decay is $t' \to t r'$, thus leading to a $t\bar{t} +
\slash{\!\!\!\!E}_{T}$ signal.  This topology was studied in a
model-independent way in~\cite{Meade:2006dw}.  Given the large
$\slash{\!\!\!\!E}_{T}$, which can be used for tagging, one can
consider the case where both top quarks decay hadronically, thus
allowing for their full reconstruction.  It was shown in the previous
study that a judicious set of cuts and kinematic variables can allow
for a $t'$ discovery at the LHC with less than $10~{\rm fb}^{-1}$, and
in fact that both $m_{t'}$ and $m_{r'}$ can be statistically measured.
Interestingly, within the context of our model, we can have access to
the LKP mass by studying the closely related radion mode, that behaves
in many ways like a Higgs.  For $m_{r} \approx m_{r'} \sim 300~{\rm
GeV}$, the radion will be produced by gluon fusion and has a
significant branching fraction into
$ZZ$~\cite{Csaki:1999mp,Giudice:2000av,Bae:2000pk}, thus allowing for
a clean mass measurement in the fully charged $4l$ channel.  We can
therefore get an interesting hint for the expected degeneracy between
$r$ and $r'$, and the relatively precise knowledge of the radion mass
can be used to study in more detail the properties of the missing
energy signal.

It would also be extremely interesting to measure the radion decay
constant, which sets the total width for both $t'$ and the radion.
For the above masses, and for $\Lambda_{r} \sim 3-4~{\rm TeV}$, as
suggested by the relic density scenario discussed in
Section~\ref{sec:DM}, these widths are of order a fraction of a GeV.
This may be too small to be measured at a hadron collider, but would
be accessible in a lepton machine.  Nevertheless, one might be able to
infer $\Lambda_{r}$ from a measurement of the radion production cross
section.  This could be a first test of the required properties to
account for the DM relic density.  Eventually, we expect that a
compelling case for the identity of DM could be made based on collider
measurements.

In previous sections, we also considered scenarios with radion decay
constants much larger than the TeV scale.  In the coannihilation
scenario of Section~\ref{sec:coannihilation}, with $\Lambda_{r}$ as
large a $10^{4}~{\rm TeV}$, the $t'$ still decays promptly but the
produced tops (and even the W's) are expected to be off-shell.  The
analysis of this signal is expected to be challenging.  It may also be
possible for the $t'$ to decay, via intergenerational mixing, into the
lighter families.  On the other hand, the non-thermal scenario studied
in Section~\ref{sec:sWIMP} can potentially lead to a $t' $ that is
long-lived on collider time scales, if $\Lambda_{r} \gtrsim
10^{10}~{\rm GeV}$.  The same applies to the ``inert'' Higgs doublet
(which has a charged component) if it turns out to be the NLKP. These
stable, charged states would lead to spectacular signals at the LHC.

\section{Conclusions}
\label{sec:conclusions}

In this work we have considered warped extra-dimensional scenarios
with a KK-parity symmetry that makes the lightest KK-parity odd
particle stable and hence a DM candidate.  The LKP is expected to be
the first KK excitation of the radion field, which is assumed to be
stabilized at tree-level by the Goldberger-Wise mechanism.  The radion
mass is expected to be of order a few hundred GeV, but its
interactions -which are controlled by the radion decay constant
$\Lambda_{r}$- can vary over orders of magnitude.  We considered a
number of scenarios.  \smallskip \\
\noindent
\textbf{Thermal KK dark matter}
\begin{enumerate}
\item
The desired KK-radion relic abundance can naturally arise when
$\Lambda_{r}$ is in the multi-TeV range.  EWSB plays an important role
here in that it induces strong mixing in the top KK tower, and as a
result a relatively large effective coupling between the KK-radions
and a $t\bar{t}$ pair that partially offsets the suppression from
$\Lambda_{r}$.

\item The presence of light fermion resonances (e.g. the first KK
excitation of the RH top) can, via coannihilation effects, deplete the
number density of thermal KK radions down to acceptable levels for
even larger radion decay constants of order $10^4~{\rm GeV}$.
Provided the required degree of degeneracy is present (of order
10-15\%), the relevant annihilation cross section may be purely
controlled by the QCD interactions.

\item Alternatively, a small KK-parity odd Higgs component of the DM
candidate, can be rather efficient in enhancing the DM
self-annihilation cross section.  This is due to the relatively strong
interactions of the CP-even (but KK-parity odd) KK Higgs with the
fermion KK states.  If this Higgs component is sizable, one expects
that these KK particles would constitute only a fraction of the
observed DM density.  However, it is possible that a combination of
radion-Higgs mixing and a large radion decay constant can result in
the required annihilation cross-section to obtain the DM relic
abundance via the freeze-out paradigm.

\end{enumerate}
\medskip
\noindent
\textbf{Non-thermal KK dark matter}
\begin{enumerate}
\item[4.]
In UED-like scenarios, where the bulk curvature is small compared to
the 5D Planck mass, the radion decay constant can be as large as the
Planck scale.  If the KK-radion interactions are so weak that the
KK-radions never reach thermal equilibrium, a lower bound on the
radion decay constant can be set from the requirement that these
non-thermal KK radions do not overclose the universe, for a given
reheat temperature.  In such a case, there is a population of
long-lived \textit{radions} and NLKPs that can decay around the time
of Big Bang nucleosynthesis.  Consistency with BBN puts an upper bound
on $\Lambda_{r}$, which controls their lifetimes.  This constraints
implies $\Lambda_{r} \lesssim {\cal O}(10^{15}~{\rm GeV})$, while the
reheat temperature should be around the EW scale or somewhat below.
We also noted that there may be another upper bound on the reheat
temperature, of order TeV, to avoid a dangerous high-temperature
deconfinement/confinement phase transition.  The point is that in the
scenario with very large radion decay constants, the rate of bubble
nucleation of low-temperature vacua would be extremely suppressed,
and furthermore the associated reheat temperature after the transition
may not be high enough to allow for a successful BBN. These
constraints can be avoided if the universe never reheated above the
critical temperature, which can nevertheless be high enough to allow
the non-thermal DM KK-radion picture above to be realized.
\end{enumerate}

We also explored the prospects for direct and indirect detection of
KK-radion DM. If the scenario has a relatively low radion decay
constant, direct detection is feasible and a region of parameter space
is already excluded by XENON100. A small KK-Higgs component is essential.
For indirect detection experiments, annihilation into positrons or
neutrinos is expected to give rather small fluxes (both from direct
annihilations, as well as in decay chains of the dominant
$t\bar{t}$/gauge/Higgs channels).  However, there may be an
interesting photon signal that includes a prominent photon line.  This
line signal can naturally fall in the energy range of sensitivity of
Fermi/LAT, and the total flux may be measurable for cuspy DM halo
models.

We also point out that warped scenarios with KK-parity would lead to
interesting collider signals, both with and without missing energy.
In particular, one may be able to probe the characteristic degeneracy
between the DM candidate and the radion mode (which leads to signals
similar to the Higgs), at least if the first KK-parity odd excitation
of the top is in the expected few hundred GeV range.  The scenarios
with very large radion decay constant can lead to highly ionizing
charged tracks.

\section*{Acknowledgments}

We would like to thank Hsin-Chia Cheng, Hooman Davoudiasl, Piyush
Kumar and Jessie Shelton for discussions.  A.M. is supported by the US
department of Energy under contract DE-FG02-91ER406746.  E.P. is
supported by DOE grant DE-FG02-92ER40699.

\appendix


\section{Radion-fermion interactions and EWSB}
\label{App:RadionFermionGeneral}

The most general interaction Lagrangian of a single $r_{i}$ and two
fermions can be written as
\bea
\mathcal{L}_{\rm r\psi \psi} &=& - \sum_{ijk=0}^\infty \, \frac{r_i}{\Lambda_{r}} \left\{
g_{ijk}^{LL} \, \bar{\psi}_L^{j} \, i \! \stackrel{\leftrightarrow}{\slash{\!\!\!\partial}} \! \psi_L^k +
g_{ijk}^{RR} \, \bar{\psi}_R^{j} \, i \! \stackrel{\leftrightarrow}{\slash{\!\!\!\partial}} \! \psi_R^k
- \left( m^{RL}_{ijk} \,
\bar{\psi}_R^{j} \psi_L^k + \mbox{h.c.} \right) \right\}~,
\label{Lrff}
\eea
where $\bar{\psi} \! \stackrel{\leftrightarrow}{\slash{\!\!\!\partial}} \!\! \chi \equiv \frac{1}{2} \left[ \bar{\psi} \gamma^{\mu} \partial_{\mu} \chi - (\partial_{\mu} \bar{\psi}) \gamma^{\mu} \chi \right]$, the $g_{ijk}^{LL}$ and $g_{ijk}^{RR}$ are dimensionless, and the $m^{RL}_{ijk}$ have dimensions of mass (these coefficients satisfy the symmetry properties $g_{ijk}^{LL} = g_{ikj}^{LL}$, $g_{ijk}^{RR} = g_{ikj}^{RR}$ and $m^{LR}_{ijk} = m^{RL}_{ikj}$). The sum runs over \textit{all} fermionic \textit{mass eigenstates} (as well as the $r_{i}$). In general, the $\psi^j$ are not gauge eigenstates, since the KK towers associated with different bulk fields mix when the Higgs gets a VEV, thus inducing EWSB mixing mass terms of the form
\bea
\mathcal{L}^{\rm EWSB}_{m} &=& - \sum_{jk=0}^\infty
\left(\tilde{m}^{RL}_{jk} \bar{\chi}_{1,R}^{j} \chi_{2,L}^k + {\rm h.c.} \right)~.
\label{EWSBMass}
\eea
Here $\chi_{1}$ and $\chi_{2}$ stand for two different bulk fields in
the gauge eigenbasis (the $SU(2)_{L}$ singlets and doublets of the 5D
SM), while the $\tilde{m}^{RL}_{jk}$ are proportional to the Higgs
VEV, $v$.  For a Higgs localized on the IR brane, these mixing mass
terms are given by $\tilde{m}^{RL}_{jk} = v \, (Y_{5D}/L) \frac{1}{2}
\!  \left[1+(-1)^{j+k} \right] f_{1,R}^j(L) f_{2,L}^k(L)$, where $v =
174~{\rm GeV}$, $f_{1,R}^j$ and $f_{2,L}^k$ are the properly
normalized fermion wavefunctions evaluated on the IR boundary at
$y=L$, and the factor in square brackets reflects the KK parity
symmetry of our setup.  Perturbatively, the 5D Yukawa coupling,
$Y_{5D}$, is related to the 4D Yukawa coupling by $Y_{4D} = (Y_{5D}/L)
f^{0}_{L}(L)f^{0}_{R}(L)$, where $m_{\psi^{0}} = Y_{4D} v$ is the
corresponding SM fermion mass.  However, one should match to the
observed fermion mass after diagonalization of the full KK mass
matrix.  This is especially true in ``anarchic'' scenarios where the
5D Yukawa couplings are large and mixing with the KK states can be
important.

There are linear interactions of the $r_{i}$ that have the same origin
as the EWSB mass terms of Eq.~(\ref{EWSBMass}) and that take the form
\bea
\mathcal{L}^{\rm EWSB}_{r\psi\psi} &=& + \sum_{ijk=0}^\infty \, \frac{r_i}{\Lambda_{r}}
\left[ v \tilde{X}^{RL}_{ijk} \, \bar{\chi}_{1,R}^{j} \chi_{2,L}^k + {\rm h.c.} \right]~.
\label{EWSBRadionff}
\eea
For Higgs doublets localized on the IR branes, one has
$\tilde{X}^{RL}_{ijk} = 2(Y_{5D}/L) \!  \left[1+(-1)^{i+j+k} \right]
\times$ $F_{i}(L) f_{1,R}^j(L) f_{2,L}^k(L)$, where $F_{i}$ is the
$r_{i}$ wavefunction.  The coefficients $v \tilde{X}^{RL}_{ijk}$,
after rotation to the mass eigenbasis, can be assumed to be part of
the $m^{RL}_{ijk}$ of Eq.~(\ref{Lrff}).  The remaining contributions
to the coefficients in Eq.~(\ref{Lrff}) arise from the bulk terms in
the 5D action, and are obtained from the overlap integrals given in
Ref.~\cite{Medina:2010mu},~\footnote{Some representative values were
given in Tables~1 and 2 of that reference.} after properly applying
the unitary transformations that diagonalize the mass matrix,
$m^{RL}_{\rm KK} + \tilde{m}^{RL} = U_{R} \, m_{\rm phys}
U_{L}^{\dagger}$, where $m^{RL}_{\rm KK}$ contains the KK masses,
$\tilde{m}^{RL}$ is the EWSB mass matrix given in
Eq.~(\ref{EWSBMass}), and $m_{\rm phys}$ is the diagonal matrix of
physical masses (chosen to be real and positive).

Also, from Eq.~(\ref{EWSBRadionff}) one can obtain $h{\rm -}r_{i}{\rm
-}\psi^{j}{\rm -}\psi^{k}$ contact interactions involving the Higgs
field, $h$.  In the mass eigenbasis, these are controlled by the
dimensionless couplings
\bea
X^{RL}_{ijk} &=& (U_{R}^{\dagger})_{jn} \, \tilde{X}^{RL}_{inm} ( U_{L} )_{mk}~.
\label{RadionHiggsCoupl}
\eea
with $X^{LR}_{ijk} = X^{RL}_{ikj}$.

For self-annihilation processes, there are also relevant terms
involving two radion modes and a fermion pair.  Following
Ref.~\cite{Medina:2010mu}, we write these as
\bea
\mathcal{L}_{\rm rr\psi \psi} = - \sum_{i_{1} i_{2}=0}^\infty \sum_{jk=0}^\infty \, \frac{3r_{i_{1}} r_{i_{2}}}{2\Lambda^{2}_{r}} \left\{
g_{i_{1} i_{2}jk}^{LL} \, \bar{\psi}_L^{j} \, i \! \stackrel{\leftrightarrow}{\slash{\!\!\!\partial}} \! \psi_L^k +
g_{i_{1} i_{2}jk}^{RR} \, \bar{\psi}_R^{j} \, i \! \stackrel{\leftrightarrow}{\slash{\!\!\!\partial}} \! \psi_R^k
+ \frac{4}{3} \left( m^{RL}_{i_{1} i_{2}jk} \,
\bar{\psi}_R^{j} \psi_L^k + \mbox{h.c.} \right) \right\}~,
\nonumber
\eea
and
\bea
\mathcal{L}^{\rm EWSB}_{rr\psi\psi} &=& -\sum_{i_{1} i_{2}=0}^\infty \sum_{jk=0}^\infty \, \frac{r_{i_{1}} r_{i_{2}}}{\Lambda^{2}_{r}}
\left[ v \tilde{X}^{RL}_{i_{1} i_{2}jk} \, \bar{\chi}_{1,R}^{j} \chi_{2,L}^k + {\rm h.c.} \right]~.
\label{EWSBXX}
\eea
For a Higgs field localized on the IR branes, one has
$\tilde{X}^{RL}_{i_{1} i_{2} jk} = 4(Y_{5D}/L) \!  \left[1+(-1)^{i_{1}
+ i_{2} + j+k} \right] \times$ $F_{i}(L) F_{j}(L) f_{1,R}^j(L)
f_{2,L}^k(L)$, where $F_{i}$ is the $r_{i}$ wavefunction.  In the same
way as for the single radion interactions, the coefficients $v
\tilde{X}^{RL}_{i_{1} i_{2}jk}$, after rotation to the mass
eigenbasis, can be assumed to be part of the $m^{RL}_{i_{1} i_{2}jk}$
above.

The Feynman rules arising from Eq.~(\ref{Lrff}) were given in
Ref.~\cite{Medina:2010mu}.  We find it convenient to express various
decay rates and cross sections involving the radion/KK-radion in terms
of the dimensionless quantities
\bea
G^{RL}_{i_{1} \cdots i_{n} jk} &\equiv& n! \times \frac{M^{RL}_{i_{1} \cdots i_{n} jk}}{m_{j}}~, \hspace{1cm} G^{LR}_{i_{1} \cdots i_{n} jk} ~\equiv~ n! \times \frac{M^{LR}_{i_{1} \cdots i_{n} jk}}{m_{j}}~,
\label{GRL}
\eea
where $m_{j}$ is the physical mass of the $j$-th KK fermion
excitation, and we defined
\bea
M^{RL}_{i_{1} \cdots i_{n} jk} & \equiv& \frac{1}{2} \left( c_{n}g^{LL}_{i_{1} \cdots i_{n} jk} m_{j} + c_{n}g^{RR}_{i_{1} \cdots i_{n} jk} m_{k} \right) - d_{n}m^{RL}_{i_{1} \cdots i_{n} jk}~,
\label{MRL}
\eea
with $c_{n} = (-3)^{n} \left( 2n/3 - 1 \right)/n!$ and $d_{n} =
(-4)^{n-1}/n!$.  One also has $M^{LR}_{i_{1} \cdots i_{n} jk} =
M^{RL}_{i_{1} \cdots i_{n} kj}$.

In the limit that EWSB is neglected, one has
\bea
m^{RL}_{i_{1} \ldots i_{n}jk} &=& 2 g^{LL}_{i_{1} \ldots i_{n}jk} \, m_{j}
+ 2 g^{RR}_{i_{1} \ldots i_{n}jk} \, m_{k} + m^{\rm scalar}_{i_{1} \ldots i_{n}jk}~,
\eea
where $m^{\rm scalar}_{i_{1} \ldots i_{n}jk}$ is a contribution coming
from the Yukawa coupling involving the stabilizing scalar field, which
is in general numerically negligible (the expression is given in
\cite{Medina:2010mu}).  In this case, the $g^{LL}_{i_{1} \ldots
i_{n}jk}$ and $g^{RR}_{i_{1} \ldots i_{n}jk}$ are simply overlap
integrals of $n$ radion modes and two fermion wavefunctions.

\section{Formulas for processes involving the Radion Tower}
\label{App:Processes}

In the following, we use extensively the notation
\bea
\epsilon_{i,j} \equiv \frac{m_{i}}{m_{j}}~.
\label{eps}
\eea
Also, in the following formulas, the $G^{RL}_{ijk}$, $G^{LR}_{ijk}$,
$G^{RL}_{iijk}$ and $G^{LR}_{iijk}$ are as defined in Eqs.~(\ref{GRL})
and (\ref{MRL}), and the $g_{ijk}^{LL}$ and $g_{ijk}^{RR}$ are as
defined in Eq.~(\ref{Lrff}).

\subsection{Self-Annihilations}
\label{Annihiations}

For annihilation into fermion pairs, we are only interested in the
process $r' r' \to ff$, but we give the result for the slightly more
general case $r_{i} r_{i} \rightarrow f^{k} \bar{f}^{k}$.  As shown in
Fig.~\ref{fig:rprpAnnihilationff}, there are two types of diagrams:
those involving a ``heavy'' $f^{j}$ exchange in the $t$- and
$u$-channels, and a contact interaction.  The contribution due to the
$t$- and $u$-channel exchange can be written as
\bea
{\rm v} \sigma^{\rm t+u}_{r_{i} r_{i} \rightarrow f^{k} \bar{f}^{k}} &\approx& \sum_{j} \frac{N_{c}}{16\pi} \,\frac{m^{2}_{i}}{\Lambda_{r}^{4}} \,
\left( G^{\rm eff}_{ijk} \right)^2  \, \left( 1 - \epsilon_{k,i}^{2} \right)^{3/2}~,
\label{sigmarprptt}
\eea
where we defined
\bea
G^{\rm eff}_{ijk} &\equiv& \frac{1}{ \left( 1 + \epsilon_{j,i}^{2} - \epsilon_{k,i}^{2} \right) } \,
\left\{ \rule{0mm}{6mm}
2 \left[ \left( G^{RL}_{ijk} \right)^{2} + \left( G^{LR}_{ijk} \right)^{2} \right] \epsilon_{j,i}^{2} \epsilon_{k,i} + 4 G^{RL}_{ijk} G^{LR}_{ijk} \epsilon_{j,i}^{3} \right.
\nonumber \\[0.5em]
&& \hspace{-2cm}
\mbox{} \left. - \left( 1 + \epsilon_{j,i}^{2} - \epsilon_{k,i}^{2} \right) \left( \frac{1}{2} \left[ \left( g_{ijk}^{LL} \right)^{2} + \left( g_{ijk}^{RR} \right)^{2} \right] \epsilon_{k,i} + \left[ 2 g_{ijk}^{LL} G^{LR}_{ijk} - g_{ijk}^{LL} g_{ijk}^{RR} + 2 g_{ijk}^{RR} G^{RL}_{ijk} \right] \epsilon_{j,i} \right) \right\}~,
\nonumber \\
\label{Geff}
\eea
and $N_{c} = 3$ ($N_{c} = 1$) for quarks (leptons), while the
$\epsilon$'s are the mass ratios defined in Eq.~(\ref{eps}), i.e. we
normalize with respect to the $r_{i}$ mass.  Eq.~(\ref{sigmarprptt})
holds in the deep non-relativistic limit (${\rm v}$ is the relative
velocity of the initial state).

The non-relativistic contribution due to the contact interaction is
\bea
{\rm v} \sigma^{\rm Contact}_{r_{i} r_{i} \rightarrow f^{k} \bar{f}^{k}} &\approx& \frac{N_{c}}{16\pi} \,\frac{m^{2}_{k}}{\Lambda_{r}^{4}} \, \sqrt{1 - \epsilon_{k,i}^2} \,
\left\{ 2 \left[ \left( G^{RL}_{iikk} \right)^{2} + \left( G^{LR}_{iikk} \right)^{2} \right] - \left( G^{RL}_{iikk} + G^{LR}_{iikk} \right)^2 \epsilon_{k,i}^2 \right\}~,
~~
\label{sigmarprpttContact}
\eea
while the crossed-term is given by
\bea
{\rm v} \sigma^{\rm Crossed}_{r_{i} r_{i} \rightarrow f^{k} \bar{f}^{k}} &\approx& - \sum_{j} \frac{N_{c}}{8\pi} \,\frac{m_{i} m_{k}}{\Lambda_{r}^{4}} \,
G^{\rm eff}_{ijk}  \left( G^{RL}_{iikk} + G^{LR}_{iikk} \right) \,
\left( 1 - \epsilon_{k,i}^{2} \right)^{3/2}~.
\label{sigmarprpttCrossed}
\eea

Annihilation into the KK-parity even Higgs degrees of freedom (see
Fig.~\ref{fig:rprpAnnihilationHH}), gives (see~\cite{Medina:2010mu}
for the relevant Feynman rules)
\bea
{\rm v} \sigma_{r_{i} r_{i} \rightarrow HH} &\approx& \frac{N_{H} F_{i}(L)^4}{4 \pi } \,
\frac{m_{i}^{2}}{\Lambda_{r}^4} \frac{\sqrt{1 - \epsilon_{h,i}^2}}{(1 + \epsilon_{H,i}^2 -
\epsilon_{h,i}^2)^2} \, \left[\epsilon_{H,i}^2(2 + 3 \epsilon_{h,i}^2) -
11 \epsilon_{h,i}^2 - 21 \epsilon_{h,i}^4 \right]^2~,
\label{sigmarprpHH}
\eea
where $\epsilon_{H,i} = m_{H}/m_{i}$ and $\epsilon_{h,i} =
m_{h}/m_{i}$, with $m_{H}$ the mass of the ``inert'' Higgs doublet and
$m_{h}$ the mass of the SM-like Higgs.  $N_{H} = 4$ takes into account
the 4 real Higgs d.o.f.~in the limit that $m_{i}$ is large.  The
$F_{i}(L)$ are the KK-radion wavefunctions evaluated on the IR
boundaries, which are order one numbers, and in particular, $F_{r'}(L)
\approx 1$ for the first KK-radion mode.

We also quote the tree-level annihilation cross section into identical
massless gauge bosons, via a massive KK gauge boson exchange (of mass
$m_{j}$) in the t- and u-channels:
\bea
{\rm v} \sigma_{r_{i} r_{i} \rightarrow VV} &\approx& \sum_{j} \frac{g_{ij0}^4}{\pi} \frac{m_{i}^2}{\Lambda_{r}^4} \frac{\epsilon_{i,j}^4}{(1 + \epsilon_{i,j}^2)^2} \left(4 + 2 \epsilon_{i,j}^2 + \epsilon_{i,j}^4 \right)~,
\label{sigmarprpVV}
\eea
where $\epsilon_{i,j} = m_{i}/m_{j}$, and $g_{ij0}$ is the overlap
integral of the wavefunctions for a radion, a massive gauge KK mode
and a massless gauge boson.  This dimensionless coupling is of order
$1/\sqrt{2k_{\rm eff}L}$.

\subsection{Co-Annihilations}
\label{Coannihiations}

For coannihilations with the NLKP, involving gluons in the final
state, we are interested in the process $r' t' \to g t$, but we give
the result for the more general case $r_{i} t^{j} \rightarrow g
t^{k}$.  Including the $s$- and $t$-channel diagrams, as well as a
quartic contact interaction that is required by gauge invariance (see
Ref.~\cite{Medina:2010mu}), we have:
\bea
{\rm v} \sigma_{r_{i} t^{j} \rightarrow g t^{k}} &\approx& \frac{(N^{2}_{c} - 1) \alpha_{s}}{16N_{c}\Lambda_{r}^{2}} \, \left[ \left( G^{RL}_{ijk} \right)^{2} + \left( G^{LR}_{ijk} \right)^{2} \right] \, \left( \frac{1}{ 1 + \epsilon_{j,i} } \right) \left[ 1 - \left( \frac{ \epsilon_{k,i} }{ 1 + \epsilon_{j,i} } \right)^{2} \right]~,
\label{vsigmaritjTogtk}
\eea
which holds in the ultra non-relativistic limit (${\rm v}$ is the
relative velocity of the initial state).  The $\epsilon$'s are the
mass ratios defined in Eq.~(\ref{eps}), i.e. we normalize with respect
to the $r_{i}$ mass.

\medskip
There are also channels with a Higgs in the final state.  We are
interested in $r' t' \to h t$, but we quote the more general case (in
the ultra non-relativistic limit, and neglecting a subdominant diagram
with a $s$-channel $t^{k}$):
\bea
{\rm v} \sigma_{r_{i} t^{j} \rightarrow h t^{k}} &\approx& \frac{1}{64\pi \Lambda_{r}^{2}} \, \frac{\sqrt{\left[ \left( 1 + \epsilon_{j,i} + \epsilon_{h,i} \right)^{2} - \epsilon_{k,i}^{2} \right]
\left[ \left( 1 + \epsilon_{j,i} - \epsilon_{h,i} \right)^{2} - \epsilon_{k,i}^{2} \right]}}{(1+\epsilon_{j,i})^{3}} \times
\nonumber \\[0.5em]
& & \hspace{-1cm}
\left\{ \rule{0mm}{5mm} \!
\left[ (X_{ijk}^{RL})^{2} + (X_{ijk}^{LR})^{2} \right] \left[ \left( 1 + \epsilon_{j,i} \right)^{2} + \epsilon_{k,i}^{2} - \epsilon_{h,i}^{2} \right]
+ 4 X_{ijk}^{RL} X_{ijk}^{LR} \, \epsilon_{k,i} \left( 1 + \epsilon_{j,i} \right)
\right\}~,
\label{vsigmaritjTohtk}
\eea
where $X_{ijk}^{RL}$ and $X_{ijk}^{LR}$ were defined in
Eq.~(\ref{RadionHiggsCoupl}).  The $\epsilon$'s are the mass ratios
defined in Eq.~(\ref{eps}), and we also defined $\epsilon_{h,i} \equiv
m_{h}/m_{i}$, where $m_{h}$ is the Higgs mass (i.e. we normalize all
the masses to the $r_{i}$ mass).

\subsection{$f^j$ Decays}
\label{tpDecays}

The production of $r_{i}$ in $f^{j}$ decays is governed by
\bea
\Gamma^{\rm CM}_{f^j \to r_{i} f^{k}} &=&
\frac{m^3_{j}}{16 \pi \Lambda_{r}^2} \, \sqrt{ \rule{0mm}{3.5mm}
\left[1 - (\epsilon_{k,j} + \epsilon_{i,j})^2 \right] \left[1 - (\epsilon_{k,j} - \epsilon_{i,j})^2 \right] }
\nonumber \\[0.5em]
& &
\times \left\{\frac{1}{2} \left[ (G_{ijk}^{RL})^{2} + (G_{ijk}^{LR})^{2} \right] \left( 1 + \epsilon_{k,j}^{2} - \epsilon_{i,j}^{2} \right) + 2 G_{ijk}^{RL} G_{ijk}^{LR} \, \epsilon_{k,j} \right\}~,
\label{GammaCMtj}
\eea
where the $\epsilon$'s are the mass ratios defined in Eq.~(\ref{eps}),
i.e. we normalize with respect to the mass of the decaying particle.

\subsection{$V^j_{\mu}$ Decays}
\label{GpDecays}

The production of $r_{i}$ in $V_{\mu}^j$ decays, when the final gauge
boson has mass $m_{k} \neq 0$, is governed by
\bea
\Gamma^{\rm CM}_{V^j_{\mu} \to V^k_{\mu} r_i} \approx
\frac{(G^{\rm eff,V}_{ijk})^2 m_{j}^{5}}{192\pi m_{k}^{2} \Lambda_r^2} \sqrt{ \rule{0mm}{3.5mm}
\left[1 - (\epsilon_{k,j} + \epsilon_{i,j})^2 \right] \left[1 - (\epsilon_{k,j} - \epsilon_{i,j})^2 \right]} \left[
\left(1 + \epsilon_{k,j}^2 - \epsilon_{i,j}^2\right)^2 + 8 \epsilon_{k,j}^2 \right]~,
\label{GammaCMgj}
\eea
where
\bea
G^{\rm eff,V}_{ijk} &\equiv& 4g^m_{ijk} \epsilon_{k,j} + g_{ijk} \left( 1 + \epsilon_{k,j}^{2} - \epsilon_{i,j}^{2} \right)~,
\eea
and
\bea
g_{ijk}=\frac{1}{2L}\int^{L}_{-L} \! dy \, F_{i}f^j_{V}f^{k}_{V}~,
\hspace{1cm}
g^m_{ijk}=\frac{1}{2L}\int^{L}_{-L} \! dy \, e^{-2A} F_{i}f^j_{5}f^{k}_{5}~,
\label{VVrpgijk}
\eea
with $f^{j}_{5}(y)$ the wavefunction corresponding to $V^{j}_{5}$
(which is related to the $V^{j}_{\mu}$ wavefunction by
$f_5^{j}=\partial_y f^j_{V}/m_j$).

When $m_{k} = 0$, we have instead
\bea
\Gamma^{\rm CM}_{V^j_{\mu} \to V^0_{\mu} r_i} &\approx&
\frac{g_{ijk}^{2}m_{j}^{3}}{16\pi\Lambda_{r}^{2}} \left( 1 - \epsilon_{i,j} \right)^{3}~,
\eea
where $g_{ijk}$ is suppressed by order $1/\sqrt{2k_{\rm eff}L}$ due to
the massless wavefunction.

\subsection{KK-parity odd Higgs Decays}
\label{HDecays}

The states in the KK-parity odd Higgs doublet can decay into $r'$
plus a KK-parity even Higgs.  Parameterizing the KK-parity odd and
even Higgs doublets as $H_{-} = \{H^{+}, \frac{1}{\sqrt{2}} (h_{-}
+ i a)\}$ and $H_{+} = \{G^{+}, \tilde{v} + \frac{1}{\sqrt{2}}
(h_{+} + i G^{0})\}$, respectively, where $\tilde{v} = 174~{\rm
GeV}$, $h_{+}$ is the SM-like Higgs and $G^{0}$, $G^{\pm}$ are the
would-be Nambu-Goldstone bosons, we have
\bea
\Gamma^{\rm CM}_{h_{-} \to h_{+} r'} &=&
\frac{F_{r'}(L)^2 m_{h_{-}}^{3}}{16\pi \Lambda_r^2}
\left(1 - 7 \epsilon_{h_{+},h_{-}}^2 - \epsilon_{r',h_{-}}^2\right)^2~,
\nonumber \\[0.5em]
&& \mbox{} \times
\sqrt{ \rule{0mm}{3.5mm}
\left[1 - (\epsilon_{h_{+},h_{-}} + \epsilon_{r',h_{-}})^2 \right] \left[1 - (\epsilon_{h_{+},h_{-}} - \epsilon_{r',h_{-}})^2 \right]}
\label{Gammahm}
\eea
for the CP-even Higgs, and
\bea
\Gamma^{\rm CM}_{a \to G^{0} r'} &=&
\frac{F_{r'}(L)^2 m_{a}^{3}}{16\pi \Lambda_r^2}
\left(1 - \epsilon_{r',a}^2\right)^3~,
\hspace{1cm}
\Gamma^{\rm CM}_{H^{\pm} \to G^{\pm} r'} ~=~
\frac{F_{r'}(L)^2 m_{H^{\pm}}^{3}}{16\pi \Lambda_r^2}
\left(1 - \epsilon_{r',H^{\pm}}^2\right)^3
\label{GammaHpma}
\eea
for the CP-odd and charged Higgses. Here $F_{r'}(L) \approx 1$.

\subsection{Production of $r_{i}$ through Scattering}
\label{Scattering}

The $r_{i}$ production cross section (times flux) via the process
$gq^{j} \to r_{i} q^{k}$ (which includes $s$- and $t$-channel
diagrams, as well as contact interactions required by gauge
invariance) can be written as
\bea
F \sigma_{gq^{j} \to r_{i} q^{k}} &=& \frac{\alpha_{s}}{64 N_{c}} \, \frac{m_{j}^{2}}{\Lambda_{r}^{2}} \, \frac{\theta(z,\Delta \epsilon, \epsilon_{k,j})}{(1-z)^2} \times
\nonumber \\[0.5em]
& & \left\{ \rule{0mm}{5mm}
\left[ (G_{ijk}^{RL})^{2} + (G_{ijk}^{LR})^{2} \right] {\cal F}_{1}(z,\Delta \epsilon, \epsilon_{k,j}) + G_{ijk}^{RL} G_{ijk}^{LR} \, {\cal F}_{2}(z,\Delta \epsilon, \epsilon_{k,j}) \right\}~,
\label{scatteringXS}
\eea
where
\bea
\theta(z,\Delta \epsilon, \epsilon_{k,j}) &=& \sqrt{(1 + \Delta \epsilon \, z)^2 - 4 \epsilon_{k,j}^{2} z}~,
\\[0.3em]
{\cal F}_{1}(z,\Delta \epsilon, \epsilon_{k,j}) &=& 4 \left[ 1 + 2 \Delta \epsilon \, z + \left( 1 + 2 \Delta \epsilon + 2 \Delta \epsilon^2 \right) z^{2} \right] \frac{\tan^{-1} \! \left[ \frac{\theta(z,\Delta \epsilon, \epsilon_{k,j})}{1 + \Delta \epsilon \, z} \right]}{\theta(z,\Delta \epsilon, \epsilon_{k,j})}
\nonumber \\[0.5em]
& & \mbox{} - \left[3 + (2 + 7 \Delta\epsilon) z + (3 + 2 \Delta\epsilon) z^2 - \Delta\epsilon \, z^3 \right]~,
\\[0.5em]
{\cal F}_{2}(z,\Delta \epsilon, \epsilon_{k,j}) &=& 32 \epsilon_{k,j} \, z \left[ (1 + \Delta \epsilon \, z) \frac{\tan^{-1} \! \left[ \frac{\theta(z,\Delta \epsilon, \epsilon_{k,j})}{1 + \Delta \epsilon \, z} \right]}{\theta(z,\Delta \epsilon, \epsilon_{k,j})} - 1 \right]~.
\eea
Here $z = m^{2}_{j}/s$, with $\sqrt{s}$ the CM energy, $\Delta
\epsilon \equiv \epsilon_{k,j}^{2} - \epsilon_{i,j}^{2}$, and the
$\epsilon$'s are the mass ratios defined in Eq.~(\ref{eps}), i.e. we
normalize with respect to the incident $t^{j}$ mass.  The flux factor
is given in the CM frame by $F = \sqrt{s} |\vec{p}|$, where
$|\vec{p}|$ is the momentum of any of the initial state particles, and
replaces the relative velocity appropriate for the non-relativistic
scattering of massive particles.

The singularity at $z = 1$ in Eq.~(\ref{scatteringXS}) arises from the
$s$-channel diagram, and is regularized by the $q^{j}$ width,
$\Gamma_{j}$.  For instance, in the non-relativistic regime, $z
\approx 1$, we have
\bea
F \sigma_{gq^{j} \to r_{i} q^{k}} &\approx& \frac{\alpha_{s}}{8 N_{c}} \, \frac{m_{j}^{2}}{\Lambda_{r}^{2}} \, \frac{m^{4}_{j}}{(s-m^{2}_{j})^2 + m^{2}_{j} \Gamma_{j}^{2} }
\left[ (1 + \Delta \epsilon) \, \frac{\tan^{-1} \! \left[ \frac{\theta(1,\Delta \epsilon, \epsilon_{k,j})}{1 + \Delta \epsilon} \right]}{\theta(1,\Delta \epsilon, \epsilon_{k,j})} ~ - ~ 1 \right]
\nonumber \\[0.5em]
& & \mbox{} \times \theta(1,\Delta \epsilon, \epsilon_{k,j}) \,
\left\{ \rule{0mm}{5mm}
\left[ (G_{ijk}^{RL})^{2} + (G_{ijk}^{LR})^{2} \right] (1 + \Delta \epsilon) + 4 G_{ijk}^{RL} G_{ijk}^{LR} \, \epsilon_{k,j} \right\}~.
\eea
In the ultra-relativistic regime, Eq.~(\ref{scatteringXS}) reduces to
\bea
F \sigma_{gq^{j} \to r_{i} q^{k}} &\approx& \frac{\alpha_{s}}{64 N_{c}} \, \frac{m_{j}^{2}}{\Lambda_{r}^{2}} \left[ (G_{ijk}^{RL})^{2} + (G_{ijk}^{LR})^{2} \right] \left[ 2\log \left( \frac{s}{m^{2}_{k}} \right) - 3 \right]~.
\eea
%

\subsection{Radion Decays}
\label{RadionDecays}

We also summarize the partial decay widths for the radion (KK-parity
even) decays into fermion pairs, massive and massless gauge bosons,
and the CP-even Higgs.  These are used in Section~\ref{sec:results}.
For the decays into fermion pairs, we have
\bea
\Gamma_{r \to f \bar{f}} &=& \frac{m_{r} m_{f}^2}{16 \pi \Lambda_{r}^2} \sqrt{
 1 - 4 \epsilon_{f,r}^2} \left\{ \left[ \left(G^{RL}_{000}\right)^2 + \left(G^{LR}_{000}\right)^2 \right] \left( 1 - 2 \epsilon_{f,r}^2 \right) - 4 G^{RL}_{000} G^{LR}_{000} \, \epsilon_{f,r}^2 \right\}~,
\label{Gammarff}
\eea
where $\epsilon_{f,r} = m_{f}/m_{r}$, and a color factor $N_{c}$
should be included for quarks.

For the decays into (indistinguishable) massive gauge bosons, we have
\bea
\Gamma^{\rm massive}_{r \to VV} &\approx& \frac{F_{0}(L)^2 m_{r}^3}{32 \pi \Lambda_{r}^2} \sqrt{
 1 - 4 \epsilon_{V,r}^2} \left(1 - 4 \epsilon_{V,r}^2 + 12 \epsilon_{V,r}^4 \right)~,
\label{GammarVV}
\eea
where $\epsilon_{V,r} = m_{V}/m_{r}$, and $F_{0}(L) \approx 1$ is the
wavefunction of the radion evaluated on the IR brane.  Here we have
assumed that the Higgs (hence EWSB) is IR-localized, and neglected a
subdominant contribution from the bulk, present even for massless
gauge bosons.  The decay rate into massless gauge bosons is
\bea
\Gamma^{\rm massless}_{r \to VV} &=& \frac{g_{000}^2 m_{r}^3}{16 \pi \Lambda_{r}^2}~,
\label{GammarVVmassless}
\eea
where $g_{000} \approx 1/(2 k_{\rm eff} L)$ is the overlap integral
for a radion and two massless gauge boson wavefunctions (the latter
being flat).

For the decays into the CP-even Higgs, $h_{+}$, we have
\bea
\Gamma_{r \to hh} &=& \frac{F_{0}(L)^2 m_{r}^3}{32  \pi \Lambda_{r}^2} \sqrt{
 1 - 4 \epsilon_{h,r}^2} \left(1 + 2 \epsilon_{h,r}^2\right)^2~,
\label{Gammarhh}
\eea
where $\epsilon_{h,r} = m_{h}/m_{r}$, and a factor of $1/2$ for
identical particles has been included.  A similar expression applies
for the decay into the KK-parity odd Higgses, provided these channels
are kinematically open.


\end{document}